\documentclass[12pt]{article}

\usepackage{graphicx}
\usepackage{amssymb,xfrac}
\usepackage{amsthm,amsmath,dsfont}

\usepackage{lineno}

\usepackage[colorlinks=true, citecolor=blue]{hyperref}
\usepackage[algoruled,titlenotnumbered]{algorithm2e}
\usepackage{algorithmic}
\usepackage{wrapfig}
\usepackage[numbers]{natbib}
\usepackage{comment}
\usepackage{subfig}
\usepackage[left=0.9 in, right=0.9 in, top=1.0 in, bottom=1.0 in]{geometry}

\usepackage{float}
\usepackage{subfig}
\usepackage{wrapfig}
\usepackage{framed,xcolor}
\usepackage{newfloat}
\usepackage[xcolor,framemethod=TikZ]{mdframed}
\usepackage{stmaryrd}
\usepackage{mathtools}

\newcommand{\RR}{\mathbb{R}}

\newcommand{\EE}{\mathbb{E}}
\newcommand{\PP}{\mathbb{P}}

\newcommand{\dd}{\mathrm{d}}

\newcommand{\ff}{\mathrm{f}}

\theoremstyle{definition}

\mdfdefinestyle{myStyle}{roundcorner=5pt,backgroundcolor=brown!20,linecolor=red!40!black,linewidth=1.5pt}

\graphicspath{{./figures/}}

\begin{document}
	
		
		
		\title{Traffic state estimation using stochastic Lagrangian dynamics 
			\\ \vspace{0.1 in} \normalsize Fangfang Zheng$^\mathrm{a}$, Saif Eddin Jabari$^\mathrm{b,c,}$\footnote{Corresponding author, Email: \url{sej7@nyu.edu}} , Henry X. Liu$^\mathrm{d}$, DianChao Lin$^\mathrm{b,c}$}
		



\author{\small $^\mathrm{a}$ School of Transportation and Logistics, Southwest Jiaotong Univeristy, 
	\\ \small No. 111, Erhuanlu Beiyiduan, Chengdu 610031, P.R. China
	\\ \small $^\mathrm{b}$ Division of Engineering, New York University Abu Dhabi, Saadiyat Island, 
	\\ \small P.O. Box 129188, Abu Dhabi, U.A.E.
	\\ \small $^\mathrm{c}$ Tandon School of Engineering, New York University, New York, U.S.A.
	\\ \small $^\mathrm{d}$ Department of Civil and Environmental Engineering, 
	\\ \small University of Michigan Ann Arbor, 2350 Hayward,
	\\ \small 2116 GG Brown, Ann Arbor, Michigan 48109-2125, U.S.A.
}
\date{}

\maketitle		
		
\begin{abstract}
	This paper proposes a new stochastic model of traffic dynamics in Lagrangian coordinates.  The source of uncertainty is heterogeneity in driving behavior, captured using driver-specific speed-spacing relations, i.e., parametric uncertainty.  It also results in smooth vehicle trajectories in a stochastic context, which is in agreement with real-world traffic dynamics and, thereby, overcoming issues with aggressive oscillation typically observed in sample paths of stochastic traffic flow models.  We utilize ensemble filtering techniques for data assimilation (traffic state estimation), but derive the mean and covariance dynamics as the ensemble sizes go to infinity, thereby bypassing the need to sample from the parameter distributions while estimating the traffic states.  As a result, the estimation algorithm is just a standard Kalman-Bucy algorithm, which renders the proposed approach amenable to real-time applications using recursive data.  Data assimilation examples are performed and our results indicate good agreement with out-of-sample data. 
	
	\vspace{0.1in}
	
	\noindent \textit{Keywords}: Lagrangian coordinates; heterogeneous drivers; car following; mean dynamics; variability; hydrodynamic limits; uncertainty quantification; data assimilation; traffic state estimation; Kalman filtering
\end{abstract}
		
			
		
	
	
	
\section{Introduction}
\label{S:intro}
Efficient traffic operation and optimization require knowledge of prevailing traffic conditions. The Lighthill, Whitham and Richards traffic flow model \citep{lighthill1955kinematic,richards1956shock} (the LWR model) has been widely applied in estimation and prediction of traffic states on both freeways and high-speed intersections. The model is formulated using traditional spatial-temporal (Eulerian) coordinates and is suitable for state estimation with point sensor measurements (macroscopic data, e.g., traffic volume, speeds).  Data from probe vehicles or connected vehicles (microscopic data, e.g., vehicle trajectories) are becoming increasingly available. Traffic flow models that are able to effectively utilize such data are of greater interest in modern applications. A simple way of interfacing between the microscopic and the macroscopic worlds is via coordinate transformations. Indeed, this was done by Daganzo \citep{daganzo2005variationala,daganzo2005variationalb} and later extended by \cite{leclercq2007lagrangian}. The former proposes a variational formulation of the LWR model in Eulerian coordinates while the later proposes to formulate the model in Lagrangian coordinates. More recently, Hamilton-Jacobi based formulations of traffic flow have appeared in the literature \citep{claudel2010lax,friesz2013dynamic} and \cite{laval2013hamilton} applied the theory to formulate first-order models in three different coordinate systems, namely the traditional Eulerian coordinates and two variants of the Lagrangian coordinates.  Proposed solutions schemes for the deterministic Lagrangian models include both variational techniques and the Godunov scheme using a triangular fundamental diagram.  Specifically, the Godunov scheme in Lagrangian coordinates simplifies to an upwind scheme, enabling more efficient application of data assimilation methods \citep{duret2017traffic,laval2013hamilton,yuan2012real}.
	
Though deterministic traffic flow models and their solution methods have been extensively studied in the literature, stochastic models of traffic flow are still in a burgeoning stage of development and are primarily extensions of existing deterministic models.  For example, stochastic extensions of the cell transmission model \citep{daganzo1994cell,daganzo1995cell} have been proposed \citep{sumalee2011stochastic,jabari2012stochastic}; other approaches have extended the link transmission model \citep{yperman2007link}, both at the individual link level and the network level \citep{osorio2011dynamic,osorio2014capturing,osorio2017analytical,lu2017probabilistic}.  In general, there still remain issues related to the physical accuracy of the sample paths of existing stochastic traffic models, particularly those developed for purposes of traffic state estimation (see \citep{seo2017traffic,wang2017comparing} for recent reviews).  The main culprit is the dominance of time-stochasticity (or noise) in the stochastic models, mostly developed in Eulerian coordinates  \citep{gazis1971line,szeto1972application,munoz2003traffic,gazis2003kalman,wang2005real,boel2006compositional,wang2007real,work2008ensemble,di2010hybrid,sumalee2011stochastic,blandin2012sequential,jabari2013stochastic}, but also in Lagrangian coordinates \citep{yuan2012real,yuan2015mesoscopic,chu2016stochastic}.  This results in sample paths prone to aggressive oscillation in the time dimension.  The interpretation of these oscillations is (unreasonably) aggressive acceleration and deceleration dynamics.
	
This paper addresses the physical relevance issue of stochastic traffic dynamics via a new stochastic Lagrangian model of traffic flow. The source of uncertainty in the model is parametric in the same sense presented in \citep{2014probabilistic}.  The interpretation of this form of uncertainty is heterogeneity in the driving population. We utilize a stochastic version of Newell-Franklin speed spacing relation \citep{newell1961nonlinear,del1995functional}.  Unlike Newell’s simplified relation \citep{newell2002simplified}, we can derive a unique inverse function, which can be used in data assimilation applications.   Using parametric uncertainty, the sample paths of the stochastic process are smooth and do not contain the oscillatory behavior above.  Our analysis substantially extends and expands our previous work \citep{jabari2018stochastic}.

The paper focuses on application of the proposed model for traffic state estimation (TSE), which is a precursor to a variety of traffic management applications.  TSE is the fundamental tool providing situational awareness, particularly when data availability is limited.  In this context, non-linearity of traffic models renders the state estimation problem particularly challenging.  In theory, one utilizes sampling techniques (e.g., ensemble filters, particle filter, etc.).  These approaches are time consuming and cannot be applied in real-time.  To address this issue, we \textit{derive} the mean and covariance dynamics in a way that preserves the dependencies (i.e., richness) in the model, while allowing for use of standard Kalman filtering techniques.  The latter are known to be computationally tractable and amenable to real-time applications. 


This paper is organized as follows: \autoref{SS:motivation} discusses the motivation of this research. \autoref{S:model} presents the Newell’s speed-spacing relation with heterogeneous drivers along with the stochastic version of this relation.  We interpret the stochasticity as uncertainty about the driver characteristics using driver-specific stochastic parameters.  In \autoref{S:stoch}, we derive the mean and covariance dynamics of the stochastic system by applying ensemble averaging and then derive the dynamics of a deviation process, which serves as a (second-order) Gaussian approximation.  \autoref{S:estimation} demonstrates how the proposed stochastic model can be utilized in data assimilation, that is, to estimate missing information when only limited vehicle trajectory data is available. \autoref{S:numerics} presents numerical examples to show the estimation performance both on the individual level (spacing dynamics, position trajectories) and the aggregated level (queue length, speed dynamics and density dynamics). \autoref{S:conc} concludes the paper.

\section{Motivation}\label{SS:motivation}
We interpret the stochasticity in traffic flow models as one that describes uncertainty about the vehicle/driver attributes.  This type of uncertainty arises in situations where data is limited and/or noisy, e.g., when there are low probe vehicle penetration rates.  In such situations, one combines data that \textbf{is} available with models of traffic flow to fill the gaps.  The combination of data and models can be (heuristically) thought of as taking a weighted average of the two.  More weight is assigned to the predictor with lower uncertainty and vice versa.

\textbf{Example}: When the dynamics involve linear mappings, the Kalman filter is known to produce optimal solutions (both in terms of producing posterior probabilities and least squares estimates).  The Kalman gain matrix plays the role of the weight used to combine a prediction produced by the model with the measurements.  The main ingredients used to calculate the Kalman gain are the state covariance matrix (representing model uncertainty) with measurement covariance (representing measurement error).

In the context of data assimilation, there are two sources of challenges:
\begin{enumerate}
	\item Uncertainty about traffic dynamics depends on the traffic state.  The variance in a vehicle's position depends not only on their own state, but also on the positions (and speeds) of adjacent vehicles, particularly the leader.  These types of dependencies need to be considered when assessing the uncertainty about the dynamics to produce accurate estimates. 
	\item  Traffic flow models are non-linear.  This dictates the use of estimation techniques that rely on sampling to produce the estimates (e.g., ensemble Kalman filtering).  These techniques can be computationally cumbersome and preclude real-time applications.
\end{enumerate}

We address the first challenge by identifying the source of uncertainty in the model.  This is the role played by the proposed Lagrangian model with a stochastic speed-spacing relation.  The majority of traffic state estimation (TSE) papers in the literature provide little or no guidance on how to model variability (namely, model covariance).  In contrast, we provide an approach that bases the variability on uncertainty about simple driver/vehicle primitives. This endows stochasticity in the model with a physical interpretation that can be quantified.  

To address the second challenge, we carefully derive a surrogate stochastic model that is amenable to standard estimation techniques.  Essentially, we derive the mean and covariance dynamics corresponding to our Lagrangian model.  
We apply a (functional) law of large numbers to an ensemble of our Lagrangian dynamics to obtain a mean relation and a (functional) central limit theorem to obtain a model of the deviation of the ensemble from the mean, representing the covariance dynamics of the system.  These can be thought of as generalizations of the standard law of large numbers and central limit theorem.  While the latter are applied to ensembles (a.k.a. sequences) of independent and identically distributed (i.i.d.) scalar random variables, in our case the ensemble is a group of independent identical random processes, each describing the evolution of the trajectories of a group/platoon of vehicles.  These derivations are carried out in order to preserve the dependence on state both in the mean and the covariance/deviation dynamics and in order to ensure the correct approximation is used.  

\textbf{Remark}. It is notable that these types of approximations, which apply functional laws of large numbers and functional central limit theorems are widely applied in the queueing systems literature.  However, they were originally pioneered in the mid-1960s for traffic operations problems by Gordon F. Newell \citep{newell1965approximation}!  To the best of our knowledge, it was in fact Gordon F. Newell who coined the terms \textit{fluid} and \textit{diffusion} approximations, which correspond, respectively, to the mean and covariance dynamics in this paper. 

\section{The traffic dynamics}
\label{S:model}

\subsection{Heterogeneous model}
\label{SS:hetDynamics}
We assume a discrete system with $N + 1$ vehicles numbered in descending order of position; that is vehicle $n=0$ is the leader, $n=1$ is the immediate follower, and so on.   We assume a finite time horizon $T< \infty$ and that time is continuous (i.e., $T \in \RR_+$).  Let $x_n(t)$ and $v_n(t)$ denote the position and speed of vehicle $n$ at time $t \in [0,T]$, respectively.  We denote the spacing between vehicle $n$ and their leader, $n-1$, by 
\begin{linenomath*} 
\begin{align}
	s_n(t) \equiv x_{n-1}(t) - x_n(t).  
\end{align} 
\end{linenomath*}

Heterogeneity in the driver population is represented by driver-specific speed-spacing relations.  Without loss of generality, we adopt the Newell-Franklin (stationary) speed-spacing relation \citep{newell1961nonlinear,del1995functional}:
\begin{linenomath*} \begin{align}
	V_n\big( s \big) = v_{n,\ff}  - v_{n,\ff} e^{\frac{-c_n}{v_{n,\ff}}\left(s - d_n \right)}, \label{E:hetV}
\end{align} \end{linenomath*}
where the driver-specific parameters $(v_{n,\ff}, d_n,c_n)$ represent driver $n$'s desired (free-flow) speed, minimum safety distance, and the constant $c_n$ is the inverse of the reaction time of driver $n$ when their speed is restricted by the trajectory of their leader. In addition to the properties discussed in \citep{del1995functional}, this choice is inspired by the unique inverse function\footnote{Opposed, for instance, to Newell's simplified relation \citep{newell2002simplified}.}, which can be used in data assimilation applications.  The inverse is given by:
\begin{linenomath*} \begin{align}
	S_n\big(v\big) = d_n - \frac{v_{n,\ff}}{c_n} \log_e \left( 1 - \frac{v}{v_{n,\ff}} \right). \label{E:hetS}
\end{align} \end{linenomath*}

The position dynamics are given, for any $n$, by
\begin{linenomath*} \begin{align}
	x_n(t) = x_n(0) + \int_0^t v_n(\tau) \dd \tau
\end{align} \end{linenomath*}
and utilizing the speed spacing relation, we write
\begin{linenomath*} \begin{align}
	x_n(t) = x_n(0) + \int_0^t V_n\big( s_n(\tau) \big) \dd \tau.
\end{align} \end{linenomath*}
Hence, the spacing dynamics evolve according to 
\begin{linenomath*} \begin{align}
	s_n(t) = s_n(0) + \int_0^t \Big(V_{n-1}\big( s_{n-1}(\tau) \big) - V_n\big( s_n(\tau) \big) \Big) \dd \tau.
\end{align} \end{linenomath*}
The spacing dynamics can be simulated using the following recursion:
\begin{linenomath*} \begin{equation}
	s_n(t+ \Delta t) = s_n(t) + \Delta t \Big( V_{n-1}\big( s_{n-1}(t) \big) - V_n\big( s_n(t) \big) \Big).
\end{equation} \end{linenomath*}
In settings with homogeneous drivers, in which $(v_{n,\ff},d_n,c_n) = (v_{0,\ff},d_0,c_0)$ for all $n \ge 1$, $\Delta t$ is chosen so as to ensure no violations of the Courant-Friedrichs-Lewy (CFL) condition, i.e., $\Delta t \le \Delta n / c_0$ and to mitigating numerical diffusion, one chooses the largest such time discretization: $\Delta t = \Delta n / c_0$.

\subsection{Parametric uncertainty and stochastic dynamics}
\label{SS:parametricModel}
To introduce stochasticity, we let the parameters be random variables.  We interpret this as uncertainty about the driver characteristics.  To differentiate the stochastic case from the deterministic case, we write the (stochastic) parameters as functions of $\omega$, where $\Omega \ni \omega$ is the random space.  We assume the random triples (the parameters) 
constitute $n$ independent draws from identically distributed joint distributions.  That is, we define the parameter vector $\theta(\omega) \equiv (v_{\ff}, d,c)(\omega)$ with joint distribution function $F_{\theta}$ and the parameter tuple for each driver $n$, $\theta_n = (v_{n,\ff} , d_n, c_n)$, is drawn independently from this common distribution: $\theta_n \sim F_{\theta}$.  The stochastic speed-spacing relation is given by\footnote{We will use $\omega$ to distinguish between the stochastic relation and the deterministic ones $\{V_n(\cdot)\}_n$.}: 
\begin{linenomath*} \begin{align}
	V(s,\omega) = v_{\ff}(\omega)  - v_{\ff}(\omega) e^{\frac{-c(\omega)}{v_{\ff}(\omega)} (s - d(\omega) )}. \label{E:stochV}
\end{align} \end{linenomath*}
The stochastic dynamical model evolves according to
\begin{linenomath*} \begin{equation}
	s_n(t,\omega) = s_n(0,\omega) + \int_0^t \Big( V\big( s_{n-1}(\tau,\omega) \big) - V\big( s_n(\tau,\omega) \big) \Big) \dd \tau. \label{E:stoch}
\end{equation} \end{linenomath*}

To ensure that the speed-spacing relations are physically reasonable, the supports of the three distributions \textbf{must} be bounded from both above and below.  That is, we assume the existence of constants, $0 < v_{\ff}^{\min} < v_{\ff}^{\max} < \infty$, $0 < d^{\min} < d^{\max} < \infty$, and $0 < c^{\min} < c^{\max} < \infty$, such that $\PP(\theta(\omega) \in R) = 1$ where $R \equiv \{r \in \RR^3: v_{\ff}^{\min} \le r_1 \le v_{\ff}^{\max}, d^{\min} \le r_2 \le d^{\max}, c^{\min} \le r_3 \le c^{\max}\}$ is a rectangular box with extrema given by the constants.
In this stochastic setting, time discretization is chosen as: $\Delta t = \Delta n / c^{\max}$. An algorithm for simulating the sample paths of the process is given in \ref{sec:A}.

\section{Mean dynamics and variability}\label{S:stoch}
A consequence of the non-linearity in the stochastic model \eqref{E:stoch}, via the non-linearity in $V(\cdot,\omega)$, is that applications such as data assimilation and traffic control will require some form of sampling.  In this section, we \textit{derive} a surrogate stochastic model that approximates the Lagrangian model above.  

We derive two deterministic models below: the first representing the dynamics of the mean of the stochastic Lagrangian model, the second representing the dynamics of the covariance of the system.  These are achieved as mean and covariance dynamics of an infinitely sized ensemble of the proposed Lagrangian model.  \textit{This is akin to using a sampling technique (e.g., ensemble Kalman filter) with an infinitely sized sample, while circumventing the computational costs that come with the need for sampling.}  Indeed, standard Kalman filters can be applied using these two ingredients.

\subsection{Mean dynamics}
\label{SS:mean}
In this section, we demonstrate that the mean of a large ensemble converges to a (particular) deterministic mean process.  \textit{Due to the stochasticity in $V(\cdot,\omega)$ and (the resulting) stochasticity in the spacings, simply taking expectations will not deliver the desired result}.  Instead, consider the deterministic processes given, for each $n \in \{1,\cdots,N\}$, by
\begin{linenomath*} \begin{align}
	\overline{s}_n(t) = s_n(0) + \int_0^t \Big( \overline{V}\big(\overline{s}_{n-1}(\tau) \big) - \overline{V}\big(\overline{s}_n(\tau) \big) \Big) \dd \tau, \label{E:fluidSpacings}
\end{align} \end{linenomath*}
where $\overline{V}(s)$ is a deterministic speed-spacing process to be defined below.  In this section, we formally establish that the average of a large number of stochastic processes given by \eqref{E:stoch}, converges to this mean process \eqref{E:fluidSpacings}.

\medskip
\textbf{Ensemble-averaged process}. Let $M$ denote the ensemble size and $m =1 ,\cdots,M$ the index of a stochastic process in the ensemble.  We denote the $m$th spacing process by $s_n^m(\cdot,\omega)$ for $n=1,\cdots,N$\footnote{We use the notation ``$\cdot$'' to indicate that we are referring to the entire trajectory of the process.  In other words, the difference between $s_n(t,\omega)$ and $s_n(\cdot,\omega)$ is that the former is a scalar random variable while the latter is an entire random curve.}.  The ensemble averaged spacing is given by:
\begin{equation}
	s_n^M(\cdot,\omega) = \frac{1}{M} \sum_{m=1}^M s_n^m(\cdot,\omega).
\end{equation}
In essence, these are $M$ independent stochastic processes, with parameter tuples that are drawn from identical distributions.  Below, we derive the ensemble-averaged state and the deviation processes.

For an ensemble of size $M$, we denote the ensemble-averaged process by $\{s^M_n(\cdot,\omega)\}_{n=1}^N$.   which evolves according to
\begin{linenomath*} \begin{align}
	s^M_n(t,\omega) = s_n(0) + \frac{1}{M} \sum_{m=1}^M \int_0^t \Big( V^m \big( s^M_{n-1}(\tau,\omega),\omega \big) - V^m \big( s^M_n(\tau,\omega),\omega \big)  \Big) \dd \tau \label{E:ensembleAvg}
\end{align} \end{linenomath*}
for $n=1,\cdots,N$. Here, $\{V^m(\cdot,\omega)\}_{m=1}^M$ 
are 
random realizations of the stochastic relations, \eqref{E:stochV}
.  Without loss of generality, we have assumed that the initial spacings are deterministic.  

\begin{figure}[h!]
	\centering
	\resizebox{0.6\textwidth}{!}{%
		\includegraphics{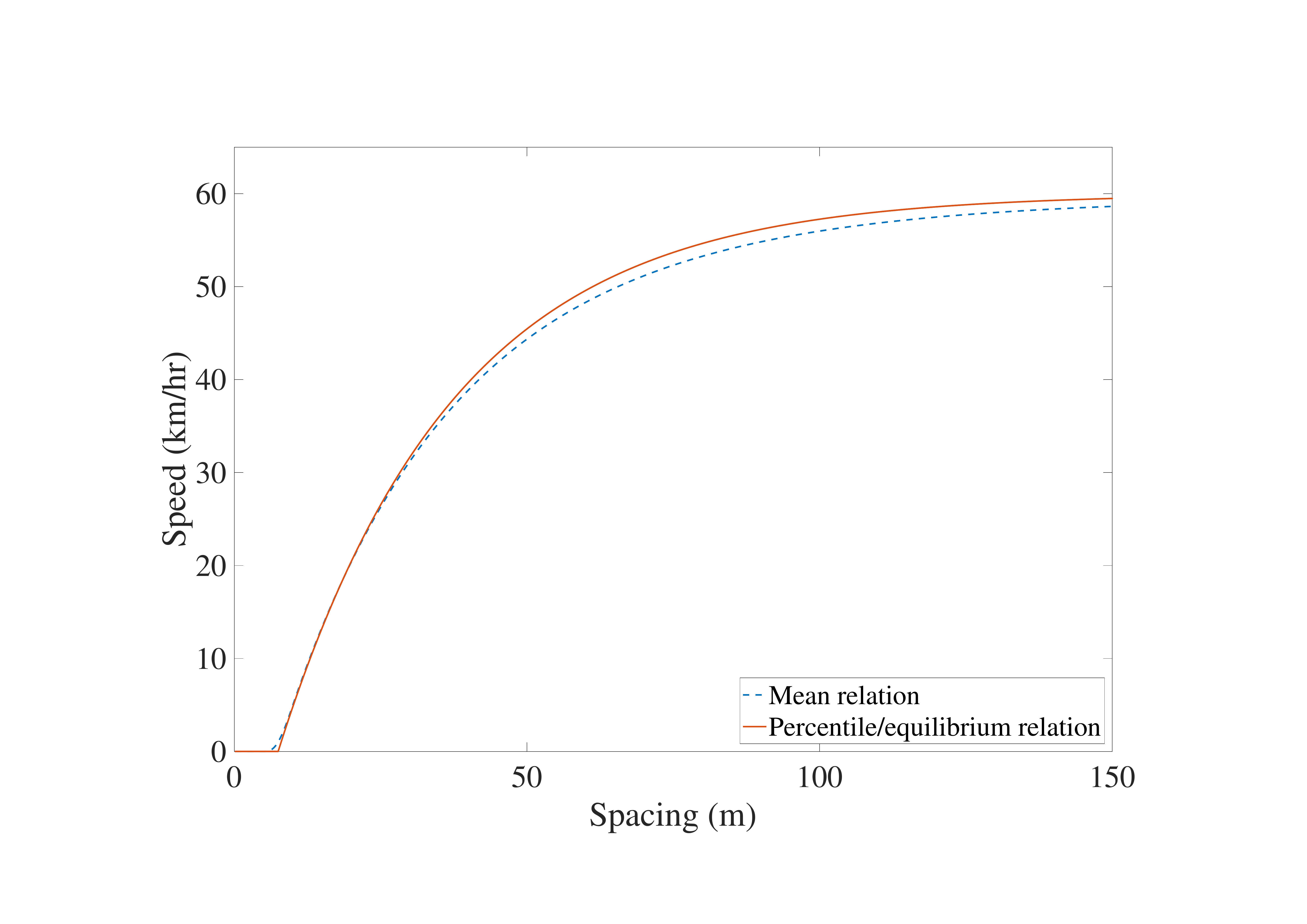}}
	\caption{Mean relation, $\overline{V}(\cdot)$ vs. percentile relation.} \label{f_Ve_vs_Vm}
\end{figure}

\medskip
\textbf{Mean speed-spacing relation}. From the strong law of large numbers we have that
\begin{linenomath*} \begin{align}
	\frac{1}{M} \sum_{i=1}^{M} V^m\big( s,\omega \big)  \underset{M \rightarrow \infty}{\longrightarrow}  \overline{V}\big(s \big)  \mbox{ almost surely}, \label{E:SLLN}
\end{align} \end{linenomath*}
where $\overline{V}(s) \equiv \EE V(s,\omega)$.  Note that $\overline{V}(s) \ne  \overline{v}_{\ff}  - \overline{v}_{\ff} \exp(-(\overline{c}/\overline{v}_{\ff})(s - \overline{d}))$, where $\overline{v}_{\ff} = \EE v_{\ff}(\omega)$,  $\overline{c} = \EE c(\omega)$, and $\overline{d} = \EE d(\omega)$.  The right-hand side is a percentile speed-spacing relation (typically, a 0.5-percentile or equilibrium relation), while $\overline{V}(s)$ is a mean speed-spacing relation; see \citep{2014probabilistic} for more details.  An example comparison is shown in \autoref{f_Ve_vs_Vm}.
We will not attempt to derive an expression for $\overline{V}(\cdot)$. Instead, we will use an empirical approximation: Let $\{\theta^j = (v_{\ff}^j,d^j,c^j), j=1,\cdots,J\}$ be a random sample of size $J$ of parameter 3-tuples.  Then, for any $s$ and a sufficiently large $J$, $\overline{V}(s)$ is very well approximated by
\begin{linenomath*} \begin{align}
\overline{V}(s) \approx \frac{1}{J}\sum_{j=1}^J V_j(s). \label{E:Vapprox}
\end{align} \end{linenomath*}
This approximation can be carried out off-line as part of a preprocessing step using historical data. Computational efficiency can be further improved by means of sparse approximations (see \citep{jabari2017GammaKernels,jabariGenGammaKernels} for example).

\medskip
\textbf{The norm}. To establish convergence we utilize the uniform norm, which for the sake of completeness we review next:  For a process (with continuous sample paths) $\mathbf{y}(\cdot): [0,T] \rightarrow \RR^N$ with components $\{y_i(\cdot)\}_{i=1}^N$, i.e., $\mathbf{y}(\cdot) = [y_1(\cdot) ~ \cdots ~ y_N(\cdot)]^{\top}$, the uniform norm is defined as
\begin{linenomath*} \begin{align}
\| \mathbf{y}(\cdot) \|_T \equiv \underset{0 \le t \le T}{\sup} ~ \underset{1 \le i \le N}{\max} |y_i(t)|.
\end{align} \end{linenomath*}
Whenever
\begin{linenomath*} \begin{align} 
\big\| \mathbf{y}^{(\nu)}(\cdot) - \mathbf{y}(\cdot) \big\|_t \underset{\nu \rightarrow \infty}{\longrightarrow} 0 \mbox{ for all } t \in [0,T]
\end{align} \end{linenomath*}
holds for a sequence of processes $\mathbf{y}^{(1)}(\cdot), \mathbf{y}^{(2)}(\cdot),\cdots$ and a limit process $\mathbf{y}(\cdot)$, the sequence is said to converge uniformly on compact sets (u.o.c.), a form of strong convergence (almost sure convergence on $[0,T]$). 

\medskip
\textbf{Vector notation}. To establish the mean dynamics, we first write \eqref{E:ensembleAvg} in vector form:
\begin{linenomath*} \begin{align}
	\mathbf{s}^M(t,\omega) = \mathbf{s}(0) + \frac{1}{M} \sum_{m=1}^M \int_0^t \mathbf{D} \mathbf{V}^m\big( \mathbf{s}^M(\tau,\omega),\omega \big) \dd \tau, \label{E:scaledSpacings4}
\end{align} \end{linenomath*}
where $\mathbf{s}^M(\cdot,\omega) \equiv [s_1^M(\cdot,\omega) \cdots s_N^M(\cdot,\omega)]^{\top}$,  $\mathbf{V}^m(\mathbf{s},\omega) \equiv [V^m(s_1,\omega) \cdots V^m(s_N,\omega)]^{\top}$, and $\mathbf{D}: \RR^N \rightarrow \RR^N$ is an affine transformation defined by $\mathbf{D} \mathbf{V}^m(\mathbf{s}(\cdot),\omega) \equiv$ $[v_0(\cdot) - V^m(s_1(\cdot),\omega)$ $\cdots V^m(s_{N-1}(\cdot),\omega)-V^m(s_N(\cdot),\omega)]^{\top}$. We may alternatively append $v_0(\cdot)$ to the vector $\mathbf{V}^m(\mathbf{s},\omega)$ (as the first element), then $\mathbf{D}$ is a linear operator.  We will treat $\mathbf{D}$ as a linear operator below.  

Next, define $\overline{\mathbf{s}}(\cdot) \equiv [\overline{s}_1(\cdot) ~ \cdots ~ \overline{s}_N(\cdot)]^{\top}$ and $\overline{\mathbf{V}}\big(\overline{\mathbf{s}}(\cdot)\big) \equiv [\overline{V}\big( \overline{s}_1(\cdot) \big) ~ \cdots ~ \overline{V}\big( \overline{s}_N(\cdot) \big)]^{\top}$, where $\overline{s}_n(\cdot)$ is the deterministic process defined by \eqref{E:fluidSpacings} and $\overline{V}\big(\overline{s}_n(\cdot)\big)$ is the mean speed relation defined by \eqref{E:Vapprox} and \eqref{E:SLLN}.  Then
\begin{linenomath*} \begin{align}
	\overline{\mathbf{s}}(t) = \mathbf{s}(0) + \int_0^t \mathbf{D}\overline{\mathbf{V}}\big(\overline{\mathbf{s}}(\tau)\big) \dd \tau. \label{E:fluidSpacingsVec}
\end{align} \end{linenomath*}	

\medskip
\textbf{Convergence result}.
We are ready to state the main result of this section: that the ensemble-average process converges to the mean dynamic.  This is stated as follows:
\begin{linenomath*} \begin{align}
	\big\| \mathbf{s}^M(\cdot,\omega) - \overline{\mathbf{s}}(\cdot)  \big\|_t \underset{M \rightarrow \infty}{\longrightarrow} 0 \mbox{ for all } t\in[0,T]. \label{E:FSLLN}
\end{align} \end{linenomath*}

\medskip
\textbf{Proof of \eqref{E:FSLLN}}.  
First, it can be easily demonstrated that $\mathbf{V}^m(\mathbf{s},\omega)$ is Lipschitz continuous (in $\mathbf{s}$); let $0 \le K < \infty$ denote its Lipschitz constant: $K$ is the smallest constant such that for all $m =1,\cdots,M$ and any $\mathbf{s}_1,\mathbf{s}_2\ge \mathbf{0}$  
\begin{linenomath*} \begin{align}
	\big\| \mathbf{V}^m(\mathbf{s}_1,\omega) - \mathbf{V}^m(\mathbf{s}_2,\omega) \big\| \le K \| \mathbf{s}_1 - \mathbf{s}_2 \|
	\end{align} \end{linenomath*}
for some appropriately chosen norm $\|\cdot\|$. Since the supports of the parameters are bounded from above and below, $K$ exists and is easy to determine.

Next, it follows immediately from \eqref{E:SLLN} that
\begin{linenomath*} \begin{align}
	\bigg\| \frac{1}{M} \sum_{m=1}^M \mathbf{V}^m\big( \mathbf{z}(\cdot),\omega \big) - \overline{\mathbf{V}}\big( \mathbf{z}(\cdot) \big) \bigg\|_t \underset{M \rightarrow \infty}{\longrightarrow} 0 \mbox{ for all } t \in [0,T] \label{E:fluidSpeed}
\end{align} \end{linenomath*}
for any (vector) process $\mathbf{z}(\cdot)$ with continuous sample paths.  

Using the triangle inequality and noting that, for any $t$, $\overline{\mathbf{V}}\big(\overline{\mathbf{s}}(t)\big) = M^{-1}\sum_{m=1}^M \overline{\mathbf{V}}\big(\overline{\mathbf{s}}(t)\big)$, we have that
\begin{linenomath*} \begin{align}
	\big\| \mathbf{s}^M(\cdot,\omega) - \overline{\mathbf{s}}(\cdot)  \big\|_t \le \Big\| \int_0^{\bullet}  \frac{1}{M} \sum_{m=1}^{M} \mathbf{D} \Big( \mathbf{V}^m\big( \mathbf{s}^M(\tau,\omega),\omega \big) - \overline{\mathbf{V}}\big(\overline{\mathbf{s}}(\tau)\big) \Big) \dd \tau \Big\|_t. \label{E:fluid}
\end{align} \end{linenomath*}
Adding and subtracting $M^{-1}\sum_{m=1}^M\mathbf{D} \mathbf{V}^m\big( \overline{\mathbf{s}}(\tau),\omega \big)$ to the right-hand side (inside the integral), we have that
\begin{linenomath*} \begin{align}
	\big\| \mathbf{s}^M(\cdot,\omega) - \overline{\mathbf{s}}(\cdot)  \big\|_t & \le \Big\| \int_0^{\bullet}  \frac{1}{M} \sum_{m=1}^{M} \mathbf{D} \Big( \mathbf{V}^m\big( \mathbf{s}^M(\tau,\omega),\omega \big) - \mathbf{V}^m\big(\overline{\mathbf{s}}(\tau),\omega\big) \Big) \dd \tau \Big\|_t \nonumber \\
& \quad + \Big\| \int_0^{\bullet}  \frac{1}{M} \sum_{m=1}^{M} \mathbf{D} \Big( \mathbf{V}^m\big(\overline{\mathbf{s}}(\tau),\omega\big) - \overline{\mathbf{V}}\big(\overline{\mathbf{s}}(\tau)\big) \Big) \dd \tau \Big\|_t. \label{E:fluidIn}
\end{align} \end{linenomath*}
To simplify notation, define
\begin{linenomath*} \begin{align}
	\epsilon^M(t) \equiv \Big\| \int_0^{\bullet}  \frac{1}{M} \sum_{m=1}^{M} \mathbf{D} \Big( \mathbf{V}^m\big(\overline{\mathbf{s}}(\tau),\omega\big) - \overline{\mathbf{V}}\big(\overline{\mathbf{s}}(\tau)\big) \Big) \dd \tau \Big\|_t. 
\end{align} \end{linenomath*}
We have that
\begin{linenomath*} \begin{align}
\epsilon^M(t) \underset{M \rightarrow \infty}{\longrightarrow} 0 \label{E:limitTerms}
\end{align} \end{linenomath*}
for all $t \in [0,T]$ from \eqref{E:fluidSpeed} and the continuous mapping theorem.  Applying the triangle inequality, the first term on the right-hand side of \eqref{E:fluidIn} is bounded from above by
\begin{linenomath*} \begin{align}
	\int_0^t \Big\|  \frac{1}{M} \sum_{m=1}^{M} \mathbf{D} \Big( \mathbf{V}^m\big( \mathbf{s}^M(\tau,\omega),\omega \big) - \mathbf{V}^m\big(\overline{\mathbf{s}}(\tau),\omega\big) \Big) \Big\|_{\tau} \dd \tau. \label{E:inter1}
\end{align} \end{linenomath*}
Let $\overline{K} < \infty$ denote the Lipschitz constant of $\mathbf{D}\mathbf{V}^m(\cdot,\omega)$, then \eqref{E:inter1} is bounded from above by
\begin{linenomath*} \begin{align}
\overline{K} \int_0^t \big\|  \mathbf{s}^M(\cdot,\omega) - \overline{\mathbf{s}}(\cdot) \big\|_{\tau} \dd \tau.
\end{align} \end{linenomath*}
Hence, 
\begin{linenomath*} \begin{align}
\big\| \mathbf{s}^M(\cdot,\omega) - \overline{\mathbf{s}}(\cdot)  \big\|_t \le \epsilon^M(t) + \overline{K} \int_0^t \big\|  \mathbf{s}^M(\cdot,\omega) - \overline{\mathbf{s}}(\cdot) \big\|_{\tau} \dd \tau.
\end{align} \end{linenomath*}
Applying the Bellman-Gr{\"o}nwall inequality, we have that
\begin{linenomath*} \begin{align}
\big\| \mathbf{s}^M(\cdot,\omega) - \overline{\mathbf{s}}(\cdot)  \big\|_t \le \epsilon^M(t) e^{\overline{K} t}.
\end{align} \end{linenomath*}
For all $t\in[0,T]$, \eqref{E:FSLLN} follows from \eqref{E:limitTerms} as $M\rightarrow \infty$.  This completes the proof.

\subsection{Hydrodynamic limit}
The result above can be generalized to any vehicle size scaling.  In essence, we have thus far assumed that $\Delta n = 1$.  This can be easily generalized to any $\Delta n$: The ensemble averaged process becomes
\begin{linenomath*} \begin{align}
	\mathbf{s}^M(t,\omega) = \mathbf{s}(0) + \frac{1}{M\Delta n} \sum_{m=1}^M \int_0^t \mathbf{D} \mathbf{V}^m\big( \mathbf{s}^M(\tau,\omega),\omega \big) \dd \tau. \label{E:scaledSpacings5}
\end{align} \end{linenomath*}
In this case, we have $\lfloor N/\Delta n \rfloor$ `vehicles' in the system ($\lfloor \cdot \rfloor$ is the floor function).  The limit spacing process can be derived using the same procedure presented above.  It is given, for $n \in \{ 1,\cdots, \lfloor N/\Delta n \rfloor \}$, by
\begin{linenomath*} \begin{align}
	\overline{s}_n(t) = s_n(0) + \frac{1}{\Delta n} \int_0^t \Big( \overline{V}\big(\overline{s}_{n-1}(\tau) \big) - \overline{V}\big(\overline{s}_n(\tau) \big) \Big) \dd \tau. \label{E:fluidSpacings1}
\end{align} \end{linenomath*}
In vector form ($\overline{\mathbf{s}}(t) \in \RR^{\lfloor N/\Delta n \rfloor}$):
\begin{linenomath*} \begin{align}
	\overline{\mathbf{s}}(t) = \mathbf{s}(0) + \frac{1}{\Delta n} \int_0^t \mathbf{D}\overline{\mathbf{V}}\big(\overline{\mathbf{s}}(\tau)\big) \dd \tau. \label{E:fluidSpacingsVec1}
\end{align} \end{linenomath*}	
This is a deterministic process that converges as $\Delta n \rightarrow 0$ to a conservation law in Lagrangian coordinates:
\begin{linenomath*} \begin{align}
\frac{\partial \overline{s}(n,t)}{ \partial t} + \frac{\partial \overline{V}\big( \overline{s}(n,t) \big)}{\partial n} = 0,
\end{align} \end{linenomath*}
where $\overline{s}(n,t)$ is a process in which $n$ is continuous and the speed relation $\overline{V}(\cdot)$ is a mean relation and not the traditional equilibrium relation used in the literature. 

\subsection{Covariance dynamics}
The mean dynamics, $\overline{\mathbf{s}}(\cdot)$, given by \eqref{E:scaledSpacings5} can be considered as a \textit{first-order approximation} of the stochastic Lagrangian model.  It essentially represents a first moment of the system.  In this section, we derive the dynamics of a second moment of the system, the covariance dynamics, to achieve (i) a second-order approximation and (ii) facilitate the use of standard Kalman filtering techniques for traffic state estimation.  

\medskip
\textbf{The deviation process}.  In this section, we derive the covariance dynamics of the stochastic spacing process $\mathbf{s}(\cdot,\omega)$.  First, consider the (amplified) deviation process
\begin{linenomath*} \begin{align}
	\boldsymbol{\delta}^M(t,\omega) \equiv \sqrt{M} \big(\mathbf{s}^M(t,\omega) - \overline{\mathbf{s}}(t) \big). \label{E:devProcess}
\end{align} \end{linenomath*}
The scaling $\sqrt{M}$ ensures that the covariance matrix pertaining to $\boldsymbol{\delta}^M(t,\omega)$ is the same as that pertaining to $\mathbf{s}(t,\omega)$ (for each $t$).  This is demonstrated as follows: Since $\mathbf{s}(t,\omega)$ is a random vector, $\mathrm{Var}\big(\mathbf{s}(t,\omega) \big)$ is the covariance matrix of $\mathbf{s}(\cdot,\omega)$ at time $t$.  We have established in \autoref{SS:mean} that $\overline{\mathbf{s}}(\cdot)$ centers $\mathbf{s}(\cdot,\omega)$.  Consequently,
\begin{equation}
	\mathrm{Var}\big(\mathbf{s}(t,\omega) \big) = \mathrm{Var}\big(\mathbf{s}(t,\omega) - \overline{\mathbf{s}}(t) \big).
\end{equation}
We compare this to the covariance matrix of the deviation process:
\begin{align}
	\mathrm{Var}\big( \boldsymbol{\delta}^M(t,\omega) \big) &= \mathrm{Var}\big( \sqrt{M} (\mathbf{s}^M(t,\omega) - \overline{\mathbf{s}}(t)) \big)
	= \mathrm{Var} \Big( \frac{1}{\sqrt{M}} \sum_{m=1}^M \big(\mathbf{s}^m(t,\omega) - \overline{\mathbf{s}}(t) \big) \Big) \nonumber \\
	 &= \frac{1}{M}\mathrm{Var} \Big( \sum_{m=1}^M \big(\mathbf{s}^m(t,\omega) - \overline{\mathbf{s}}(t) \big) \Big) =  \mathrm{Var}\big(\mathbf{s}(t,\omega) - \overline{\mathbf{s}}(t) \big),
\end{align}
where the last equality follows from $\{\mathbf{s}^m(t,\omega)\}_{m=1}^M$ being independent and identically distributed random vectors, each with the same distribution as $\mathbf{s}(t,\omega)$.  

\textit{Note that this is true for any $M$.  The remainder of this section demonstrates that when $M \rightarrow \infty$ a tractable (deterministic) closed expression for the covariance is obtained.}

\medskip
\textbf{Limiting Deviation process}. The boundedness properties of the speed-spacing relations ensures the existence of a limiting process, $\widetilde{\boldsymbol{\delta}}(\cdot,\omega)$, such that $\boldsymbol{\delta}^M(\cdot,\omega) \rightarrow \widetilde{\boldsymbol{\delta}}(\cdot,\omega)$ (weakly) as $M\rightarrow \infty$.  We derive this limiting process next.

Expanding \eqref{E:devProcess} we have
\begin{linenomath*} \begin{align}
	\boldsymbol{\delta}^M(t,\omega) = \int_0^t \frac{1}{\sqrt{M}\Delta n} \sum_{m=1}^{M} \mathbf{D} \Big( \mathbf{V}^m\big( \mathbf{s}^M(\tau,\omega),\omega \big) - \overline{\mathbf{V}}\big(\overline{\mathbf{s}}(\tau)\big) \Big)  \dd \tau.
\label{E:weakLim1}
\end{align} \end{linenomath*}
Adding and subtracting $\int_0^t M^{-\sfrac{1}{2}} \Delta n^{-1} \sum_{i=1}^{M} \mathbf{D} \mathbf{V}^m\big( \overline{\mathbf{s}}(\tau) \big) \dd \tau$ to the right-hand side of \eqref{E:weakLim1}, we get (with some rearrangement)
\begin{linenomath*} \begin{align}
	\boldsymbol{\delta}^M(t,\omega) & = \int_0^t \frac{1}{\sqrt{M}\Delta n} \sum_{m=1}^{M} \mathbf{D} \Big( \mathbf{V}^m\big( \mathbf{s}^M(\tau,\omega),\omega \big) - \mathbf{V}^m\big(\overline{\mathbf{s}}(\tau)\big) \Big)  \dd \tau \nonumber \\
	&\qquad \qquad \qquad + \int_0^t \frac{1}{\sqrt{M}\Delta n} \sum_{m=1}^{M} \mathbf{D} \Big( \mathbf{V}^m\big( \overline{\mathbf{s}}(\tau) \big) - \overline{\mathbf{V}}\big(\overline{\mathbf{s}}(\tau)\big) \Big)  \dd \tau. \label{E:weakLim2}
\end{align} \end{linenomath*}
We will derive the limiting processes for the two terms on the right-hand side of \eqref{E:weakLim2} separately.  Take the first term.  We have by definition, \eqref{E:devProcess}, that $\overline{\mathbf{s}}(\cdot) + M^{-\sfrac{1}{2}}\boldsymbol{\delta}^M(\cdot,\omega) = \mathbf{s}^M(\cdot,\omega)$.  Hence, the first term can be written as
\begin{linenomath*} \begin{align}
	\int_0^t \frac{1}{\sqrt{M}\Delta n} \sum_{m=1}^{M} \mathbf{D} \Big( \mathbf{V}^m\big( \overline{\mathbf{s}}(\tau) + M^{-\sfrac{1}{2}} \boldsymbol{\delta}^M(\tau,\omega),\omega \big) - \mathbf{V}^m\big(\overline{\mathbf{s}}(\tau)\big) \Big)  \dd \tau.
\end{align} \end{linenomath*}
Upon dividing and multiplying the terms inside the integral by $\| \boldsymbol{\delta}^M(\tau,\omega) \|$, this is equivalent to
\begin{linenomath*} \begin{align}
	\frac{1}{\Delta n} \int_0^t \frac{1}{M} \sum_{m=1}^M  \frac{\mathbf{D} \Big( \mathbf{V}^m\big( \overline{\mathbf{s}}(\tau) + M^{-\sfrac{1}{2}} \boldsymbol{\delta}^M(\tau,\omega),\omega \big) - \mathbf{V}^m\big(\overline{\mathbf{s}}(\tau)\big) \Big)}{M^{-\sfrac{1}{2}} \| \boldsymbol{\delta}^M(\tau,\omega) \|}     \| \boldsymbol{\delta}^M(\tau,\omega) \|  \dd \tau. \label{E:weaklimit2}
\end{align} \end{linenomath*}
In accord with \eqref{E:FSLLN}, $M^{-\sfrac{1}{2}}\boldsymbol{\delta}^M(t,\omega) \rightarrow \mathbf{0}$ almost surely as $M \rightarrow \infty$.  Applying the (generalized) continuous mapping theorem \citep[Theorem 3.4.4]{whitt2002stochastic}, \eqref{E:weaklimit2} converges to
\begin{linenomath*} \begin{align}
	\frac{1}{\Delta n} \int_0^t \mathbf{D} \nabla_{\widetilde{\boldsymbol{\delta}}(\tau,\omega)} \overline{\mathbf{V}}\big(\overline{\mathbf{s}}(\tau)\big) \| \widetilde{\boldsymbol{\delta}}(\tau,\omega) \| \dd \tau,
\end{align} \end{linenomath*}
where $\nabla_{\widetilde{\boldsymbol{\delta}}} \overline{\mathbf{V}}$ is the directional derivative of $\overline{\mathbf{V}}$ along the direction given by the vector $\widetilde{\boldsymbol{\delta}}$.  This simplifies to
\begin{linenomath*} \begin{align}
	\frac{1}{\Delta n} \int_0^t \mathbf{D}  \mathrm{diag} \big[ \nabla \overline{\mathbf{V}}\big(\overline{\mathbf{s}}(\tau)\big) \big] \widetilde{\boldsymbol{\delta}}(\tau,\omega)   \dd \tau, \label{weakLim0} 
\end{align} \end{linenomath*}
where $\nabla \overline{\mathbf{V}}\big(\overline{\mathbf{s}}(t)\big) = \big[(\dd / \dd s)\overline{V}(\overline{s}_1(t)) ~ \cdots ~ (\dd / \dd s)\overline{V}(\overline{s}_N(t)) \big]^{\top}$ and $\mathrm{diag}[\mathbf{s}]$ is a diagonal matrix with diagonal elements given by the vector $\mathbf{s}$.  To simplify notation, define $\mathbf{G}\big(\overline{\mathbf{s}}(t)\big) \equiv \mathrm{diag} \big[ \nabla \overline{\mathbf{V}}\big(\overline{\mathbf{s}}(\tau)\big) \big]$.

\smallskip

We now turn to derivation of the limiting process of the second term in \eqref{E:weakLim2}, which we re-write as:
\begin{linenomath*} \begin{align}
	\int_0^t \frac{1}{\sqrt{M}} \sum_{m=1}^{M} \Big( \frac{1}{\Delta n}\mathbf{D} \mathbf{V}^m\big( \overline{\mathbf{s}}(\tau) \big) - \frac{1}{\Delta n}\mathbf{D} \overline{\mathbf{V}}\big(\overline{\mathbf{s}}(\tau)\big) \Big)  \dd \tau. \label{E:secondTerm}
\end{align} \end{linenomath*}
Applying the central limit theorem, we have, for any $t$, that
\begin{linenomath*} \begin{align}
	\frac{1}{\sqrt{M}} \sum_{m=1}^{M} \Big(\mathbf{V}^m\big( \overline{\mathbf{s}}(t) \big) - \overline{\mathbf{V}}\big(\overline{\mathbf{s}}(t)\big) \Big)
\end{align} \end{linenomath*}
converges weakly (in distribution) to a zero mean Normal random vector with a diagonal covariance matrix, the diagonal elements of which are given by $\mathrm{Var}(V(\overline{s}_n(t),\omega)) = \EE\big( V(\overline{s}_n(t),\omega) - \overline{V}((\overline{s}_n(t)) \big)^2$. Denote this covariance matrix by $\boldsymbol{\varSigma}^{\sfrac{1}{2}}(\cdot)$.  We have by the continuous mapping theorem that \eqref{E:secondTerm} converges to the stochastic integral
\begin{linenomath*} \begin{align}
	\frac{1}{\Delta n}  \int_0^t \mathbf{D} \boldsymbol{\varSigma}^{\sfrac{1}{2}}(\overline{\mathbf{s}}(\tau)) \dd \mathbf{W}(\tau,\omega), \label{weakLim1}
\end{align} \end{linenomath*}
where $\mathbf{W}(\cdot,\omega)$ is an $N$-dimensional Wiener process.  The matrix valued function $\boldsymbol{\varSigma}^{\sfrac{1}{2}}\big(\overline{\mathbf{s}}(\tau)\big)$ can be calculated (off-line) using an empirical approximation as in \eqref{E:Vapprox} (using the same pseudo-random parameter sample).  

Putting these results obtained in \eqref{weakLim0} and \eqref{weakLim1} we have that the limiting deviation process, $\widetilde{\boldsymbol{\delta}}(\cdot,\omega)$, is the solution of the following stochastic integral equation
\begin{linenomath*} \begin{align}
	\widetilde{\boldsymbol{\delta}}(t,\omega) = \widetilde{\boldsymbol{\delta}}(0) + \frac{1}{\Delta n} \int_0^t \mathbf{D} \Big(   \mathbf{G}\big(\overline{\mathbf{s}}(\tau)\big) \widetilde{\boldsymbol{\delta}}(\tau,\omega) \dd \tau + \boldsymbol{\varSigma}^{\sfrac{1}{2}}(\overline{\mathbf{s}}(\tau)) \dd \mathbf{W}(\tau,\omega) \Big).
\end{align} \end{linenomath*}
This can be written (symbolically) in differential form as
\begin{linenomath*} \begin{align}
	\dd \widetilde{\boldsymbol{\delta}}(t,\omega)  = \frac{1}{\Delta n} \mathbf{D} \Big( \mathbf{G}\big(\overline{\mathbf{s}}(t)\big) \widetilde{\boldsymbol{\delta}}(t,\omega) \dd t + \boldsymbol{\varSigma}^{\sfrac{1}{2}}(\overline{\mathbf{s}}(t)) \dd \mathbf{W}(t,\omega) \Big).
\end{align} \end{linenomath*}
The latter is a linear matrix stochastic differential equation and has a closed form solution given by \citep{ludwing1974stochastic,evans2012introduction}
\begin{linenomath*} \begin{align}
	\widetilde{\boldsymbol{\delta}}(t,\omega) = \boldsymbol{\Phi}(t) \Big( \widetilde{\boldsymbol{\delta}}(0,\omega) + \frac{1}{\Delta n}  \int_0^t \boldsymbol{\Phi}^{-1}(\tau) \mathbf{D}\boldsymbol{\varSigma}^{\sfrac{1}{2}}(\overline{\mathbf{s}}(\tau)) \dd \mathbf{W}(\tau,\omega) \Big), \label{E:deviation}
\end{align} \end{linenomath*}
where $\boldsymbol{\Phi}(\cdot)$ is the \textit{fundamental matrix}, that is, it is the solution of
\begin{linenomath*} \begin{align}
\frac{\dd \boldsymbol{\Phi}(t)}{\dd t} = \frac{1}{\Delta n} \mathbf{D}\mathbf{G}\big(\overline{\mathbf{s}}(t)\big) \boldsymbol{\Phi}(t) \label{E:phi}
\end{align} \end{linenomath*}
with initial condition $\boldsymbol{\Phi}(0) = \mathbf{I}_{N \times N}$, which is an $N \times N$ identity matrix.  

\medskip
\textbf{Covariance dynamics}. 
The above implies that $\widetilde{\boldsymbol{\delta}}(\cdot,\omega)$ is a Gaussian process with (deterministic) covariance process given by $\mathbf{P}(\cdot) = \EE \widetilde{\boldsymbol{\delta}}(\cdot,\omega) \widetilde{\boldsymbol{\delta}}(\cdot,\omega)^{\top}$.
To avoid the need to compute the fundamental matrix, $\boldsymbol{\Phi}(\cdot)$, the evolution of the covariance matrix can be calculated by taking the time derivative of $\mathbf{P}(\cdot)$, which results in the following matrix differential equation describing the evolution of the covariance dynamics:
\begin{linenomath*} \begin{align}
	\frac{\dd \mathbf{P}(t)}{\dd t} = \frac{1}{\Delta n} \mathbf{D}\mathbf{G}\big(\overline{\mathbf{s}}(t)\big) \mathbf{P}(t) +  \frac{1}{\Delta n} \mathbf{P}^{\top}(t) \mathbf{G}^{\top} \big(\overline{\mathbf{s}}(t)\big) \mathbf{D}^{\top} + \frac{\dd t}{\Delta n^2} \mathbf{D} \boldsymbol{\varSigma}(\overline{\mathbf{s}}(t)) \mathbf{D}^{\top}, \label{E:covar}
\end{align} \end{linenomath*}
where $\boldsymbol{\varSigma}(\mathbf{s}) = \big\langle \boldsymbol{\varSigma}^{\sfrac{1}{2}}(\mathbf{s}),  \boldsymbol{\varSigma}^{\sfrac{1}{2}}(\mathbf{s}) \big\rangle$.

\section{Data Assimilation}
\label{S:estimation}
This section demonstrates how the model can be used to estimate missing information when (limited) vehicle trajectory data is available.  In essence, we utilize an ensemble filter for this purpose.  We first develop the system state dynamics (both mean and covariance), then the measurement dynamics, and finally the recursive estimation algorithm (Kalman-Bucy).

\subsection{System state dynamics}
Let $\mathbf{z}(t,\omega) \equiv [s_1(t,\omega) ~ \cdots ~ s_N(t,\omega) ~ x_1(t,\omega) ~ \cdots ~ x_N(t,\omega)]^{\top}$ denote the traffic state vector.  Define the matrix
\begin{linenomath*} \begin{align}
	\mathbf{D}_{\mathrm{z}} = \frac{1}{\Delta n}\begin{bmatrix}
		\mathbf{D} \\ \mathbf{I}_{N \times N}
	\end{bmatrix}
\end{align} \end{linenomath*}
and let $\overline{\mathbf{V}}_{\mathrm{z}}: \RR^{2N} \rightarrow \RR^N$ given by $\overline{\mathbf{V}}_{\mathrm{z}}(\mathbf{z}) = \overline{\mathbf{V}}( [\mathbf{I}_{N \times N} ~ \mathbf{0}_{N \times N}]\mathbf{z})$ so that  $\overline{\mathbf{V}}_{\mathrm{z}}(\mathbf{z}) = \overline{\mathbf{V}}(\mathbf{s})$ ($\mathbf{0}_{N \times N}$ is an $N \times N$ matrix of zeros). Then, the mean state dynamics are given by
\begin{linenomath*} \begin{align}
	\overline{\mathbf{z}}(t) = \mathbf{z}(0) + \int_0^t \mathbf{D}_{\mathrm{z}} \overline{\mathbf{V}}_{\mathrm{z}} \big(\overline{\mathbf{z}}(\tau)\big) \dd \tau, \label{E:fluidStateVec}
\end{align} \end{linenomath*}
where $\overline{\mathbf{z}}(t) = [\overline{\mathbf{s}}^{\top}(t) ~ \overline{\mathbf{x}}^{\top}(t)]^{\top}$.  Similarly, the ensemble averaged state evolution is given by
\begin{linenomath*} \begin{align}
	\mathbf{z}^M(t,\omega) = \mathbf{z}(0) + \frac{1}{M} \sum_{m=1}^M \int_0^t \mathbf{D}_{\mathrm{z}} \mathbf{V}_{\mathrm{z}}^m\big( \mathbf{z}^M(\tau,\omega),\omega \big) \dd \tau. \label{E:scaledState}
\end{align} \end{linenomath*}

The same derivations in \autoref{S:stoch} yield
\begin{linenomath*} \begin{align}
	\big\| \mathbf{z}^M(\cdot,\omega) - \overline{\mathbf{z}}(\cdot)  \big\|_t \underset{M \rightarrow \infty}{\longrightarrow} 0 \mbox{ for all } t\in[0,T]. \label{E:stateFSLLN}
\end{align} \end{linenomath*}
The covariance dynamics of a Gaussian approximation of the state process are given by:
\begin{linenomath*} \begin{align}
	\frac{\dd \mathbf{P}_{\mathrm{z}}(t)}{\dd t} = \mathbf{D}_{\mathrm{z}} \mathbf{G}_{\mathrm{z}}\big(\overline{\mathbf{z}}(t)\big) \mathbf{P}_{\mathrm{z}} (t) +  \mathbf{P}_{\mathrm{z}}^{\top}(t) \mathbf{G}_{\mathrm{z}}^{\top} \big(\overline{\mathbf{z}}(t)\big) \mathbf{D}_{\mathrm{z}}^{\top} + \dd t\mathbf{D}_{\mathrm{z}} \boldsymbol{\varSigma}_{\mathrm{z}}(\overline{\mathbf{z}}(t)) \mathbf{D}_{\mathrm{z}}^{\top}, \label{E:stateCovar}
\end{align} \end{linenomath*}
where $\mathbf{G}_{\mathrm{z}}(\mathbf{z}) = \mathrm{diag} \big( \nabla \overline{\mathbf{V}}_{\mathrm{z}}(\mathbf{z}) \big)$ and $\boldsymbol{\varSigma}_{\mathrm{z}}(\mathbf{z}) = \boldsymbol{\varSigma}([\mathbf{I}_N ~ \mathbf{0}_N]\mathbf{z})$.

\subsection{Measurements} \label{SS_measurements}
Suppose position and speed measurements are available for a subset of the $N$ ``probe'' vehicles.  We assume that $N$ is known.  However, this assumption can be relaxed to one where $N$ is estimated from some sampled information, such as probe vehicle/connected vehicle trajectory data.  A methodology has been proposed in the literature \citep{zheng2017estimating} for estimating traffic volumes from sample probe vehicle trajectory data for urban signalized roads.  We refer interested readers to \citep{zheng2017estimating} for more details.

Let $\mathcal{N} \subset \{1, \cdots, N\}$ denote the indices of the vehicles with measurements, i.e., $|\mathcal{N}| N^{-1}$ is the penetration rate.  We will denote by $\mathcal{N}(n)$ the index of vehicle $n$ in the set $\mathcal{N}$. There are two types of measurements, depending on whether the probe vehicles are instrumented with sensors that can measure distances to/from surrounding vehicles or are equipped with vehicle communication systems. 

\medskip
\textbf{Unequipped vehicles}.  Vehicles that are unequipped with communication systems are the more widely available sources of probe vehicle data today (e.g., taxis, ridesharing service providers, etc.).  For such systems, spacings are not measured directly.  Instead, they can be represented as noisy measurements from the measured speeds using a random version of \eqref{E:hetS}:
\begin{linenomath*} \begin{align}
	S(v, \omega) = d(\omega) - \frac{v_{\ff}(\omega)}{c(\omega)} \log_e \left( 1 - \frac{v}{v_{\ff}(\omega)} \right). \label{E:randS}
\end{align} \end{linenomath*}
For $n \in \mathcal{N}$, let $m_n^{\mathrm{s}}(t,\omega)$ denote a spacing measurement and let $m_n^{\mathrm{x}}(t)$ denote a position measurement.  Since the first of these two quantities is not directly measured (in this scenario), it is assumed to be random.  Specifically, we assume that a sufficiently large historical sample is available that $S(v,\omega)$ is well approximated by a Normally distributed random variable, so that $\EE m_n^{\mathrm{s}}(t,\omega) = \EE S(v_n(t), \omega)$.   
Define the diagonal matrix $\boldsymbol{\Omega}^{\mathrm{s}}(t)$ with diagonal elements $\Omega^{\mathrm{s}}_{n,n}(t) =  \mathrm{Var} \big( S(v_n(t), \omega) \big)$, which can also be estimated using historical data.  The measurement vector is denoted by $\RR^{2|\mathcal{N}|} \ni \mathbf{m}(t,\omega)= [m_{\mathcal{N}_1}^{\mathrm{s}}(t,\omega)$ $~ \cdots ~$ $m_{\mathcal{N}_{|\mathcal{N}|}}^{\mathrm{s}}(t,\omega) ~ m_{\mathcal{N}_1}^{\mathrm{x}}(t) ~ \cdots ~ m_{\mathcal{N}_{|\mathcal{N}|}}^{\mathrm{x}}(t)]^{\top}$.  The measurement equation is given by
\begin{linenomath*} \begin{align}
	\mathbf{m}(t,\omega) = \mathbf{H} \mathbf{z}(t,\omega) + \boldsymbol{\Omega}^{\sfrac{1}{2}}(t) \boldsymbol{\epsilon}(\omega),
\end{align} \end{linenomath*} 
where $\mathbf{H} \in \{0,1\}^{2|\mathcal{N}|\times N}$ is a measurement-state variable incidence matrix, $\boldsymbol{\epsilon}(\omega)$  is a $2|\mathcal{N}|$-dimensional standard Normal random vector and
\begin{linenomath*} \begin{align}
	\boldsymbol{\Omega}(t) = \begin{bmatrix}
		\boldsymbol{\Omega}^{\mathrm{s}}(t) & \mathbf{0}_{|\mathcal{N}| \times N-|\mathcal{N}|} \\
		\mathbf{0}_{|\mathcal{N}| \times |\mathcal{N}|} & \mathbf{0}_{|\mathcal{N}| \times N-|\mathcal{N}|}
	\end{bmatrix}.
\end{align} \end{linenomath*}

\medskip
\textbf{When spacing measurements are available}.  With the vast advances in vehicle technologies, it is reasonable to expect that (probe) vehicles will be able to measure not only their own positions and speed, but also those pertaining to vehicles in their immediate surroundings.  It is, therefore, reasonable to expect that such probe vehicles can measure spacings between them and both their leaders and followers.  In this case, for $n \in \mathcal{N}$, the spacing measurements $m_n^{\mathrm{s}}(t)$ and $m_{n+1}^{\mathrm{s}}(t)$ are available in addition to their corresponding position measurements: $m_{n-1}^{\mathrm{x}}(t)$, $m_n^{\mathrm{x}}(t)$, and $m_{n+1}^{\mathrm{x}}(t)$.  Note that, in this case, we have dropped the `$\omega$' notation to indicate determinism\footnote{This, of course, ignores the role of measurement errors.  But in the case described here, these tend to small enough as to be negligible.}.  In this case (when spacing measurements are available), the measurement-state variable incidence matrix, $\mathbf{H}$, is denser than in the unequipped vehicle case.  Specifically, we have five measurements per probe, i.e., $\mathbf{H} \in \{0,1\}^{5|\mathcal{N}|\times N}$ and
\begin{linenomath*} \begin{align}
	\mathbf{m}(t) = \mathbf{H} \mathbf{z}(t,\omega). 
\end{align} \end{linenomath*}
The right-hand side above is random while the left-hand side is deterministic. This should be interpreted as an assignment of deterministic measurements to (what would otherwise be) random quantities.  Again, assuming (without loss of generality) that measurements are error-free, we have that $\boldsymbol{\Omega}(t) = \mathbf{0}$ for all $t$ when spacing measurements \textit{are} available.

\subsection{Kalman-Bucy filter}
The mean and covariance dynamics represent the empirical mean and covariance dynamics of the ensemble when the ensemble size grows to infinity.  In this way, we avoid having to sample from a distribution!  With these elements in place, the estimation too used is simply a Kalman-Bucy filter (a.k.a. continuous linear filter); see \citep{jazwinski1970stochastic} for more details.

Let $\widehat{\mathbf{z}}(\cdot)$ and $\widehat{\mathbf{P}}(\cdot)$ denote the (optimal) state mean and state covariance estimates. Their evolution is given by
\begin{linenomath*} \begin{align}
	\frac{\dd \widehat{\mathbf{z}}(t)}{\dd t} = \mathbf{D}_{\mathrm{z}} \overline{\mathbf{V}}_{\mathrm{z}} \big(\widehat{\mathbf{z}}(t)\big) + \mathbf{K} (t) \big(  \mathbf{m}(t) - \mathbf{H} \widehat{\mathbf{z}}(t)\big) \label{E:meanEst}
\end{align} \end{linenomath*}
and
\begin{linenomath*} 
\begin{align}
	\frac{\dd \widehat{\mathbf{P}}(t)}{\dd t} = \mathbf{D}_{\mathrm{z}} \mathbf{G}_{\mathrm{z}}\big(\widehat{\mathbf{z}}(t)\big) \widehat{\mathbf{P}}(t) +  \widehat{\mathbf{P}}^{\top}(t) \mathbf{G}_{\mathrm{z}}^{\top} \big(\widehat{\mathbf{z}}(t)\big) \mathbf{D}_{\mathrm{z}}^{\top} + \dd t \mathbf{D}_{\mathrm{z}} \boldsymbol{\varSigma}_{\mathrm{z}}(\widehat{\mathbf{z}}(t)) \mathbf{D}_{\mathrm{z}}^{\top} - \mathbf{K}(t) \mathbf{H} \widehat{\mathbf{P}}(t), \label{E:covarEst}
\end{align} 
\end{linenomath*}
where $\mathbf{K}(t)$ is the Kalman gain matrix.  For continuous systems, the Kalman gain matrix is given by $\mathbf{K}(t) = \widehat{\mathbf{P}}(t) \mathbf{H}^{\top} \boldsymbol{\Omega}^{-1}(t)$.  Under our assumptions (regardless of whether the vehicles are capable of measuring spacings or not), $\boldsymbol{\Omega}(t)$ is singular for all $t$. We overcome this issue numerically, where \eqref{E:meanEst} and \eqref{E:covarEst} are discretized in time using a $\Delta t$ that corresponds to a mean reaction time (on the order of few seconds).  To this end, we denote by $\widehat{\mathbf{z}}(k \Delta t-)$ and $\widehat{\mathbf{z}}(k \Delta t)$ predicted and updated mean state vectors at time instant $t = k\Delta t$, respectively.  We define $\widehat{\mathbf{P}}(k \Delta t-)$ and $\widehat{\mathbf{P}}(k \Delta t)$ in a similar fashion.  \autoref{alg:KalmanBucy} is an implementation of the Kalman-Bucy filter for Lagrangian traffic state estimation.

\begin{algorithm}[h!]
	\caption{Kalman-Bucy filter}
	\label{alg:KalmanBucy}
	\algsetup{linenosize=\small}
	\small
	\begin{algorithmic}[1]
		\REQUIRE $N$, $\Delta n$, $T$, $x_0(\cdot)$, $v_0(\cdot)$, $\mathbf{z}(0)$, $\{ \theta^j = (v_{\ff}^j,d^j,c^j) \}_{j=1}^J$ (historical data)
		\STATE $\Delta t \mapsfrom  \frac{1}{J} \sum_{j=1}^J \frac{\Delta n}{c^j}$
		\STATE $k \mapsfrom 0$
		\ENSURE
		\WHILE {$k \le \lfloor T/ \Delta t \rfloor$}
		\STATE \texttt{\color{teal} /* State mean and covariance prediction */}
		\STATE $\widehat{\mathbf{z}}((k+1) \Delta t-) \mapsfrom  \widehat{\mathbf{z}}(k \Delta t) + \Delta t \mathbf{D}_{\mathrm{z}} \overline{\mathbf{V}}_{\mathrm{z}} \big(\widehat{\mathbf{z}}(k \Delta t)\big)$
		\STATE $\widehat{\mathbf{P}}((k+1) \Delta t- ) \mapsfrom \widehat{\mathbf{P}}(k \Delta t ) + \Delta t \Big( \mathbf{D}_{\mathrm{z}} \mathbf{G}_{\mathrm{z}}\big(\widehat{\mathbf{z}}(k \Delta t)\big) \widehat{\mathbf{P}}(k \Delta t )$
		\STATE \hspace{1.15in} $+ \widehat{\mathbf{P}}^{\top}(k \Delta t) \mathbf{G}_{\mathrm{z}}^{\top} \big(\widehat{\mathbf{z}}(k \Delta t)\big) \mathbf{D}_{\mathrm{z}}^{\top} + \Delta t \mathbf{D}_{\mathrm{z}} \boldsymbol{\varSigma}_{\mathrm{z}}(\widehat{\mathbf{z}}(k \Delta t)) \mathbf{D}_{\mathrm{z}}^{\top} \Big)$
		\STATE \vspace{0.05 in} \texttt{\color{teal}/* Residual mean and covariance */}
		\STATE $\mathbf{r}((k+1) \Delta t) \mapsfrom  \mathbf{m}((k+1)\Delta t) - \mathbf{H} \widehat{\mathbf{z}}((k+1) \Delta t-)$
		\STATE $\mathbf{R}((k+1) \Delta t) \mapsfrom \Delta t \mathbf{H} \widehat{\mathbf{P}}((k+1) \Delta t-) \mathbf{H}^{\top} + \boldsymbol{\Omega}((k+1) \Delta t)$
		\STATE \vspace{0.05 in} \texttt{\color{teal} /*  State mean and covariance update */}
		\STATE $\mathbf{K}((k+1) \Delta t) \mapsfrom \widehat{\mathbf{P}}((k+1) \Delta t-) \mathbf{H}^{\top} \mathbf{R}^{-1}((k+1) \Delta t)$
		\STATE $\widehat{\mathbf{z}}((k+1)\Delta t) \mapsfrom \widehat{\mathbf{z}}((k+1)\Delta t-) + \mathbf{K} ((k+1)\Delta t) \mathbf{r}((k+1) \Delta t)$
		\STATE $\widehat{\mathbf{P}}((k+1)\Delta t) \mapsfrom \big(\mathbf{I} - \mathbf{K}((k+1) \Delta t) \mathbf{H} \big) \widehat{\mathbf{P}}((k+1)\Delta t-)$
		\STATE $k \mapsfrom k+1$
		\ENDWHILE
	\end{algorithmic}
\end{algorithm}

\section{Numerical Testing}
\label{S:numerics}
\subsection{Example 1}
\begin{figure}[H]
	\centering
	\resizebox{0.8\textwidth}{!}{%
		\includegraphics{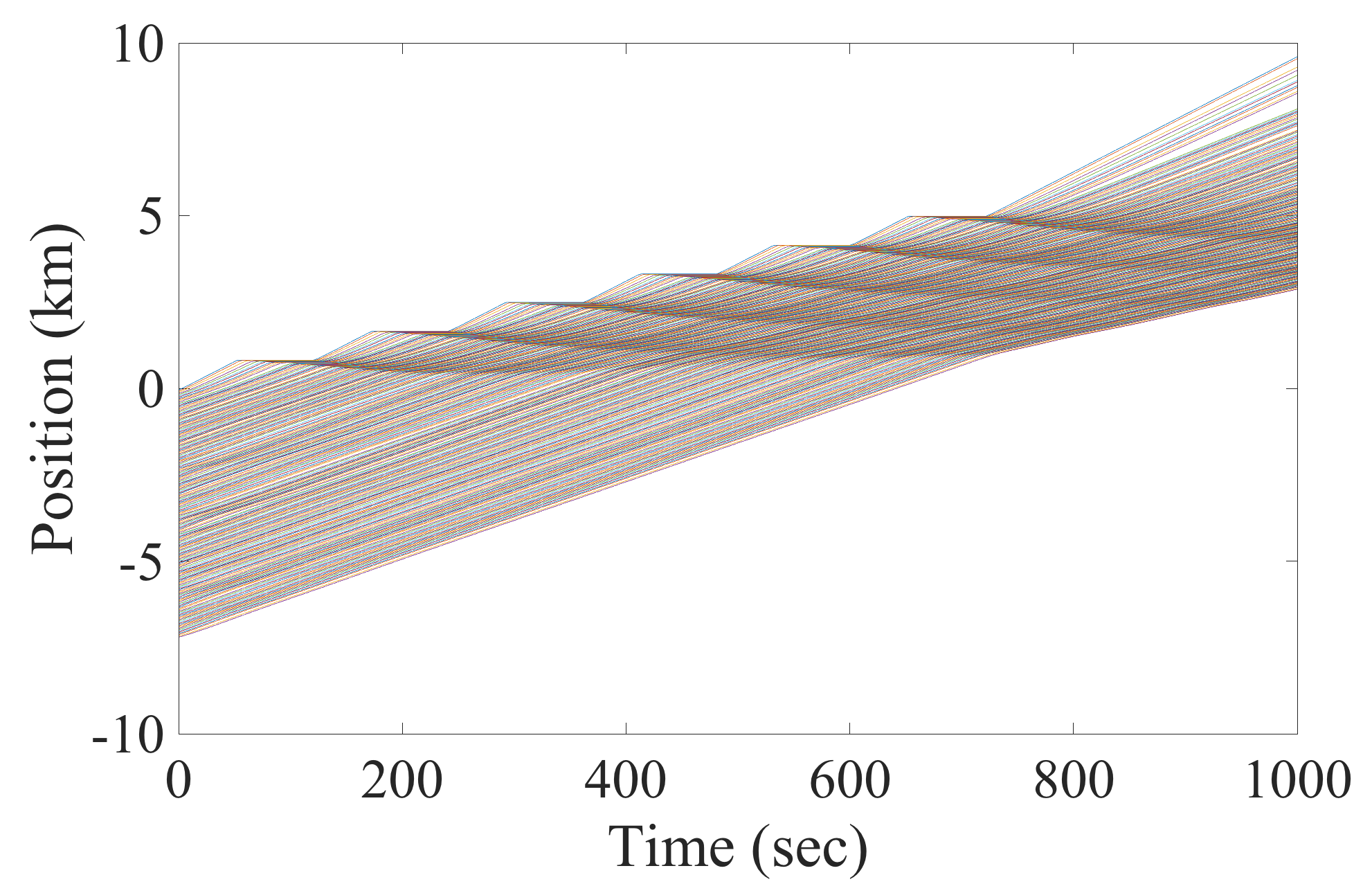}
		
	
		\includegraphics{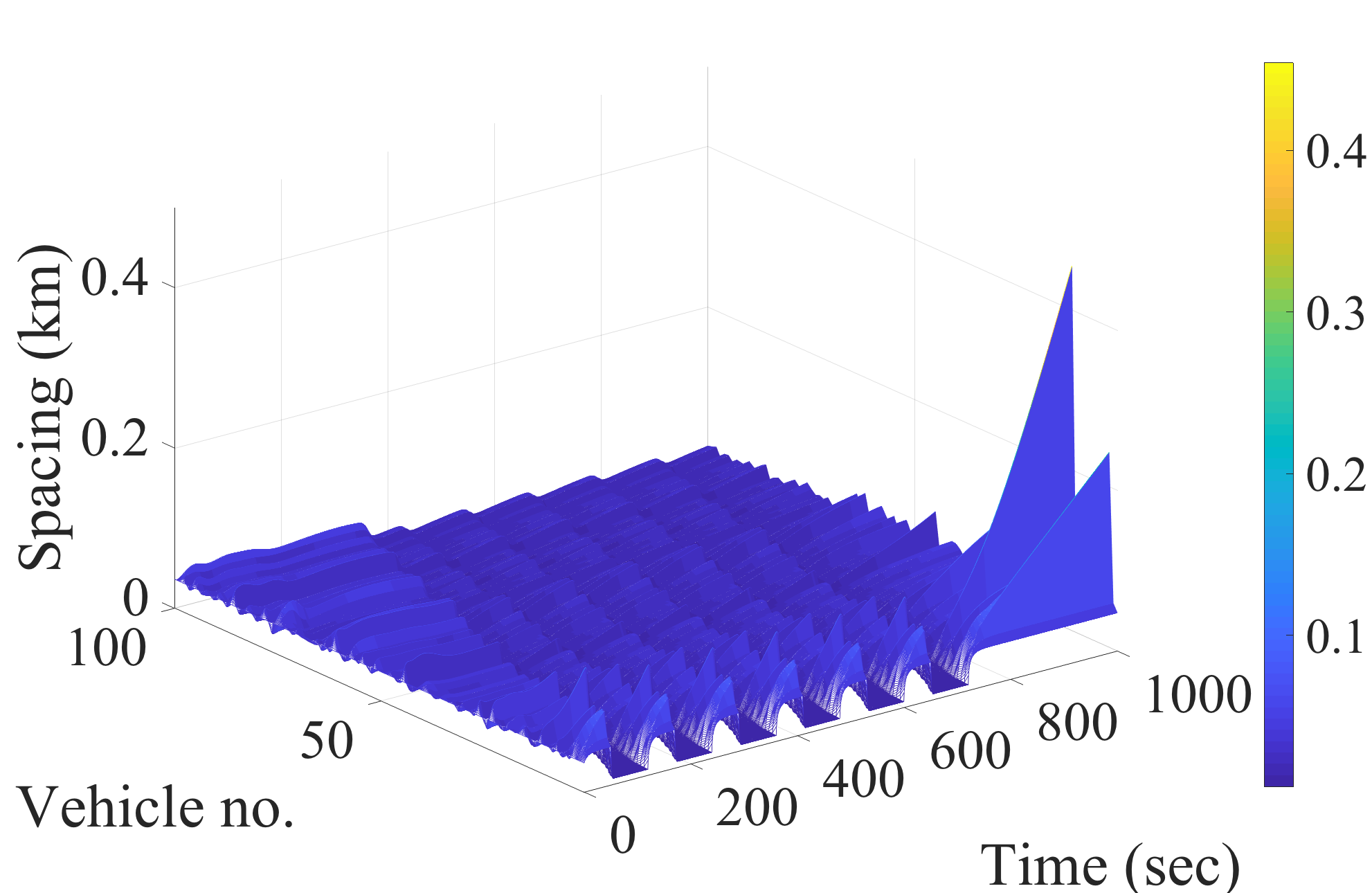}}
	
	
	(a) \hspace{2in} (b) 
	
	\caption{Simulated sample paths; (a) position trajectories $\mathbf{x}(\cdot)$, (b) spacings $\mathbf{s}(\cdot)$.} \label{f_sim1}
\end{figure}

Consider a system with $N=200$ vehicles (for example, made available by a fixed sensor in the system) and a time horizon of $T=1000$  seconds.  Assume a uniform spacing of 0.036 km at time $t=0$, that is $\mathbf{s}(0) = [0.036 ~ \cdots ~ 0.036]^{\top}$. The leader’s speed trajectory is given by: 
\begin{equation}
	v_0(t) = \left\{
	\begin{array}{cc}
		0 \mbox{ km/hr} & \mbox{if } t \in (jT_{\mathrm{r}} - T_{\mathrm{r}}, jT_{\mathrm{c}}] \mbox{ sec} \\
		60 \mbox{ km/hr} & \mbox{otherwise}
	\end{array}, \right. \label{E:leadV}
\end{equation}
where $T_{\mathrm{c}} = 120$ seconds is the cycle length, $T_{\mathrm{r}} = 70$ seconds is the red time,  and $j \in \{1,\cdots,6\}$.  The way we specify the leading vehicle’s trajectory is to create congestion such as vehicles waiting for the red signal at intersections.  We assume in this example that $v_{\ff}(\omega)$, $d(\omega)$, and $c(\omega)$ are independent Beta random variables with supports $[v_{\ff}^{\min},v_{\ff}^{\max}] = [40,80]$ km/hr, $[d^{\min},d^{\max}] = [5.88,9.09]$ meters, and $[c^{\min},c^{\max}] = [1100,5100]$ veh/hr.
We test the impact of increasing vehicle trajectory measurements on the uncertainty of traffic states in the system using \autoref{alg:KalmanBucy}.  \autoref{f_sim1} depicts a sample path of the stochastic process. 

\begin{figure}[h!]
	\centering
	\resizebox{0.8\textwidth}{!}{%
		\includegraphics{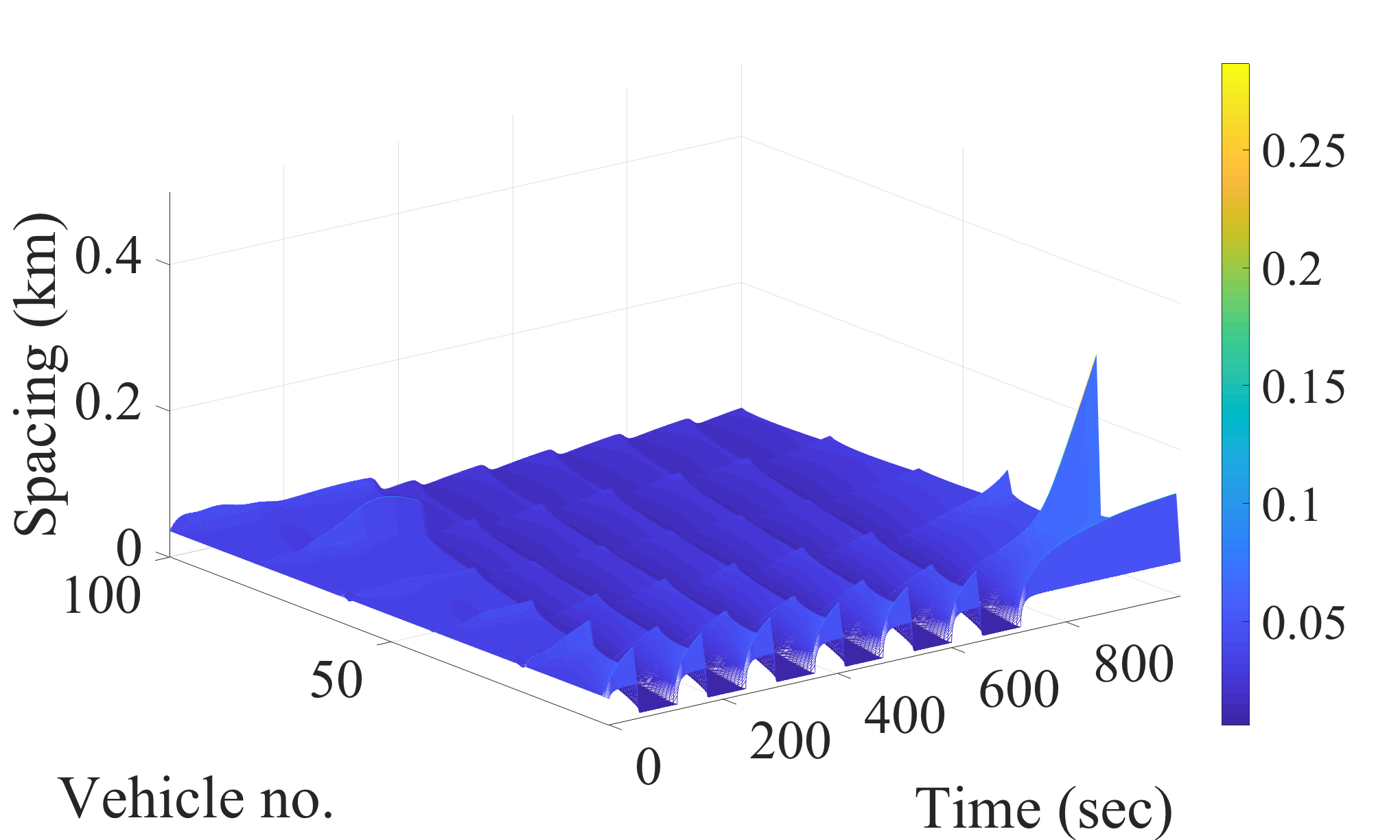}
		\includegraphics{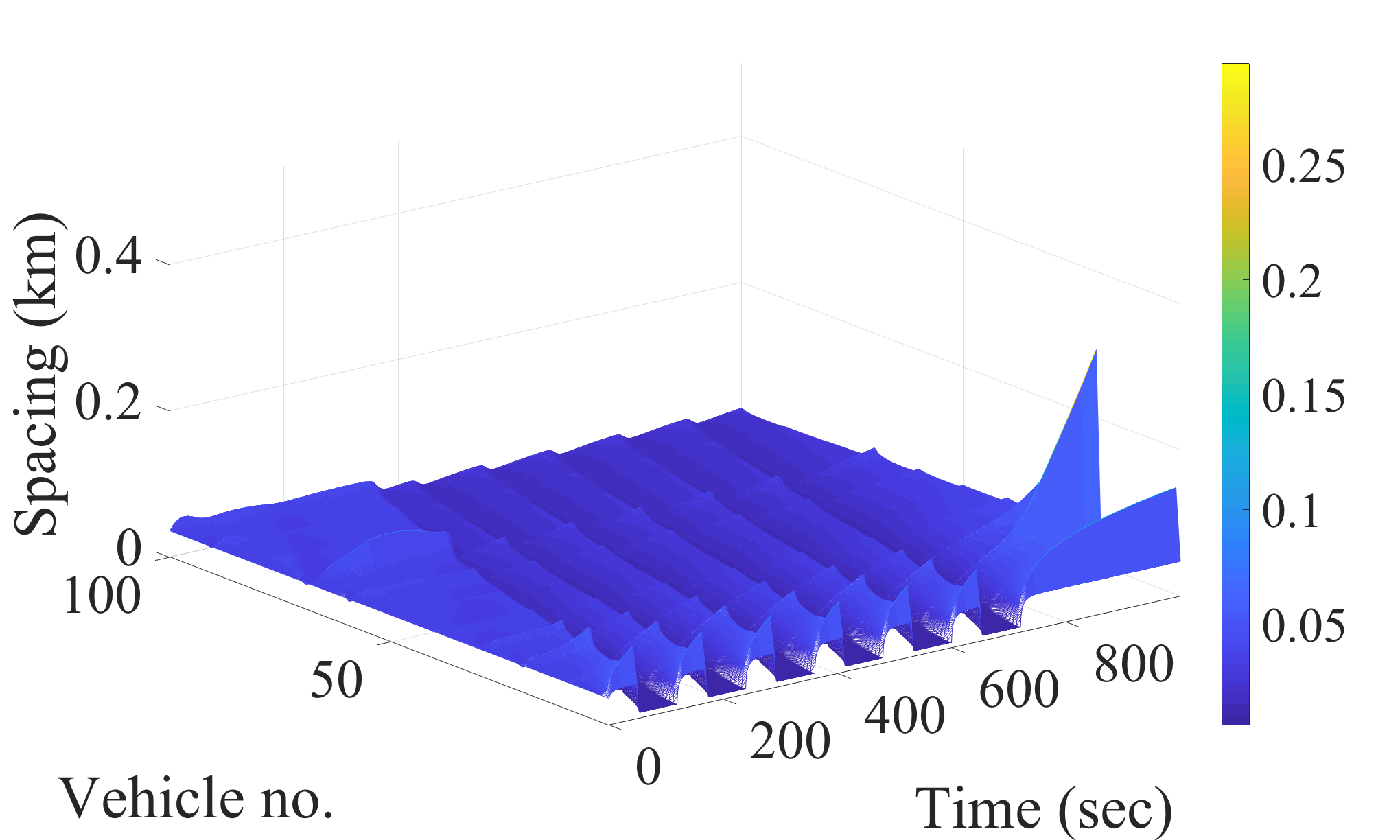}}
	
	(a) \hspace{2.4in} (b) 
	\resizebox{0.8\textwidth}{!}{%
		\includegraphics{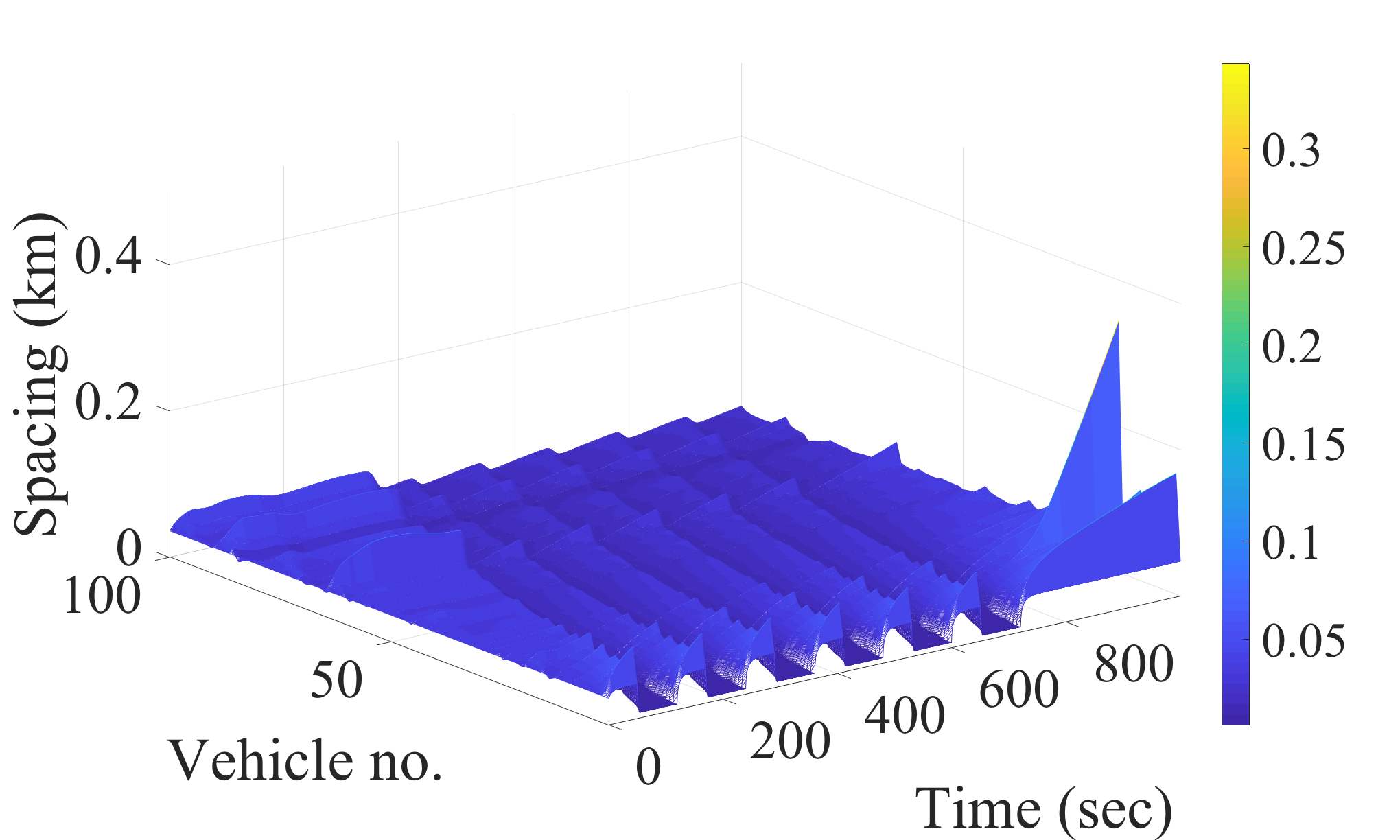}
		\includegraphics{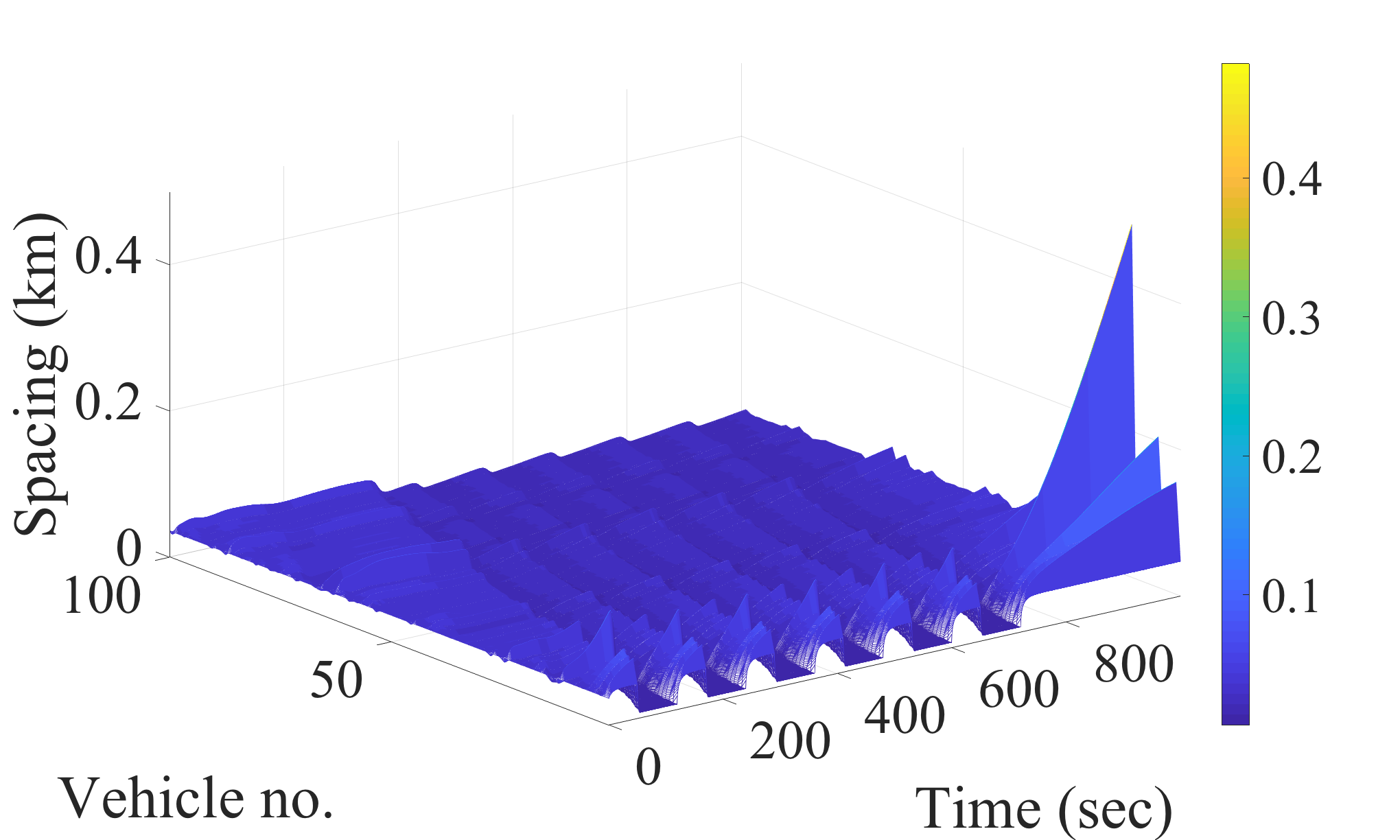}}
	
	(c) \hspace{2.4in} (d) 

	\resizebox{0.4\textwidth}{!}{%
		\includegraphics{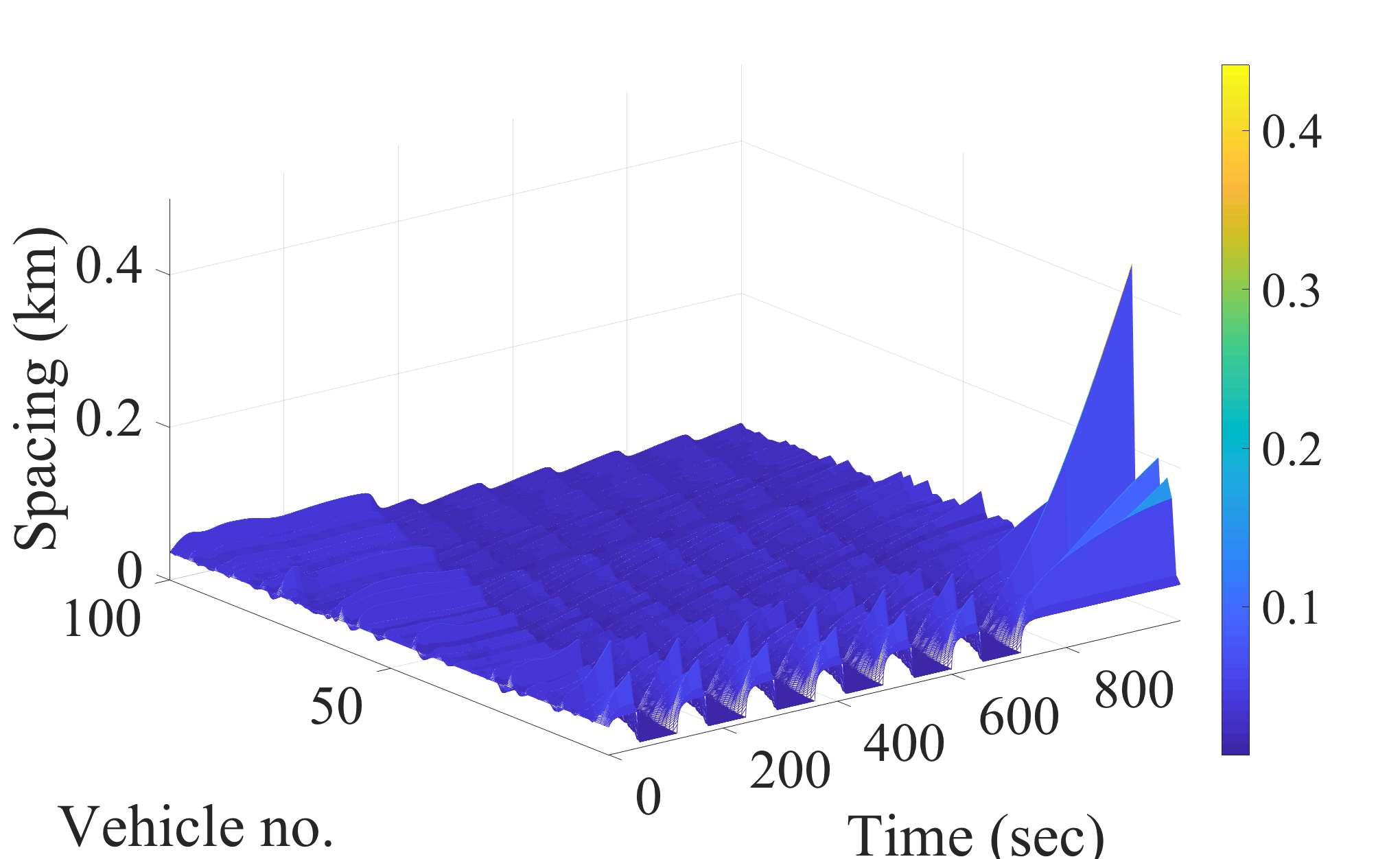}}
		
	(e)
	\caption{Estimated spacings with different penetration rates; (a) 5\%, (b) 10\%, (c) 20\%, (d) 30\%, and (e) 50\%.} \label{f_sim2}
\end{figure}

To see the impact of data availability, we consider five cases of vehicle penetration rate: 5\%, 10\%, 20\%, 30\% and 50\%. There is clear improvement in the estimate from low penetration rate (5\%) to higher penetration rates (30\%) as shown in \autoref{f_sim2}.

\autoref{t1} illustrates the estimation performance for 200 vehicles in terms of Root Mean Square Error (RMSE) in spacing and Mean Absolute Percentage Error (MAPE) in spacing. The estimation performance is improved significantly when the penetration rate is higher than 20\%.
\begin{table}[h!]
	\centering
	\small
	\caption{Estimation performance of spacings}
	\begin{tabular}{||c||c|c|c|c|c||}
		\hline \hline
		Penetration rate & 5\% & 10\% & 20\% & 30\% & 50\% \\ \hline
		RMSE (m) & 11.5 & 11.4 & 11.4 & 7.3 & 6.2 \\
		MAPE (\%) & 17.6 & 17.5 & 17.1 & 14.4 & 12.2 \\ \hline
		\hline
	\end{tabular}
\label{t1}
\end{table}

\begin{figure}[h!]
	\centering
	\resizebox{0.8\textwidth}{!}{%
		\includegraphics{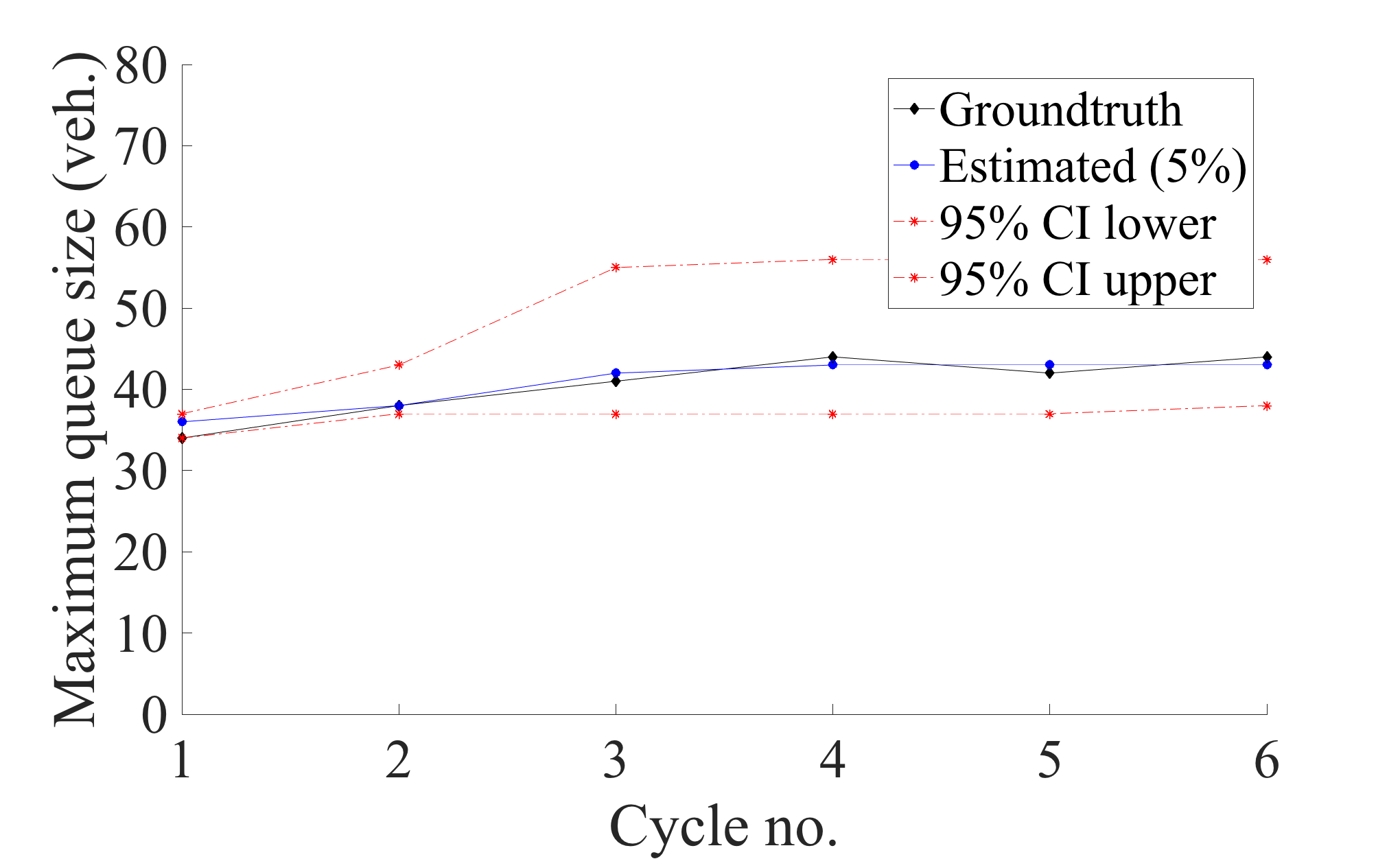}
		\includegraphics{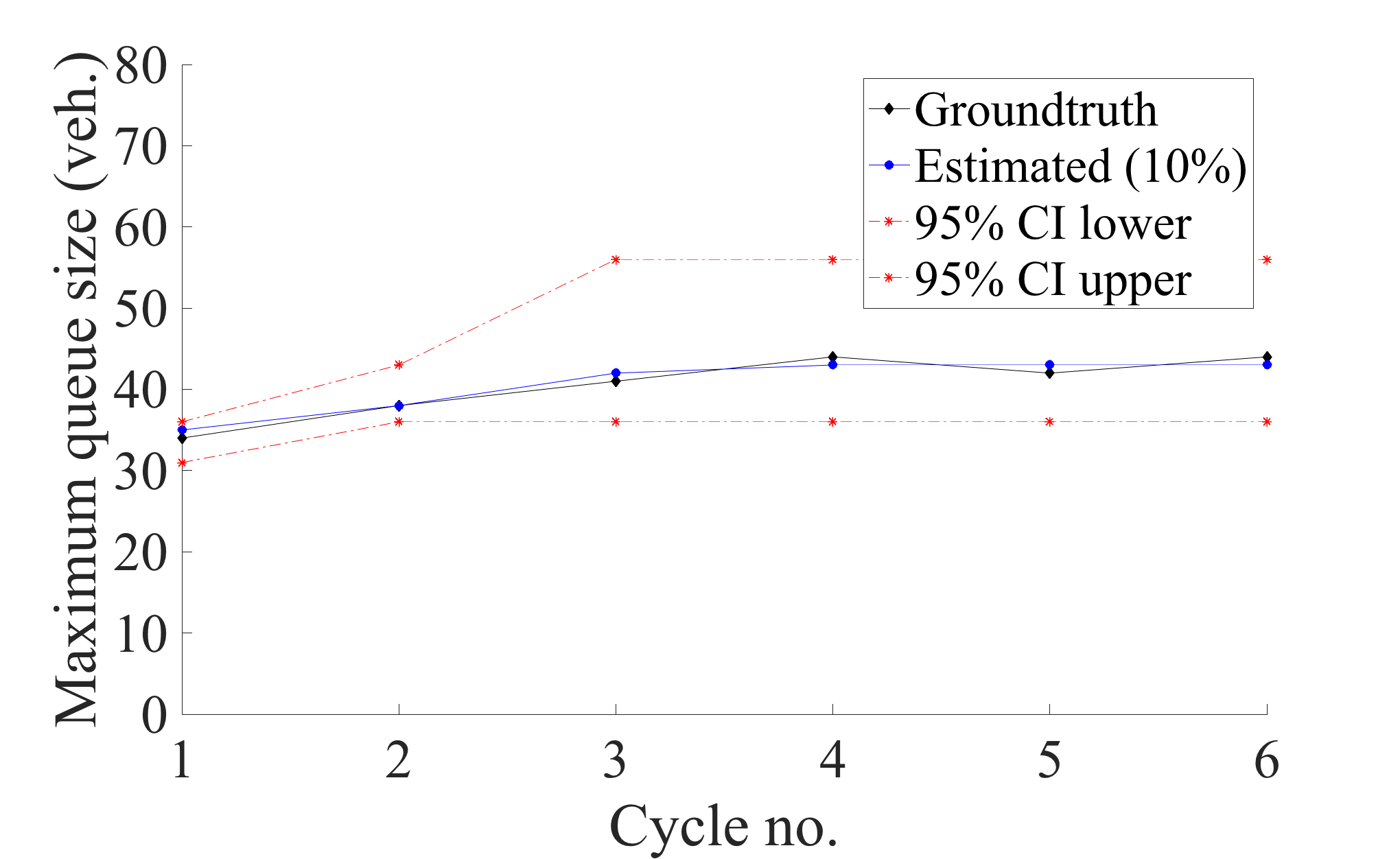}}
	
	(a) \hspace{2.4in} (b)

	\resizebox{0.8\textwidth}{!}{%
		\includegraphics{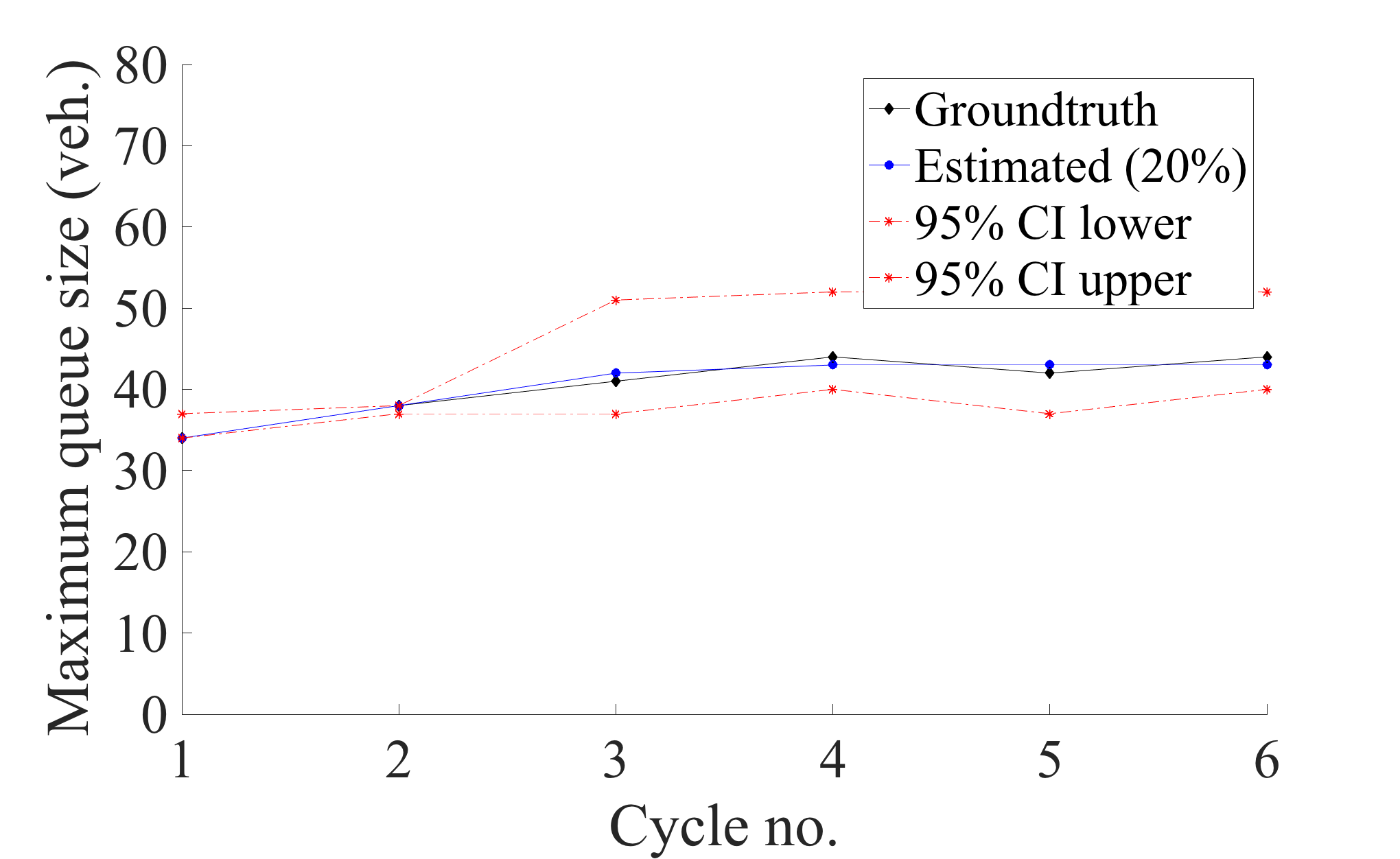}
		\includegraphics{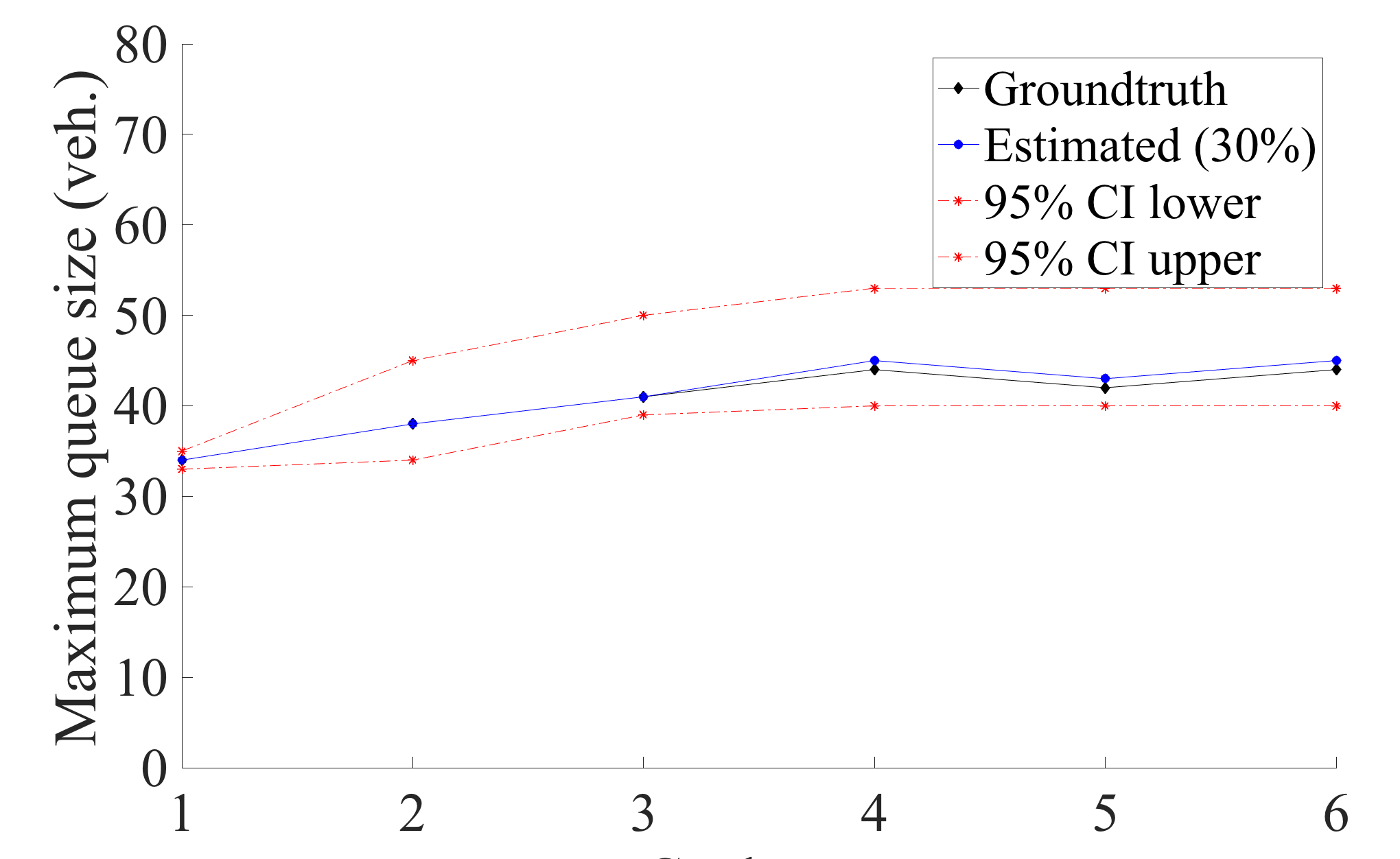}}
	
	(c) \hspace{2.4in} (d) 
	
	\resizebox{0.4\textwidth}{!}{%
		\includegraphics{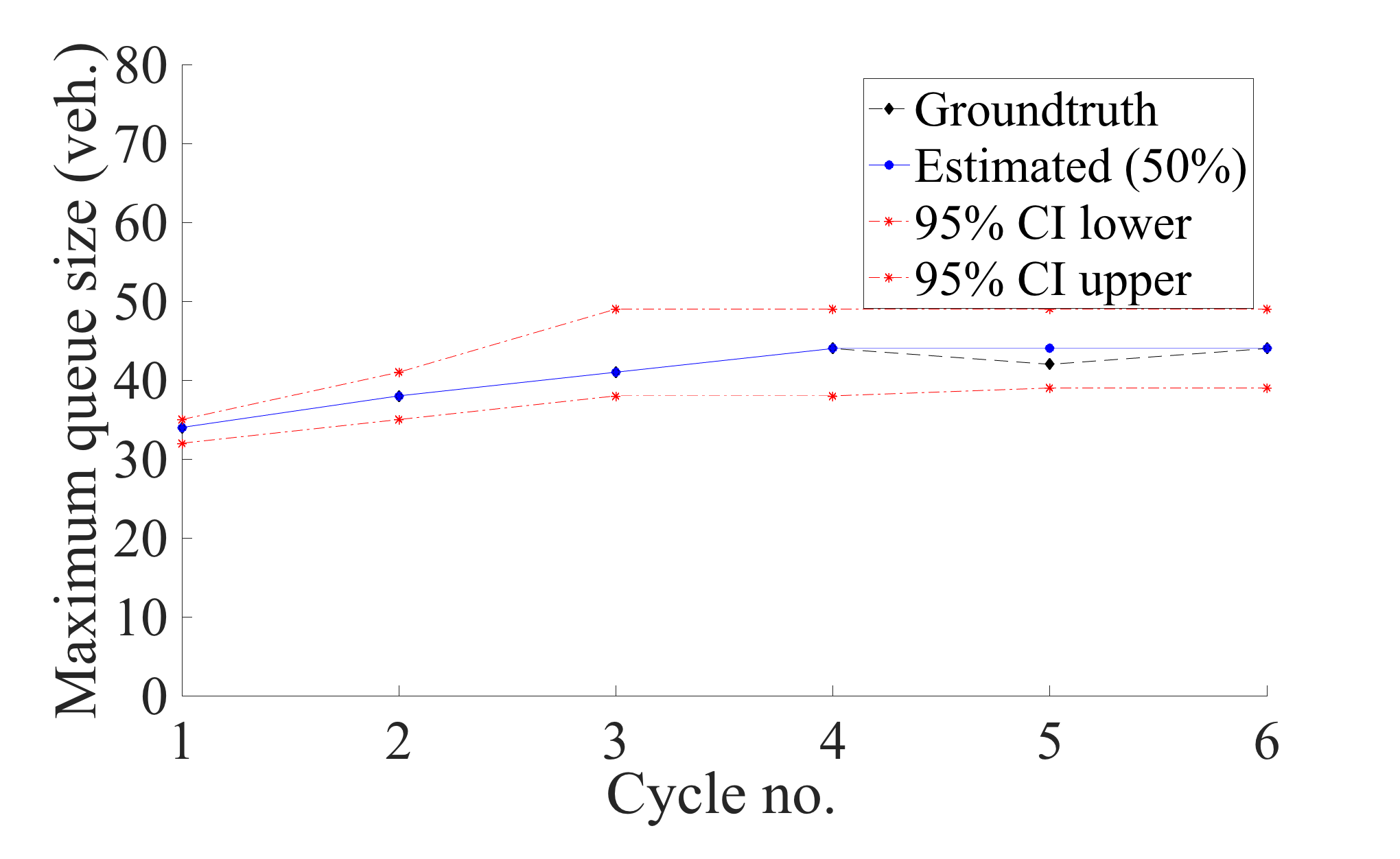}}

	
	(e)
	\caption{Maximum queue length along with 95\% confidence intervals; (a) 5\%, (b) 10\%, (c) 20\%, (d) 30\%, and (e) 50\%.} \label{f_queues}
\end{figure}

\autoref{f_queues} depicts the maximum queue sizes, along with 95\% confidence intervals. Queue sizes are not direct state variables but computed based on the estimated spacing (mean and covariance) and corresponding speed. 
\autoref{t2} lists the RMSE and MAPE for the different penetration rates. It can be seen that increasing the penetration rate from 5\% to 50\% results in the clear improvement of the estimation accuracy.
\begin{table}[h!]
	\centering
	\small
	\caption{Queue size estimation performance}
	\begin{tabular}{||c||c|c|c|c|c||}
		\hline \hline
		Penetration rate & 5\% & 10\% & 20\% & 30\% & 50\% \\ \hline
		RMSE (veh.) & 1.15 & 0.91 & 0.82 & 0.71 & 0.41 \\
		MAPE (\%) & 2.54 & 2.05 & 1.56 & 1.15 & 0.79 \\ \hline
		\hline
	\end{tabular}
	\label{t2}
\end{table}

\subsection{Example 2: Microscopic traffic simulation example}
\begin{figure}[h!]
	\centering
	\resizebox{0.7\textwidth}{!}{%
		\includegraphics{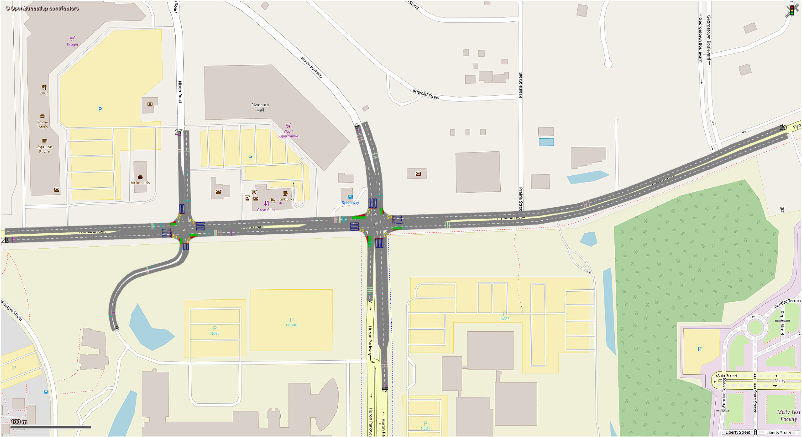}}
	\caption{The test road in Ann Arbor, Michigan.} \label{f_network}
\end{figure}

\begin{figure}[h!]
	\centering
	\resizebox{0.4\textwidth}{!}{%
		\includegraphics{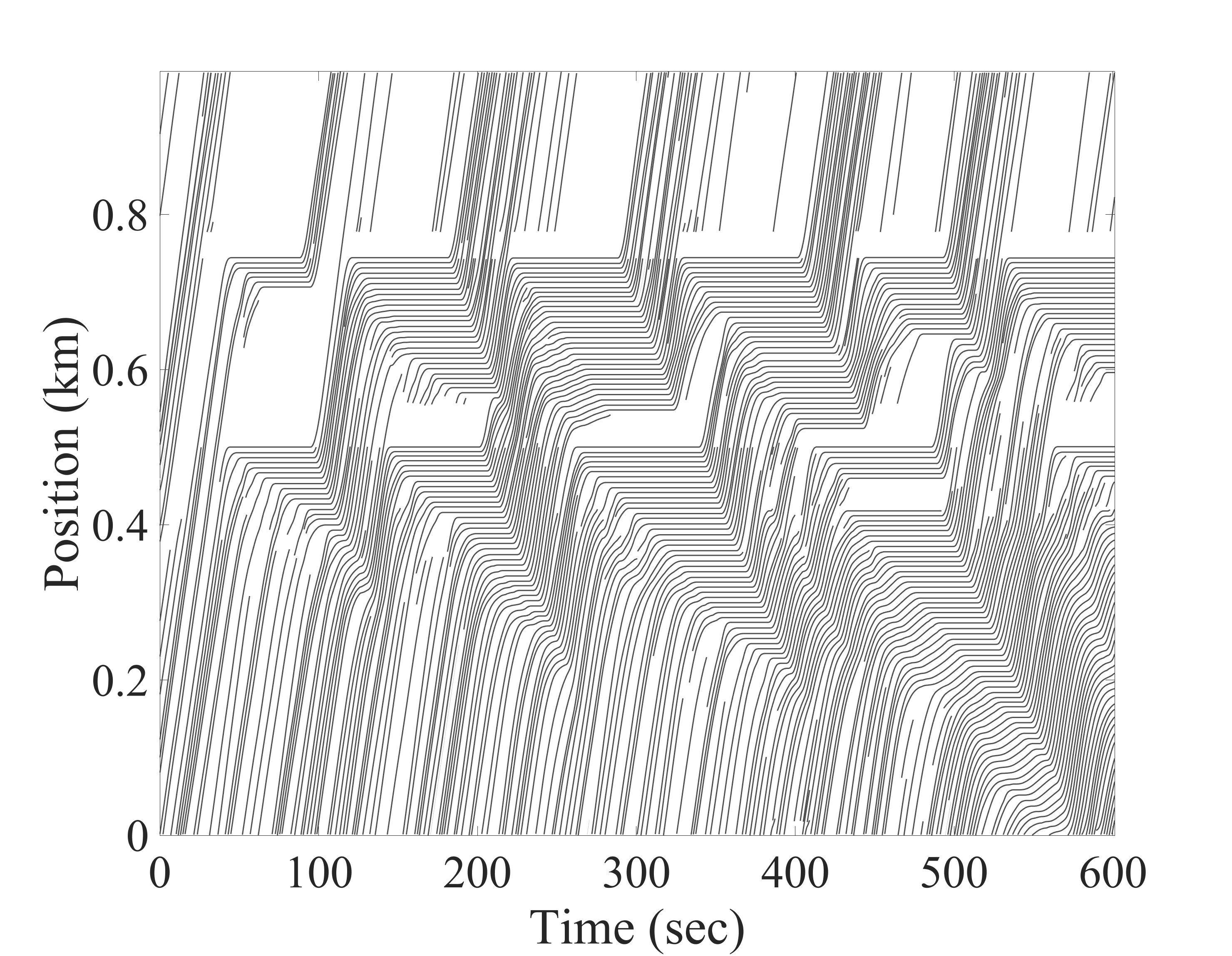}}
	\caption{Ground truth from calibrated microscopic model: position trajectories} \label{f_gTruth}
\end{figure}

\textbf{Data preparation}.  The test area we selected is Plymouth Road, which is an urban arterial road in the city of Ann Arbor, Michigan.  In order to obtain `ground truth data', we utilize a calibrated microscopic traffic simulation model of the test road.  \autoref{f_network} provides an illustration the test road (1 km in length) with two intersections (Huron Pkwy - Plymouth Rd and Nixon Rd - Plymouth Rd).  
We derive the trajectory data (positions and speeds) of through-going vehicles traveling westbound along Plymouth Road over time period of 600 seconds (see \autoref{f_gTruth}). 
The parameters $v_{\ff}(\omega)$, $d(\omega)$, and $c(\omega)$ are independent Beta random variables with supports $[v_{\ff}^{\min},v_{\ff}^{\max}] = [48,58]$ km/hr, $[d^{\min},d^{\max}] = [5.8,7.3]$ meters, and $[c^{\min},c^{\max}] = [1795,3767]$ veh/hr estimated from the simulated ground truth trajectory data. We assume the first vehicle trajectory in the system and the initial condition (spacing of vehicles at time $t = 0$) are known as well.  

\begin{figure}[h!]
	\centering
	\resizebox{0.9\textwidth}{!}{%
		\includegraphics{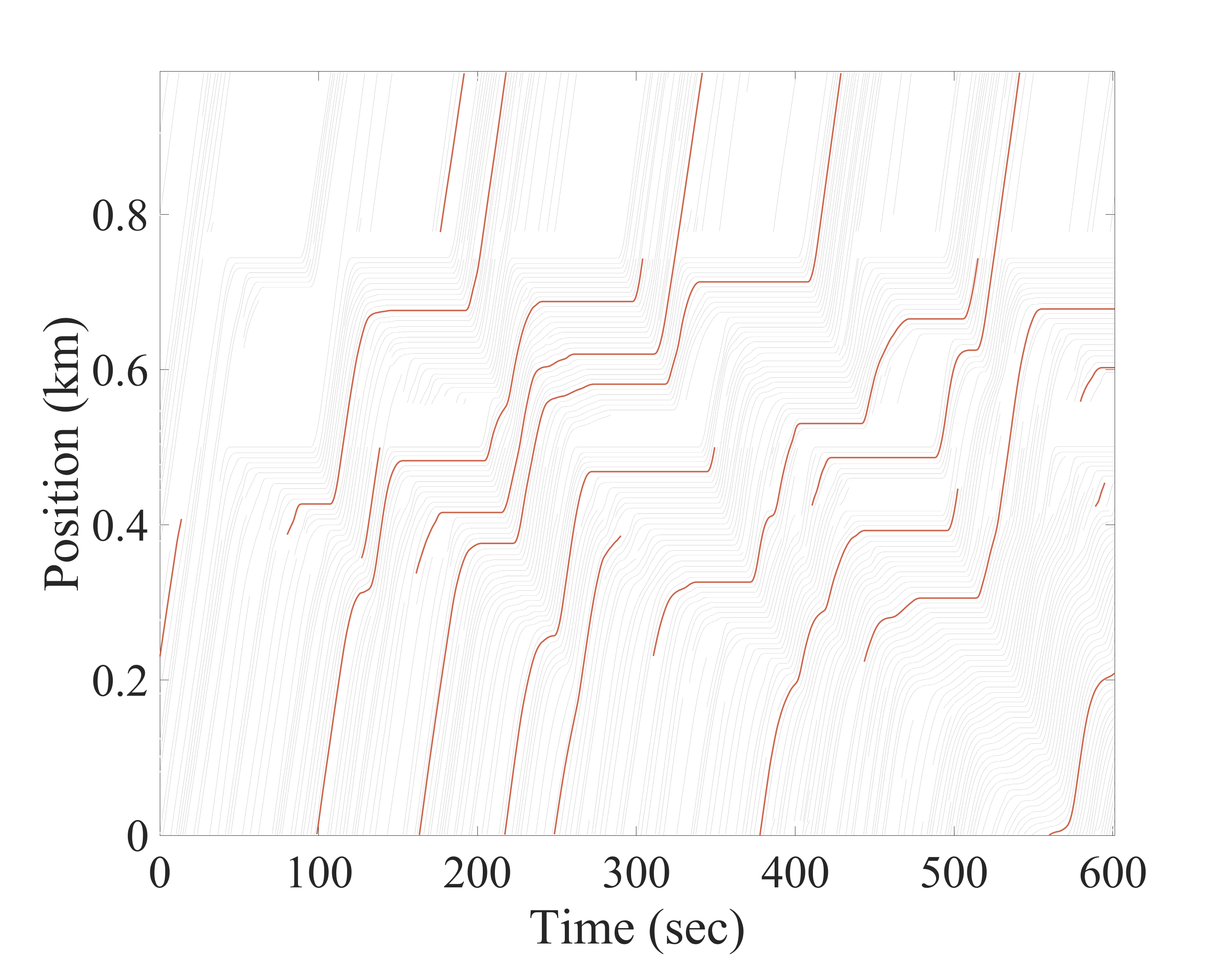}
		\includegraphics{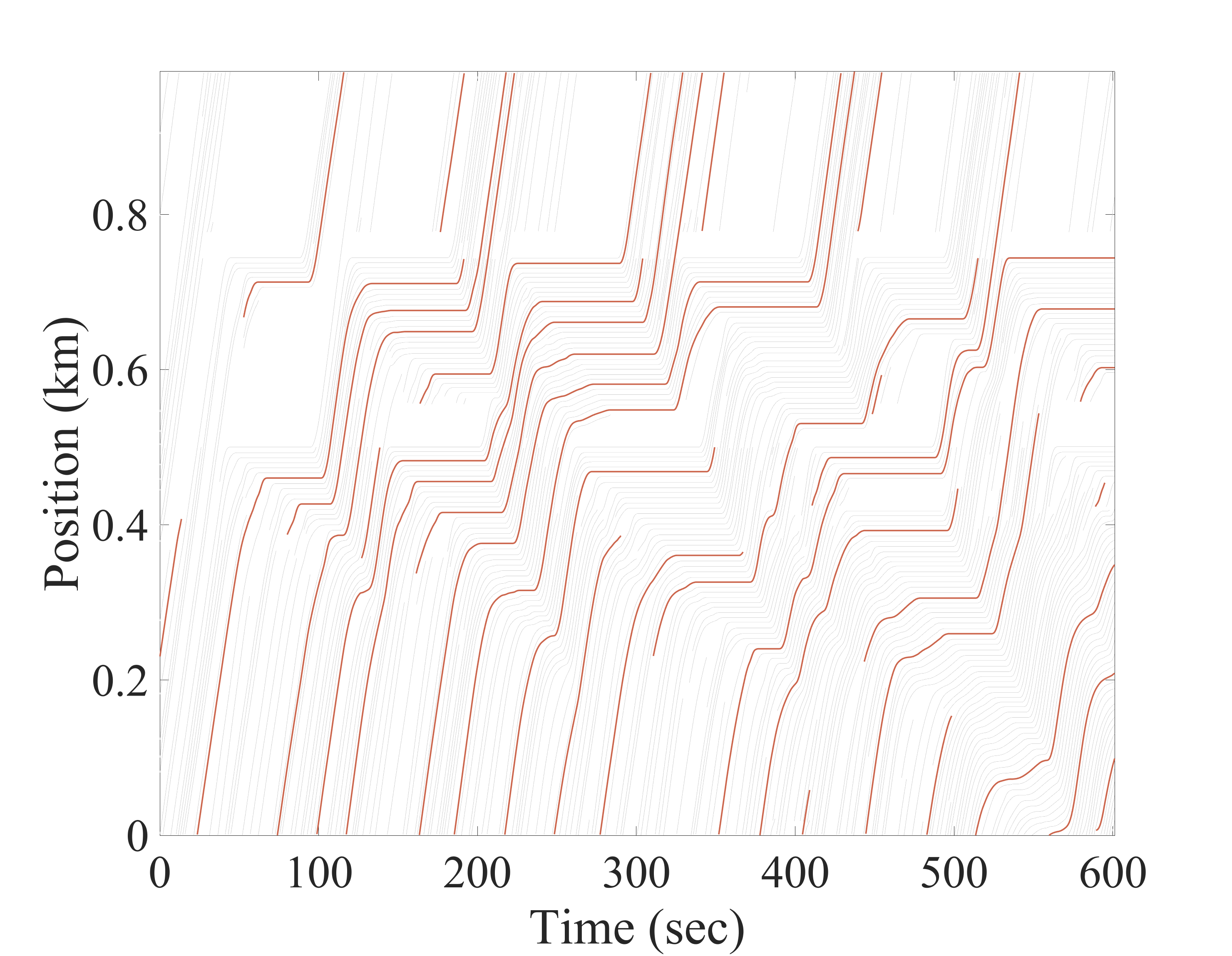}}
	
	(a) \hspace{2.4in} (b) 

	\resizebox{0.9\textwidth}{!}{%
		\includegraphics{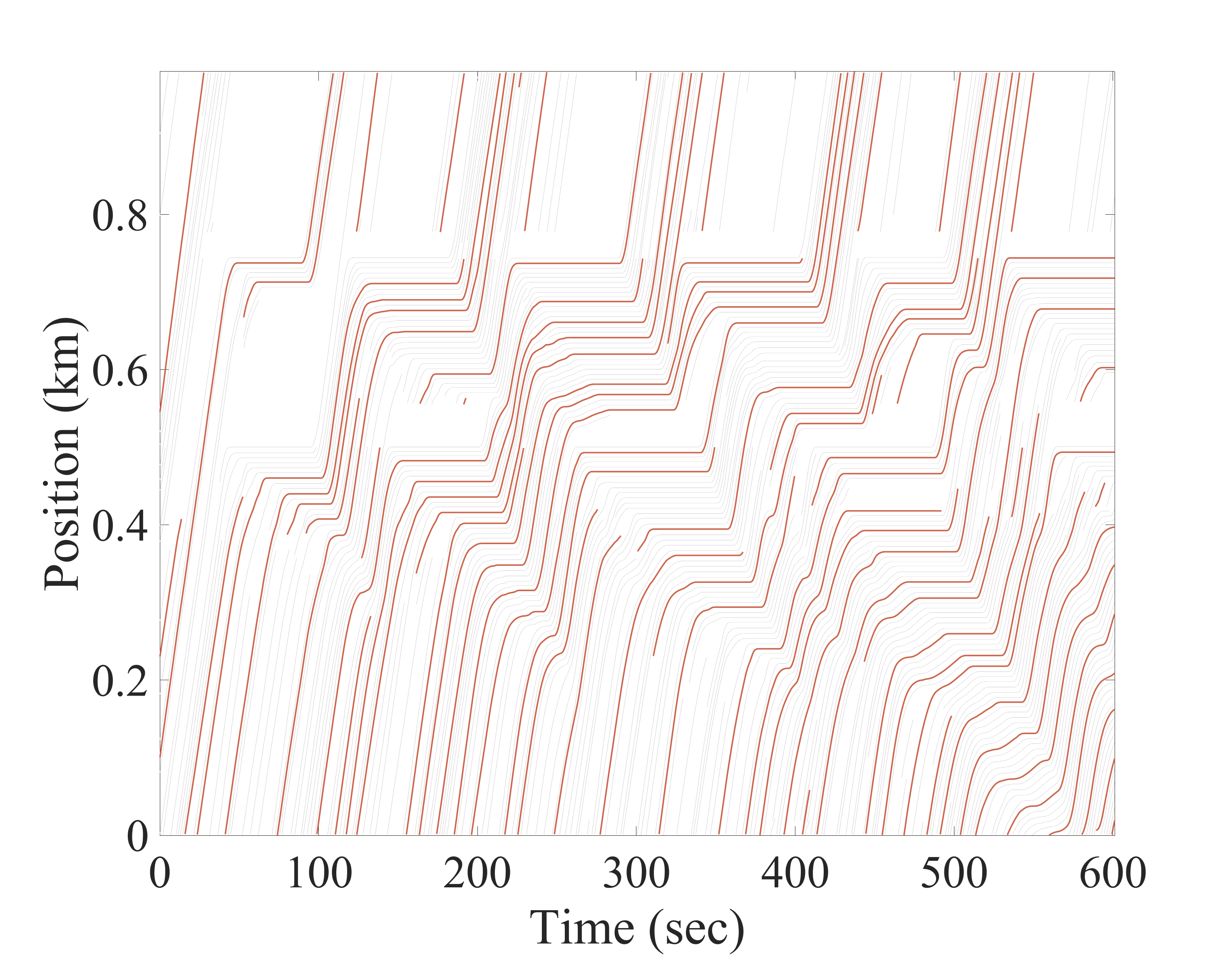}
		\includegraphics{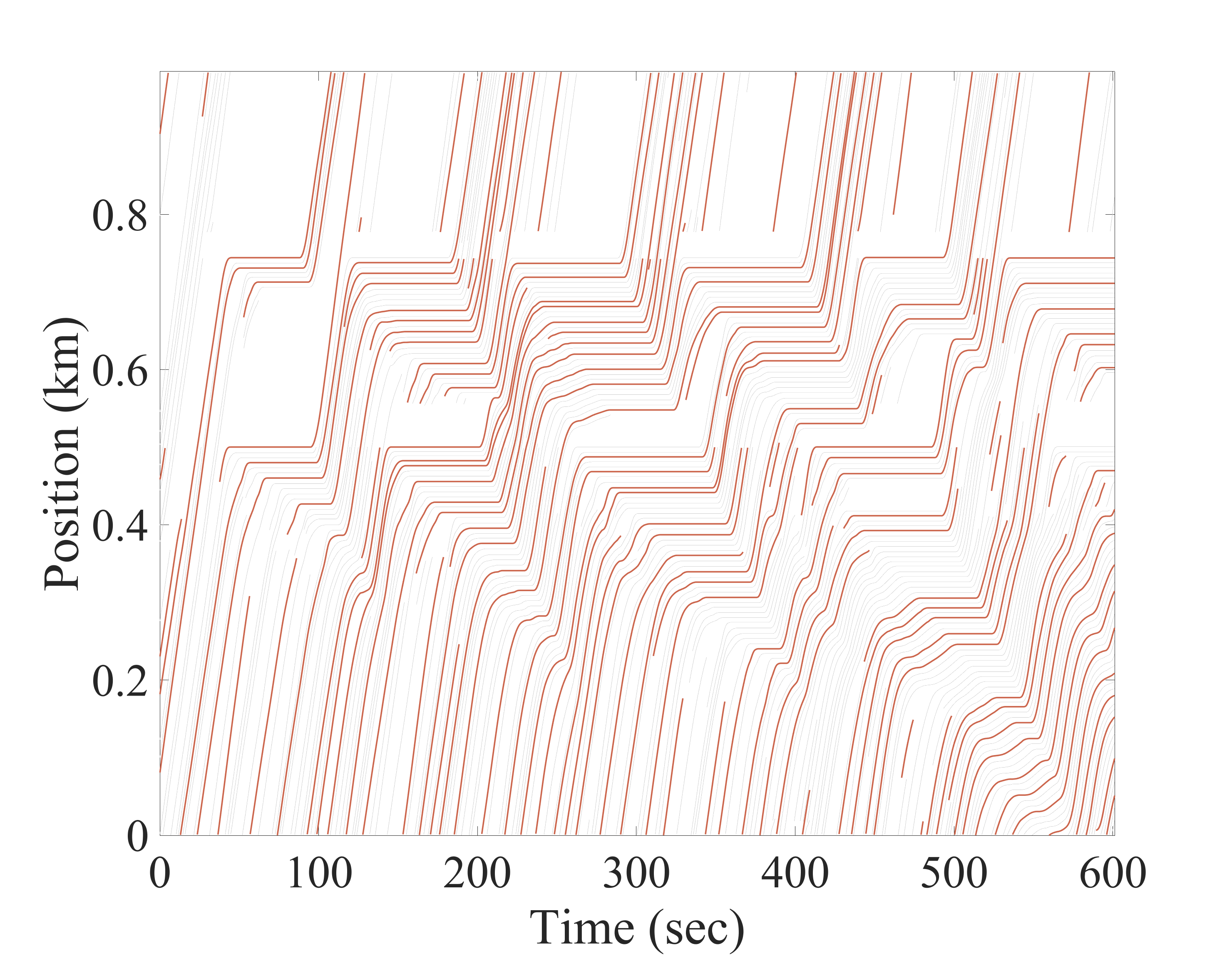}}
	

	(c) \hspace{2.4in} (d) 
	

	\resizebox{0.45\textwidth}{!}{%
		\includegraphics{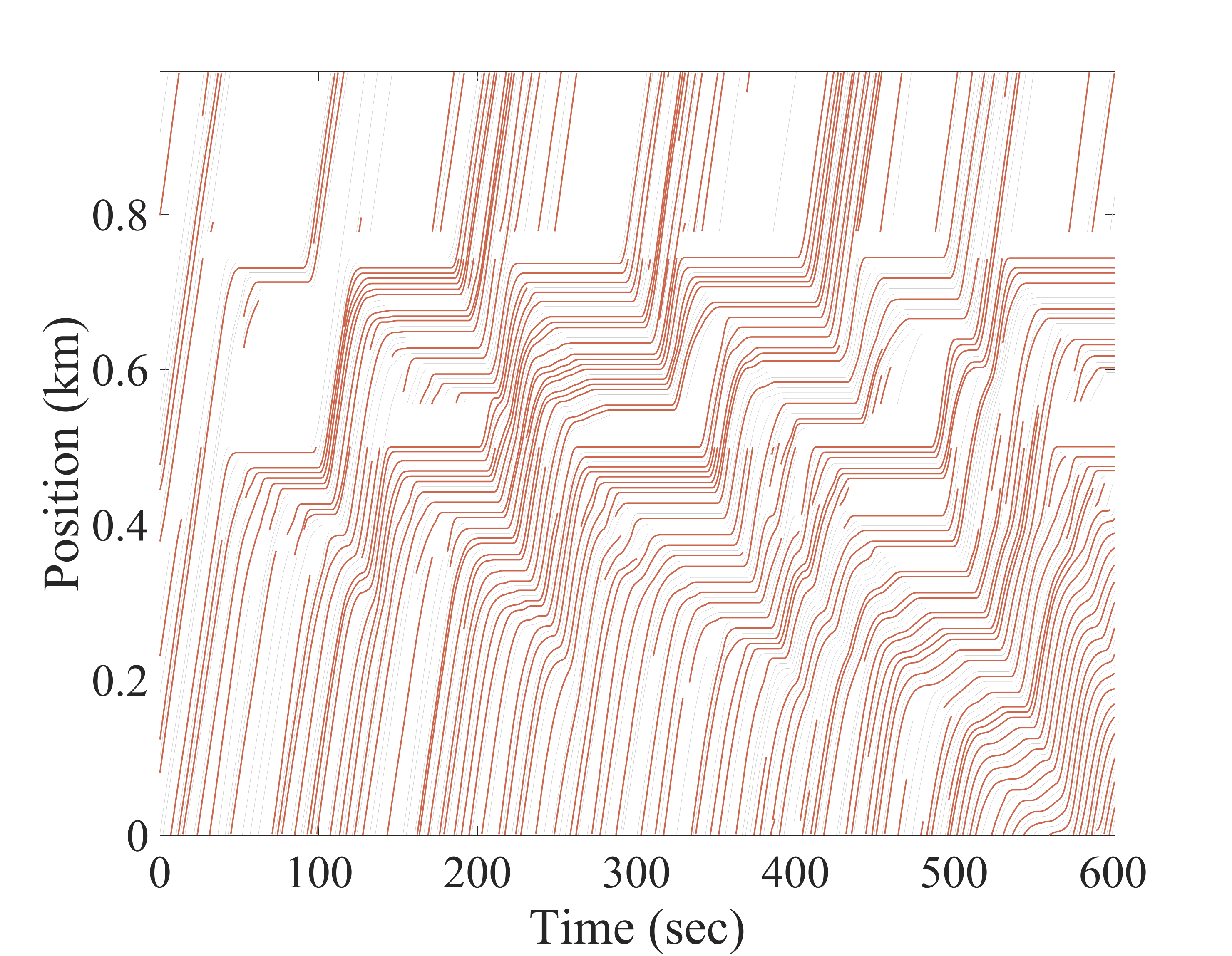}}

(e)

\caption{Measured trajectories (bold) against the ground truth trajectories for penetration rates (a) 5\%, (b) 10\%, (c) 20\%, (d) 30\%, and (e) 50\%.} \label{f_traj}
\end{figure}

To see the impact of data availability on the uncertainty of traffic state estimation, we consider five cases of penetration rate: 5\%, 10\%, 20\%, and 30\% and 50\%.  \autoref{f_traj} illustrates the position trajectories available (the measurements) for the different penetration rates used in our experiments.

\begin{figure}[h!]
	\centering
	\resizebox{0.9\textwidth}{!}{%
		\includegraphics{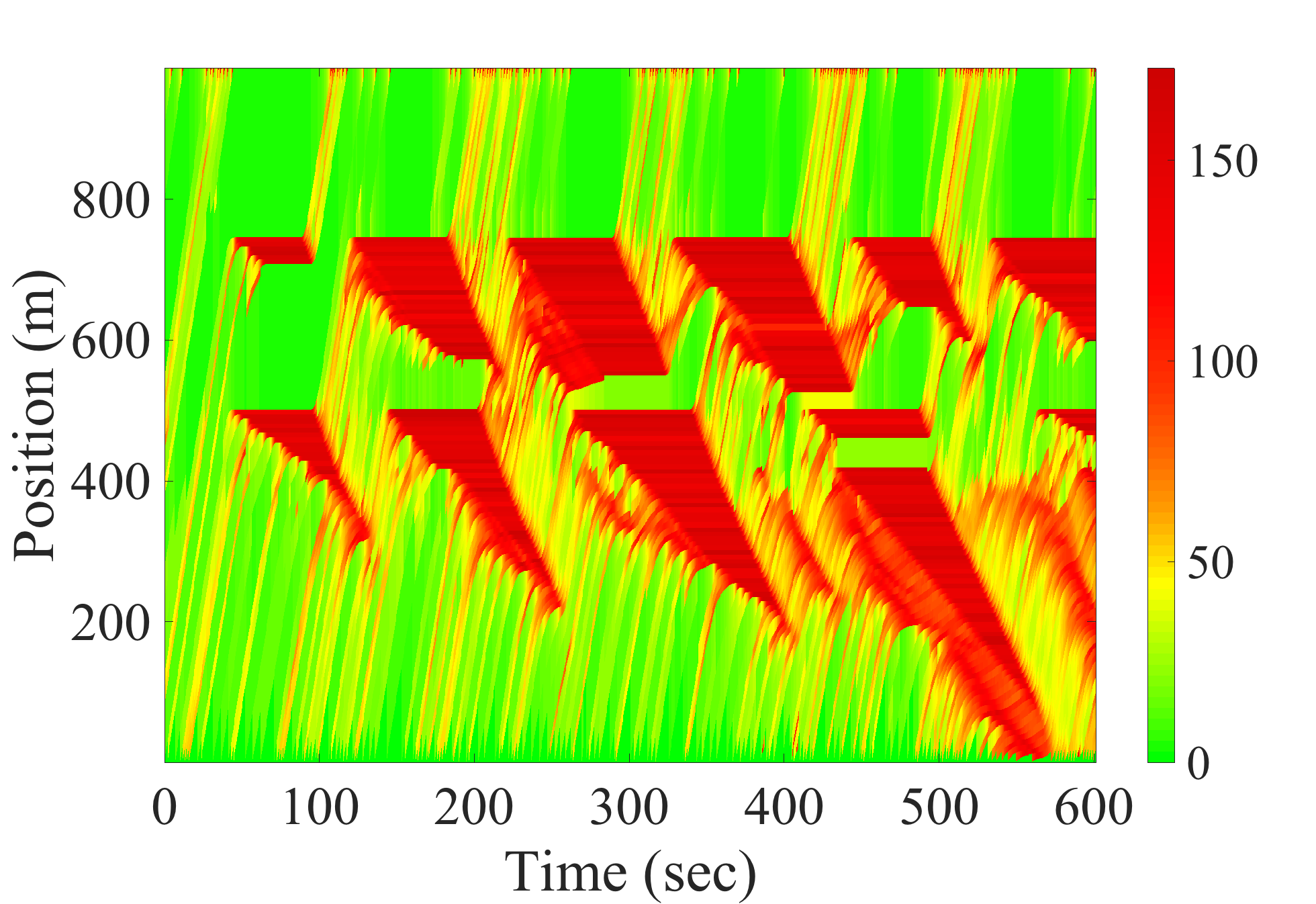}
		\includegraphics{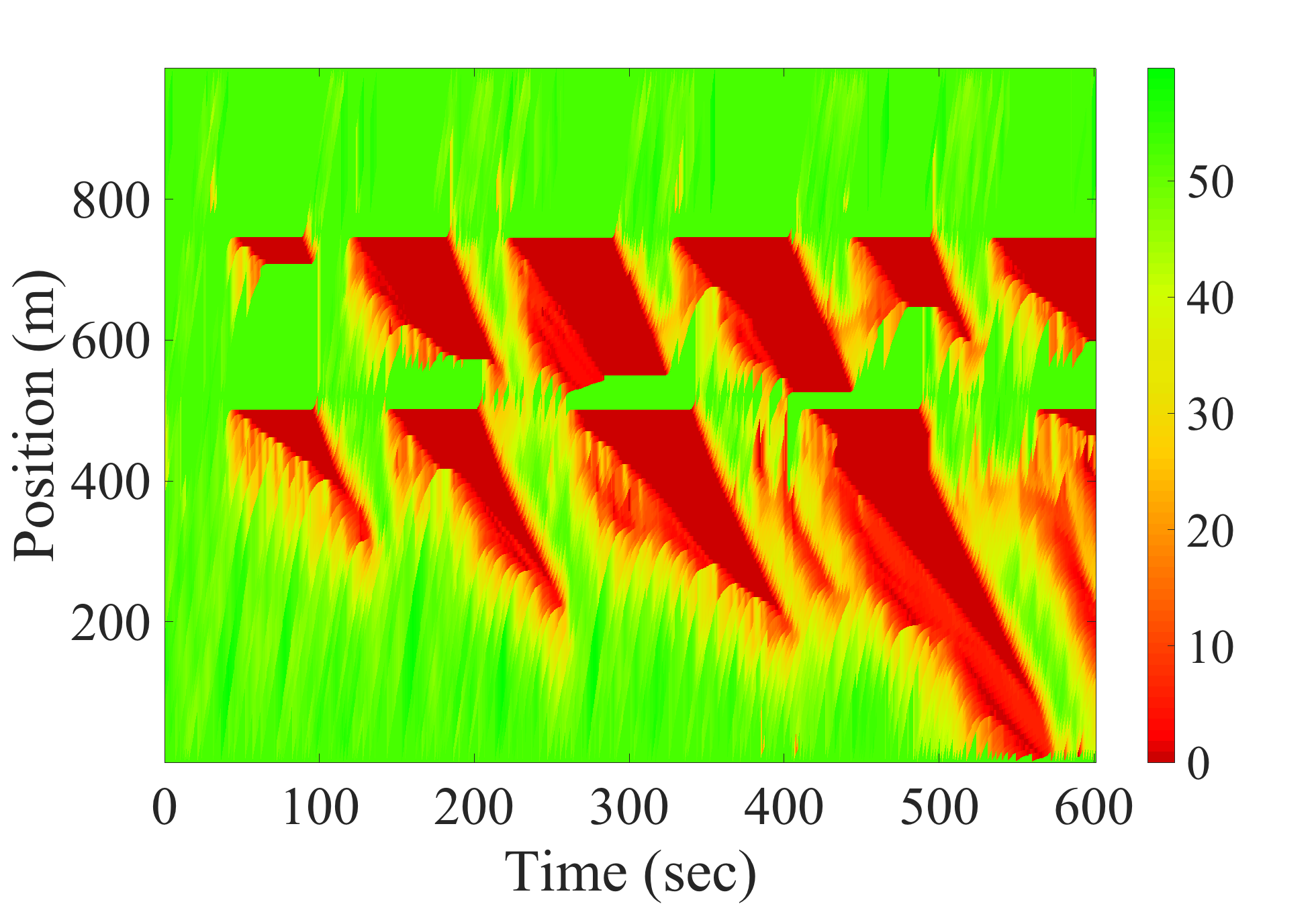}}
	
	(a) \hspace{2.4in} (b) 
	\caption{Ground truth from calibrated microscopic simulation model: (a) density dynamics (in veh/km), (b) speed dynamics (in km/hr).} \label{f_gTruth2}
\end{figure}

\textbf{Estimation results}. With estimated parameters of the model and given the first vehicle trajectory information (boundary condition) and the initial condition, the traffic state dynamics (spacing and positions) can be fully characterized.  We apply the Kalman-Bucy filter given in \autoref{alg:KalmanBucy}.  \autoref{f_gTruth2} depicts the ground truth dynamics in terms of traffic densities and speed fields.  

\begin{figure}[p]
	\centering
	\resizebox{0.8\textwidth}{!}{%
		\includegraphics{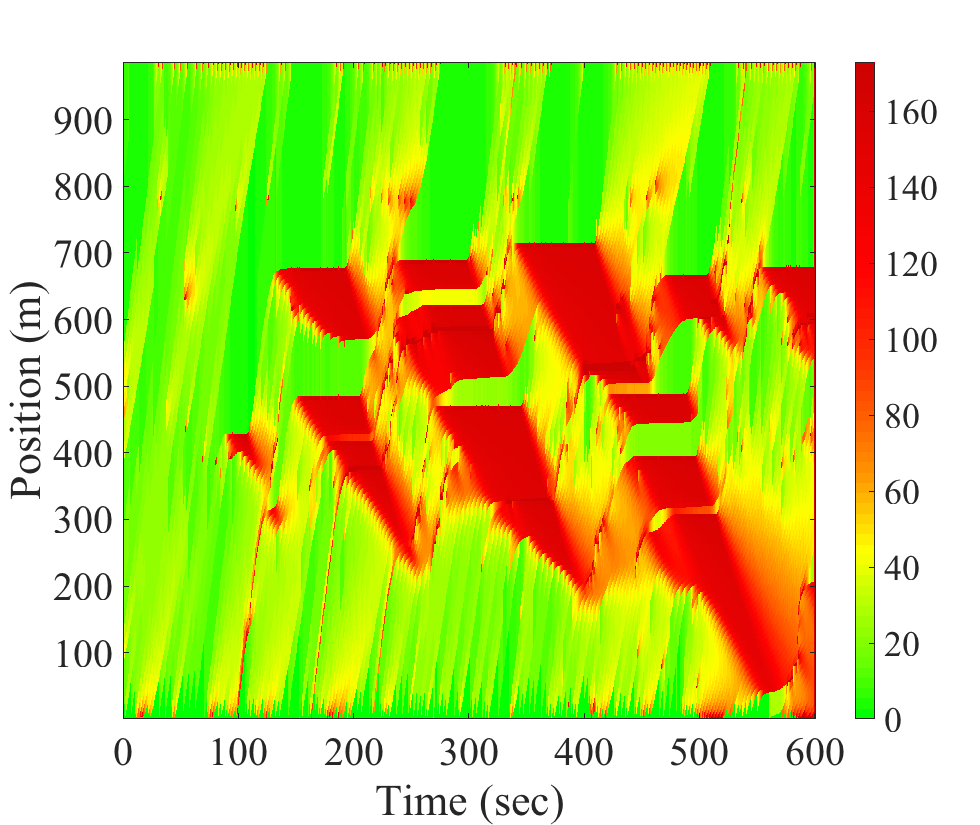}
		\includegraphics{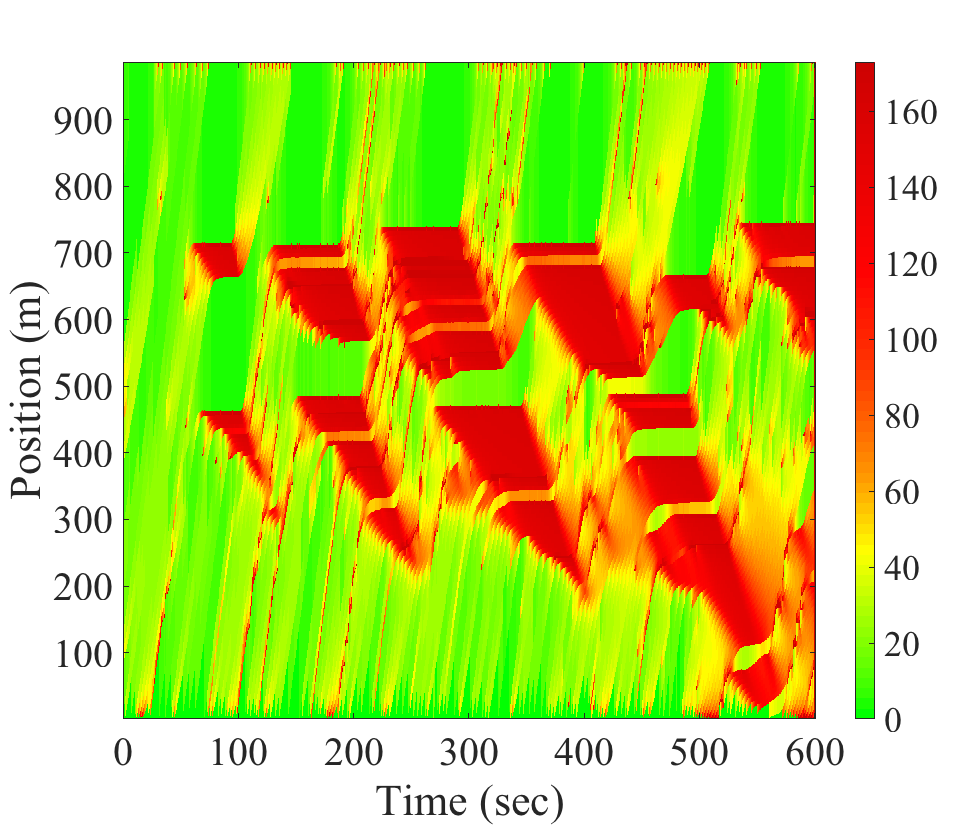}}
	
	(a) \hspace{2.0in} (b) 
	
	\resizebox{0.8\textwidth}{!}{%
		\includegraphics{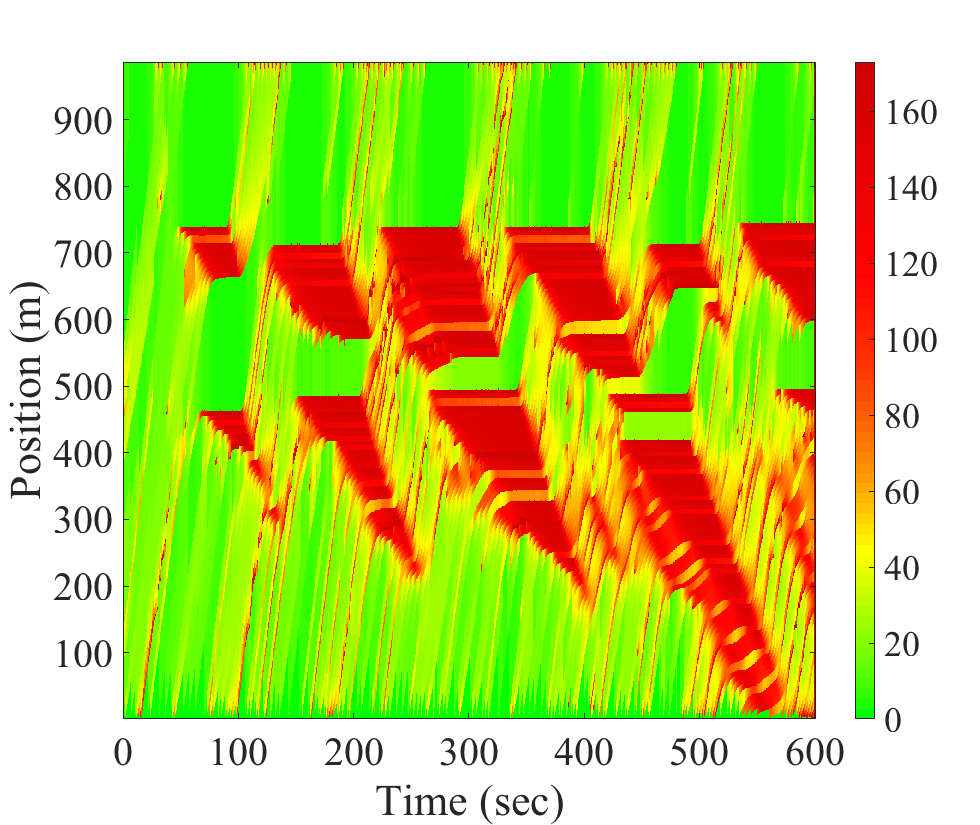}
		\includegraphics{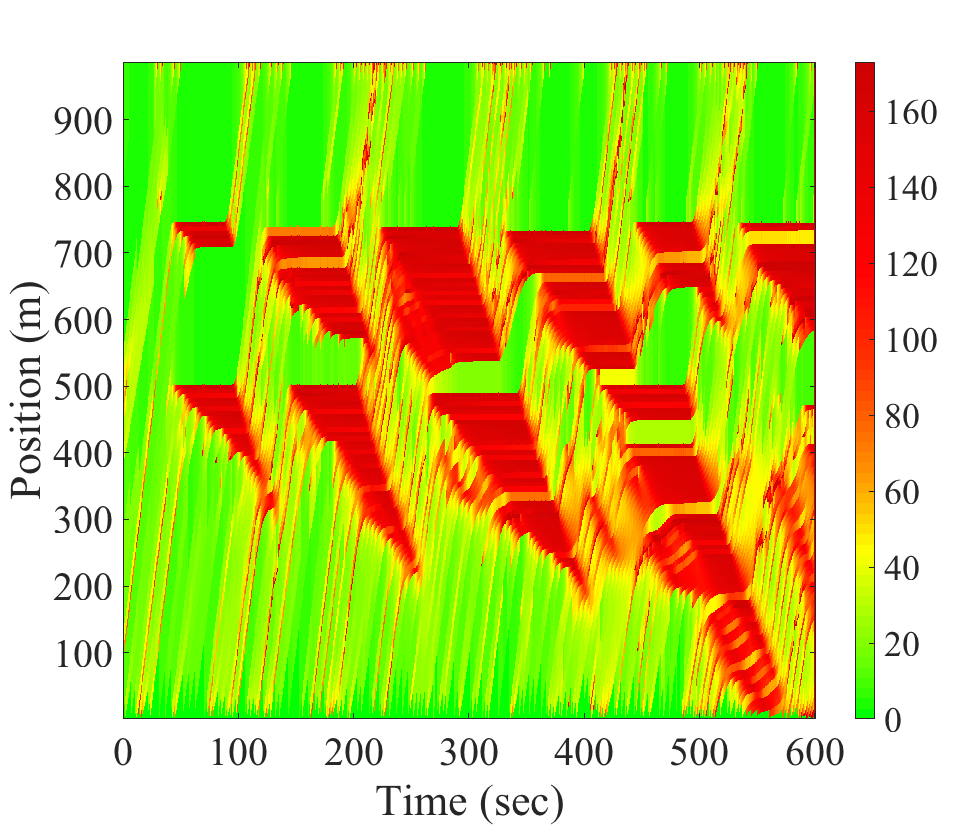}}
	
	(c) \hspace{2.0in} (d)


\resizebox{0.4\textwidth}{!}{%
	\includegraphics{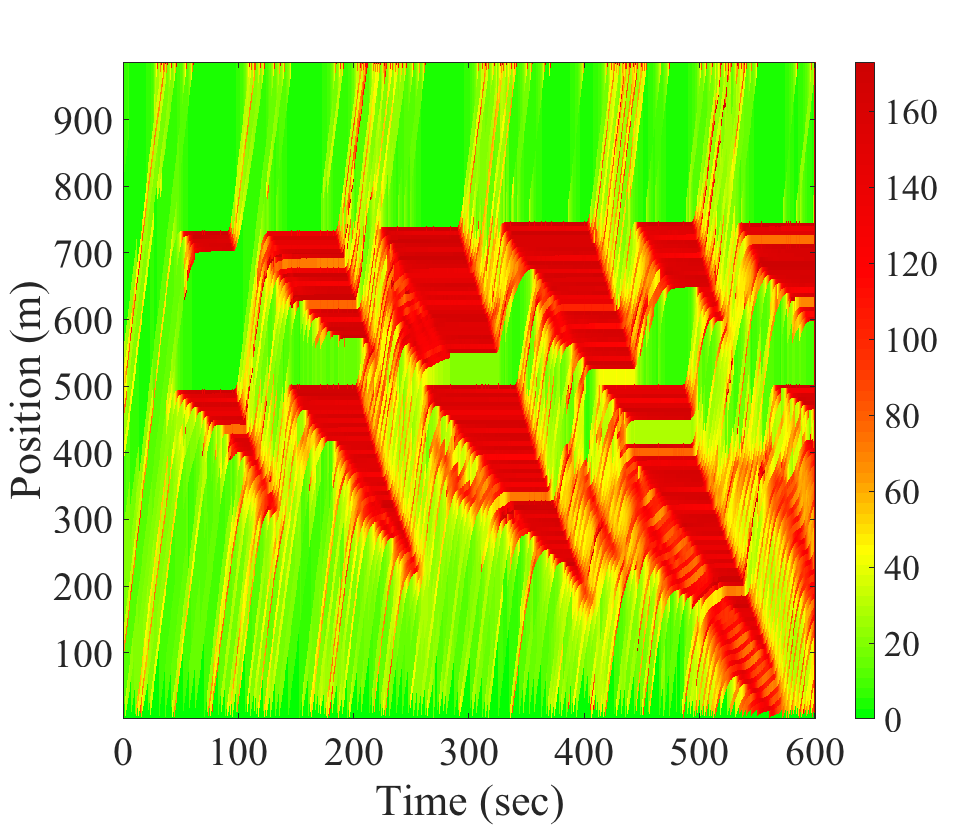}}

	(e)
	
	\caption{Estimated density fields (in veh/km) with different penetration rates:  (a) 5\%, (b) 10\%, (c) 20\%, (d) 30\% and (e) 50\%.} \label{f_den}
\end{figure}

\begin{figure}[h!]
	\centering
	\resizebox{0.8\textwidth}{!}{%
		\includegraphics{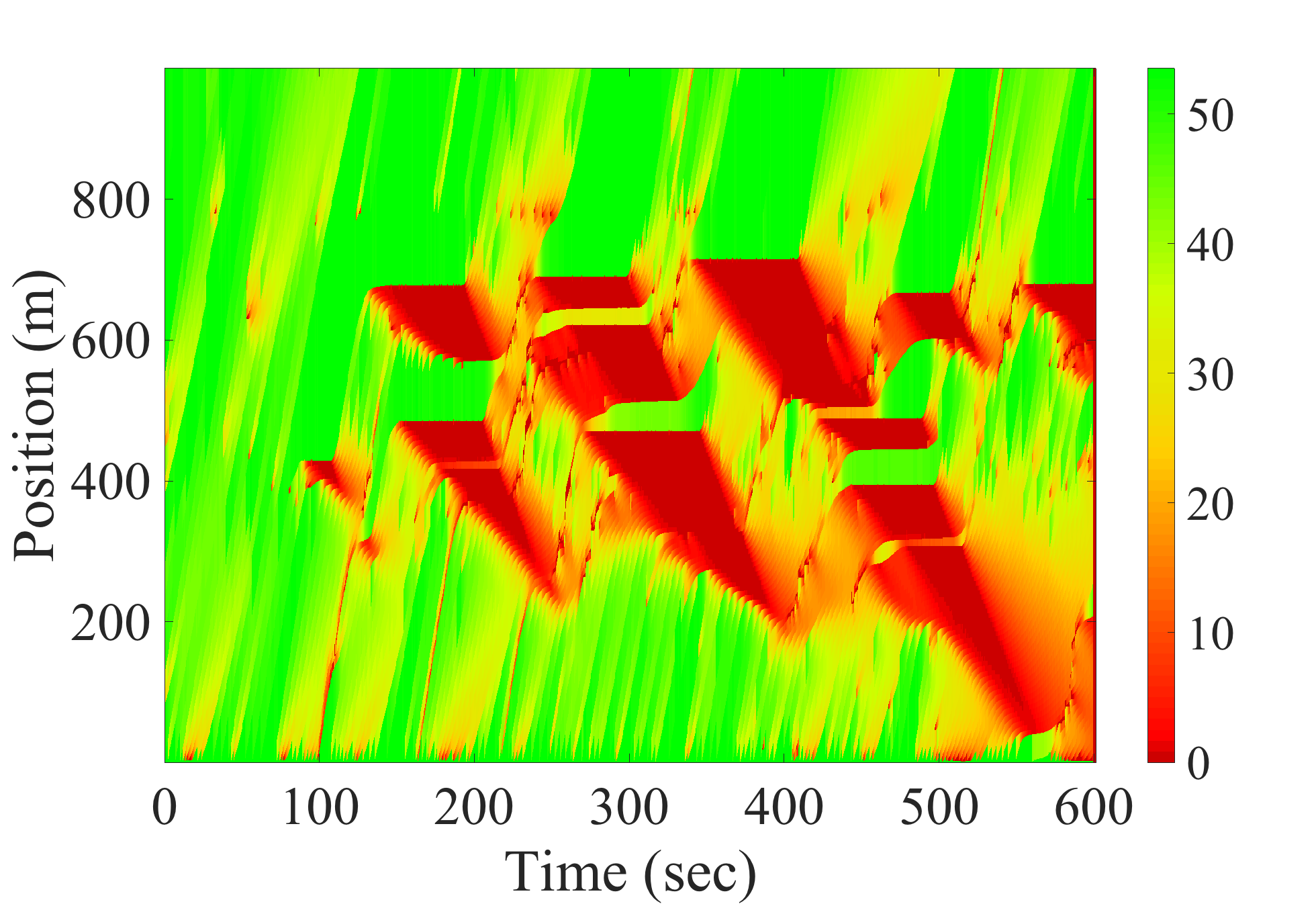}
		\includegraphics{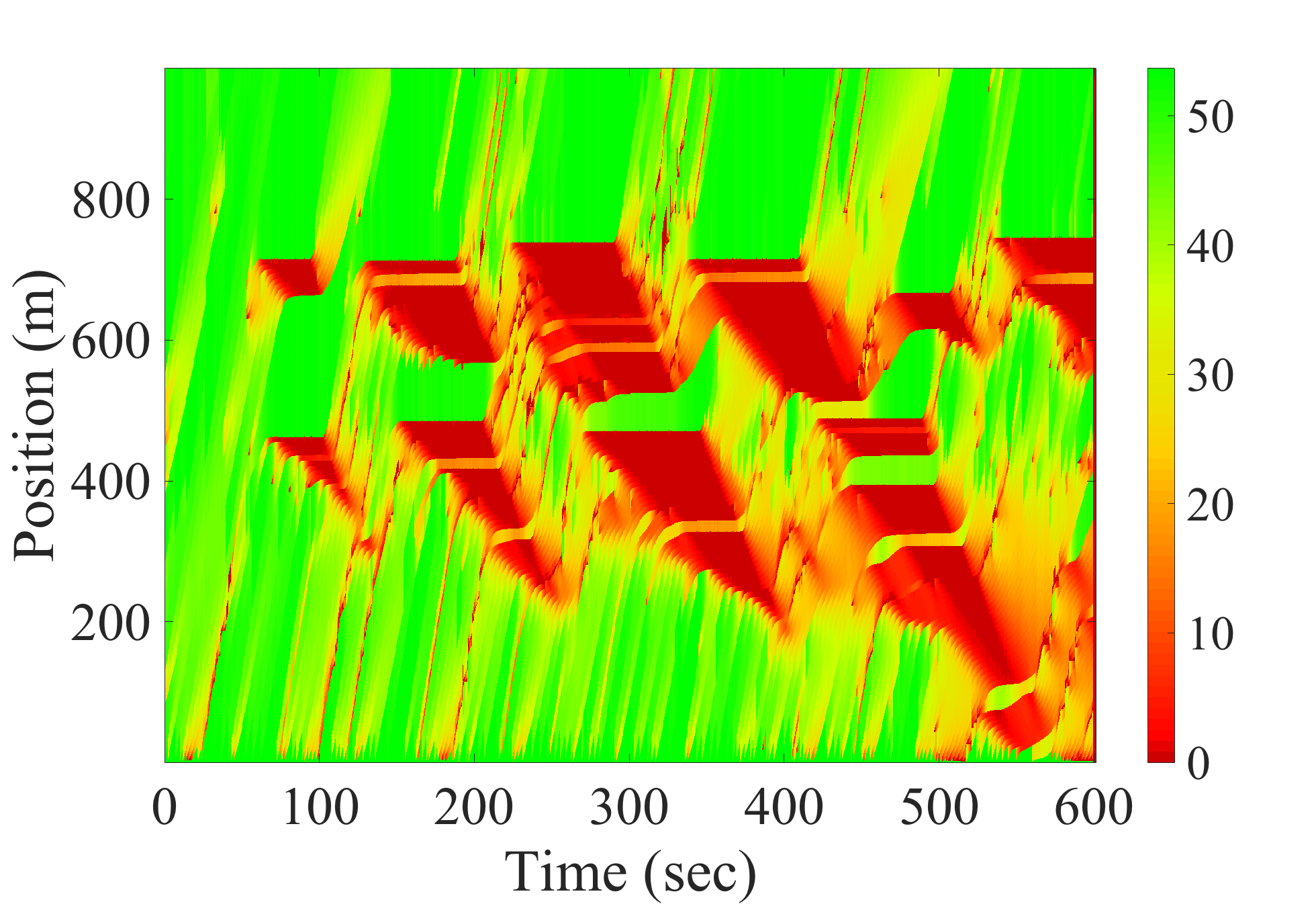}}
	
	(a) \hspace{2.0in} (b) 
	
	\resizebox{0.8\textwidth}{!}{%
		\includegraphics{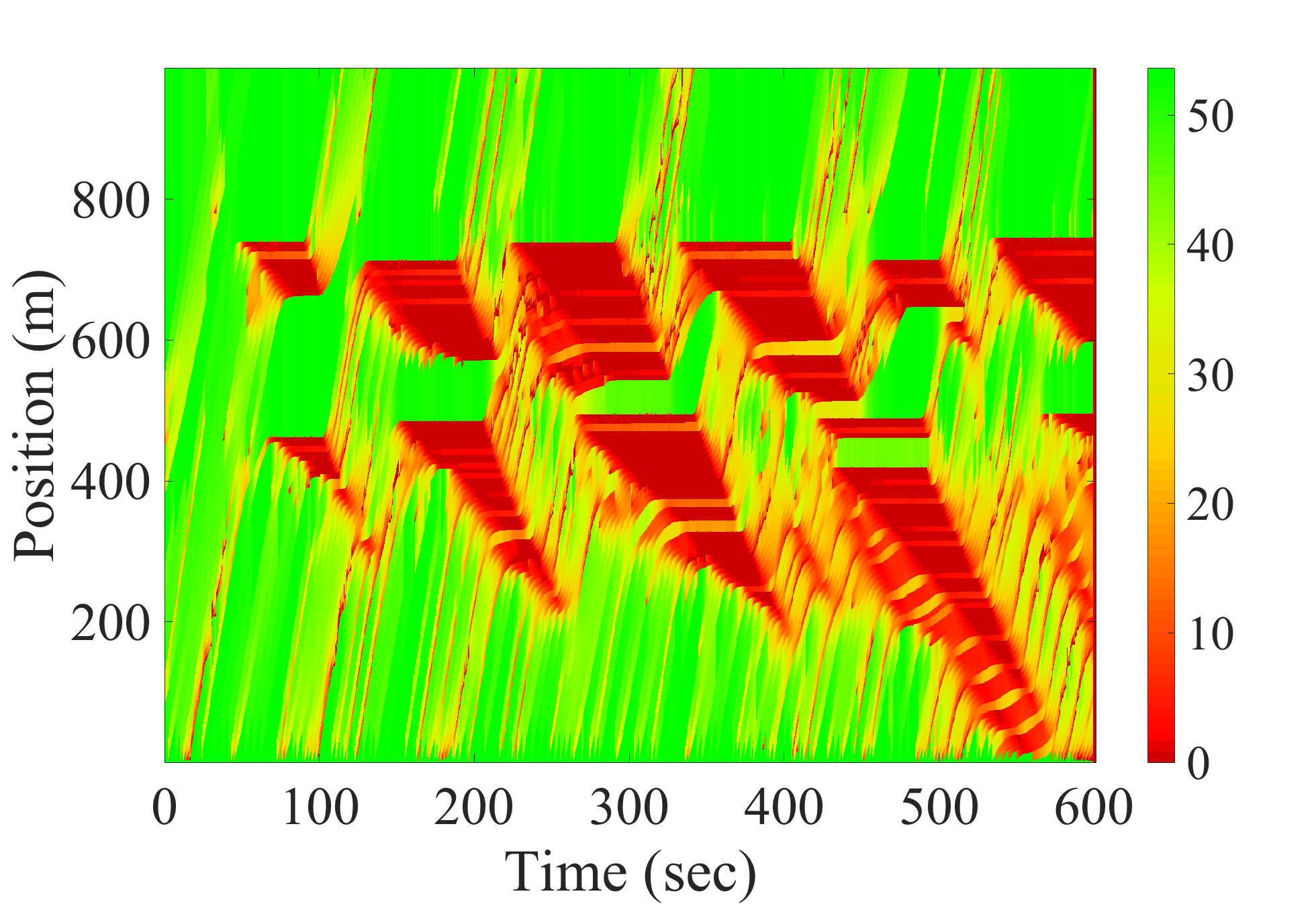}
		\includegraphics{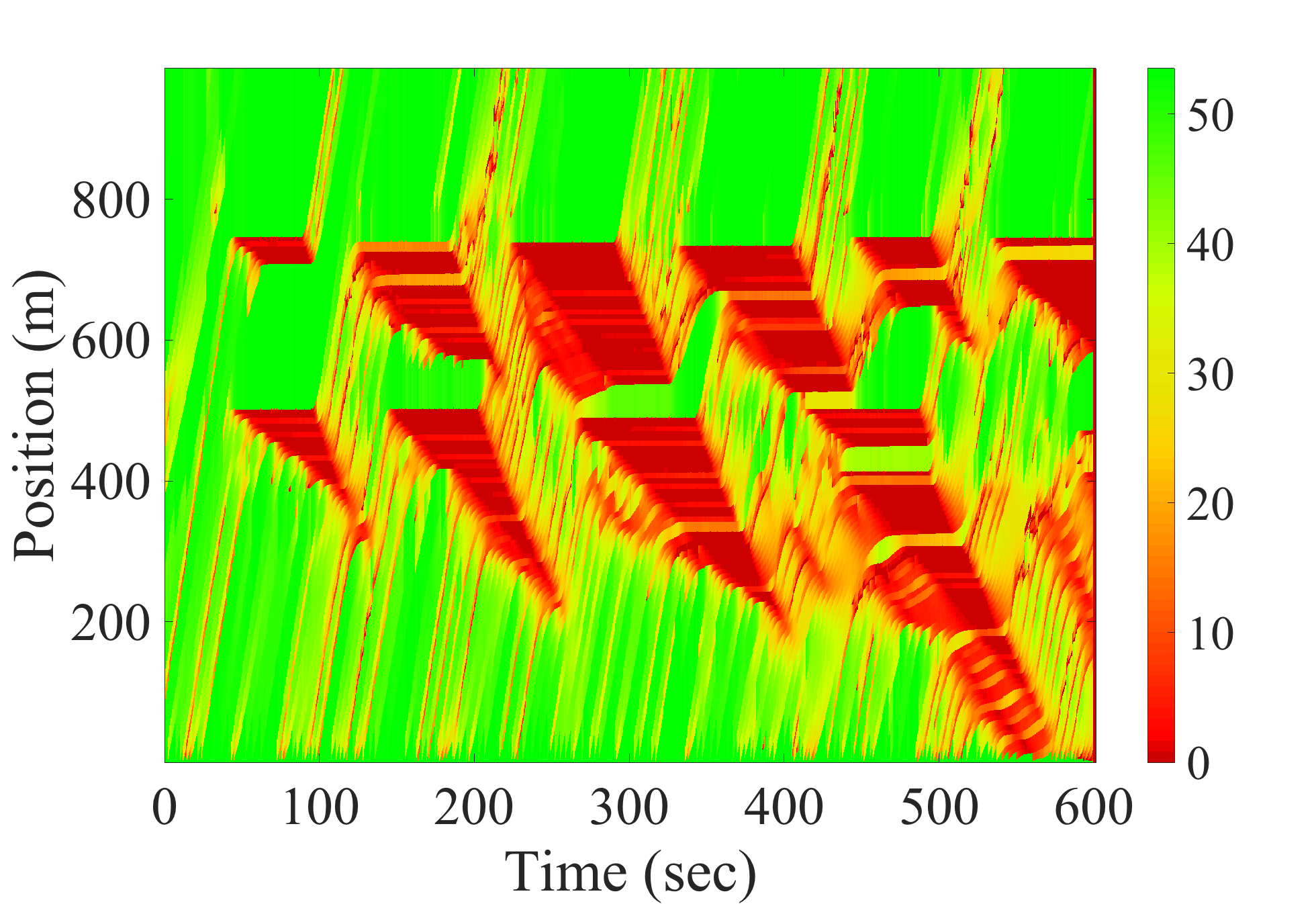}}
	
	(c) \hspace{2.0in} (d) 
	
	\resizebox{0.4\textwidth}{!}{%
		\includegraphics{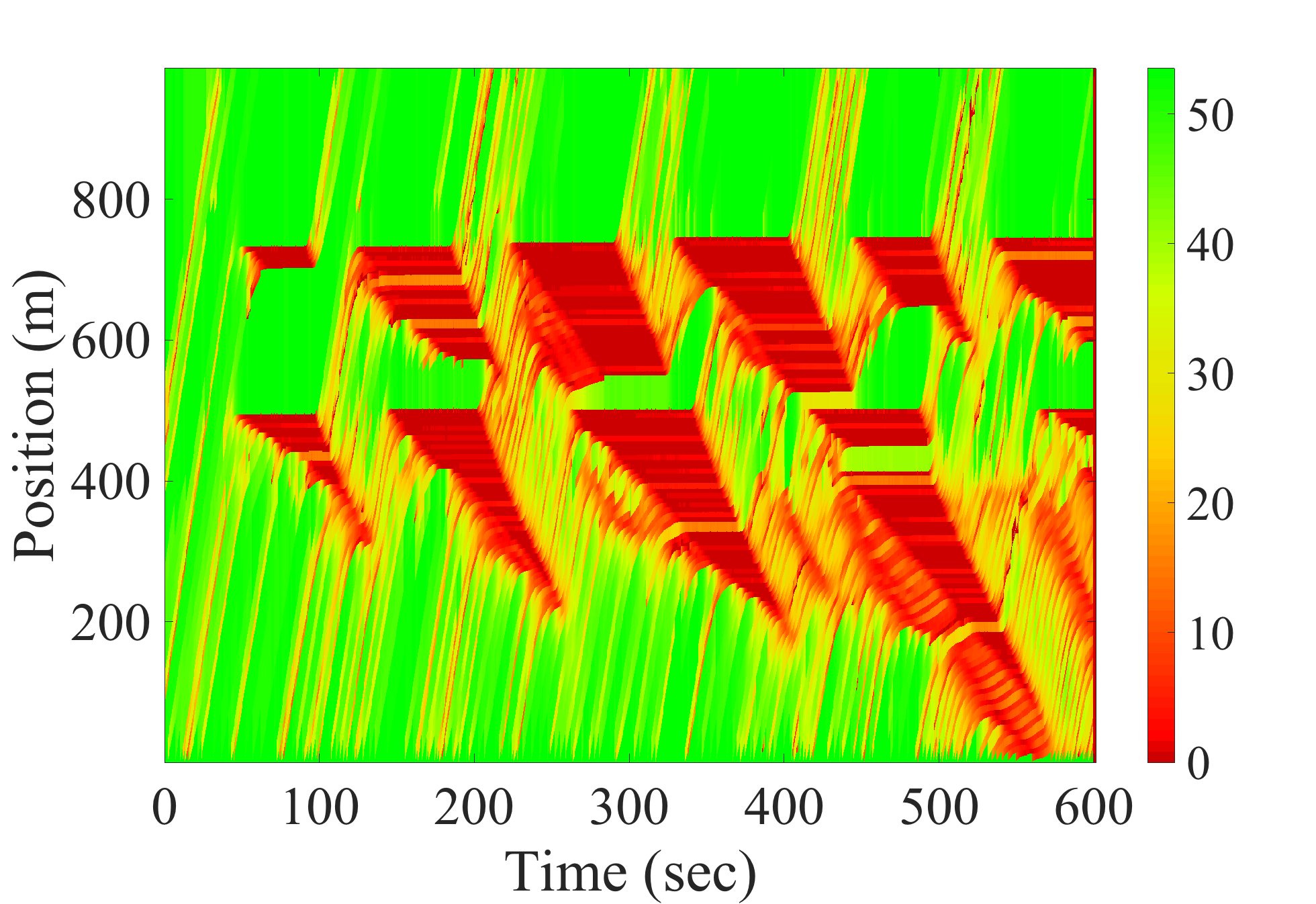}}
	
	(e)
	
	\caption{Estimated speed fields (in km/hr) with different penetration rates:  (a) 5\%, (b) 10\%, (c) 20\%, (d) 30\% and (e) 50\%.} \label{f_speed}
\end{figure}

\begin{figure}[h!]
	\centering
	\resizebox{0.65\textwidth}{!}{%
		\includegraphics{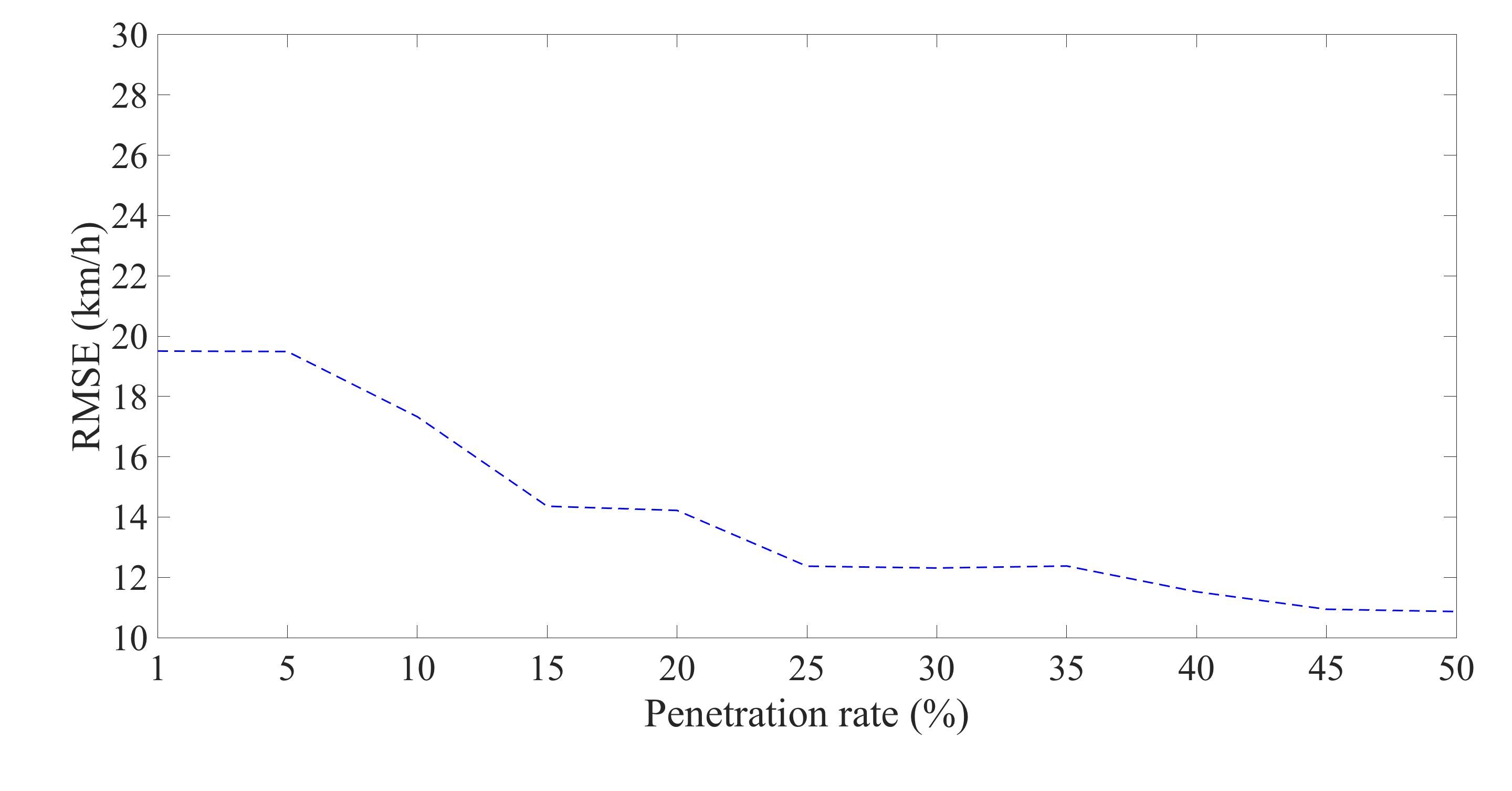}}
	\caption{RMSE in speed estimates vs. probe penetration rate.} \label{f_RMSE_SPEED_VISSIM}
\end{figure}

\autoref{f_den} and \autoref{f_speed} depict the estimated density and speed dynamics. There is clear improvement of estimation accuracy when the penetration rate increases from 5\% to 50\%.  The congestion (shockwave) and queue dynamics can be well captured when the penetration rate increases to 20\%.
As a summary of the estimation accuracy, \autoref{f_RMSE_SPEED_VISSIM} plots the RMSEs in speed for the varying penetration rates.

\subsection{Example 3: NGSIM data example}
\begin{figure}[h!]
	\centering
	\resizebox{0.65\textwidth}{!}{%
		\includegraphics{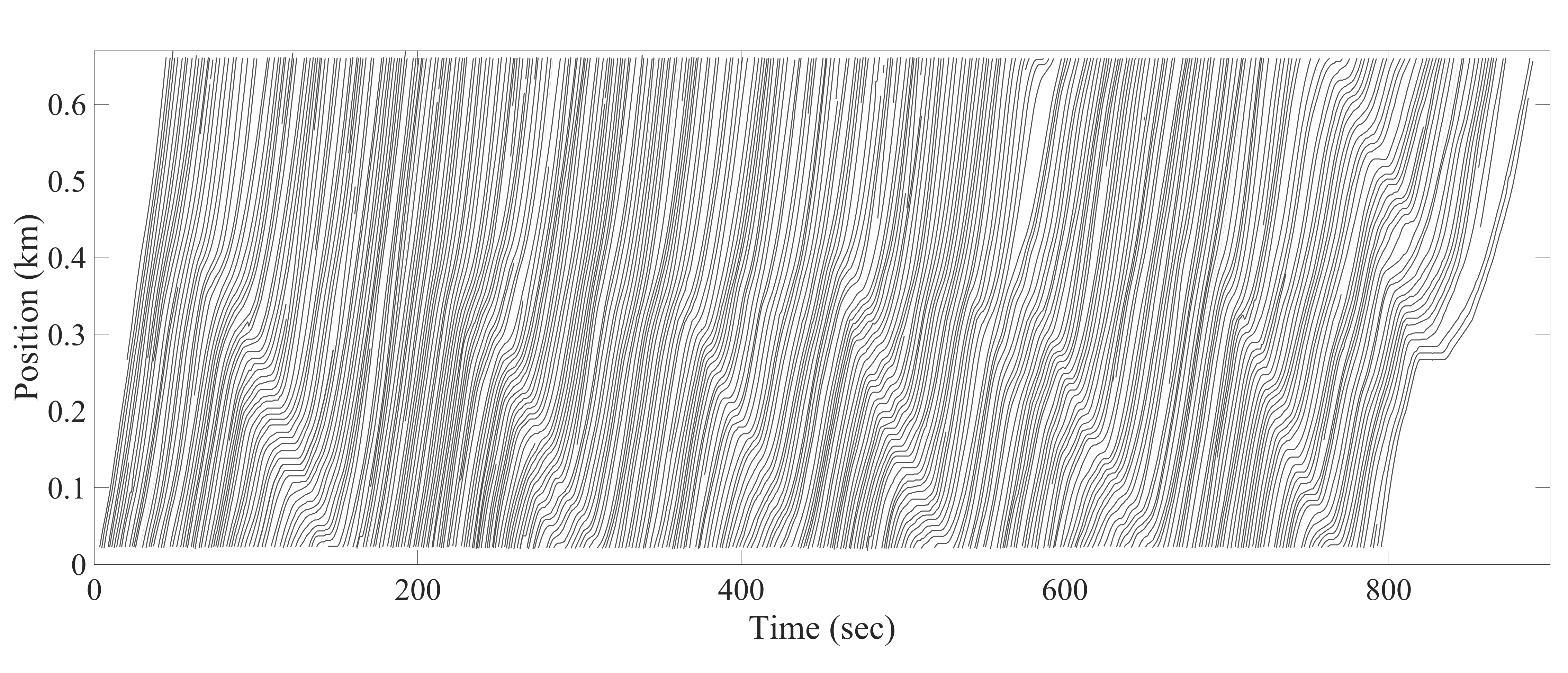}}
	\caption{Ground truth vehicle trajectories along I-80.} \label{f_i80traj}
\end{figure}

\begin{figure}[h!]
	\centering
	\resizebox{0.65\textwidth}{!}{%
		\includegraphics{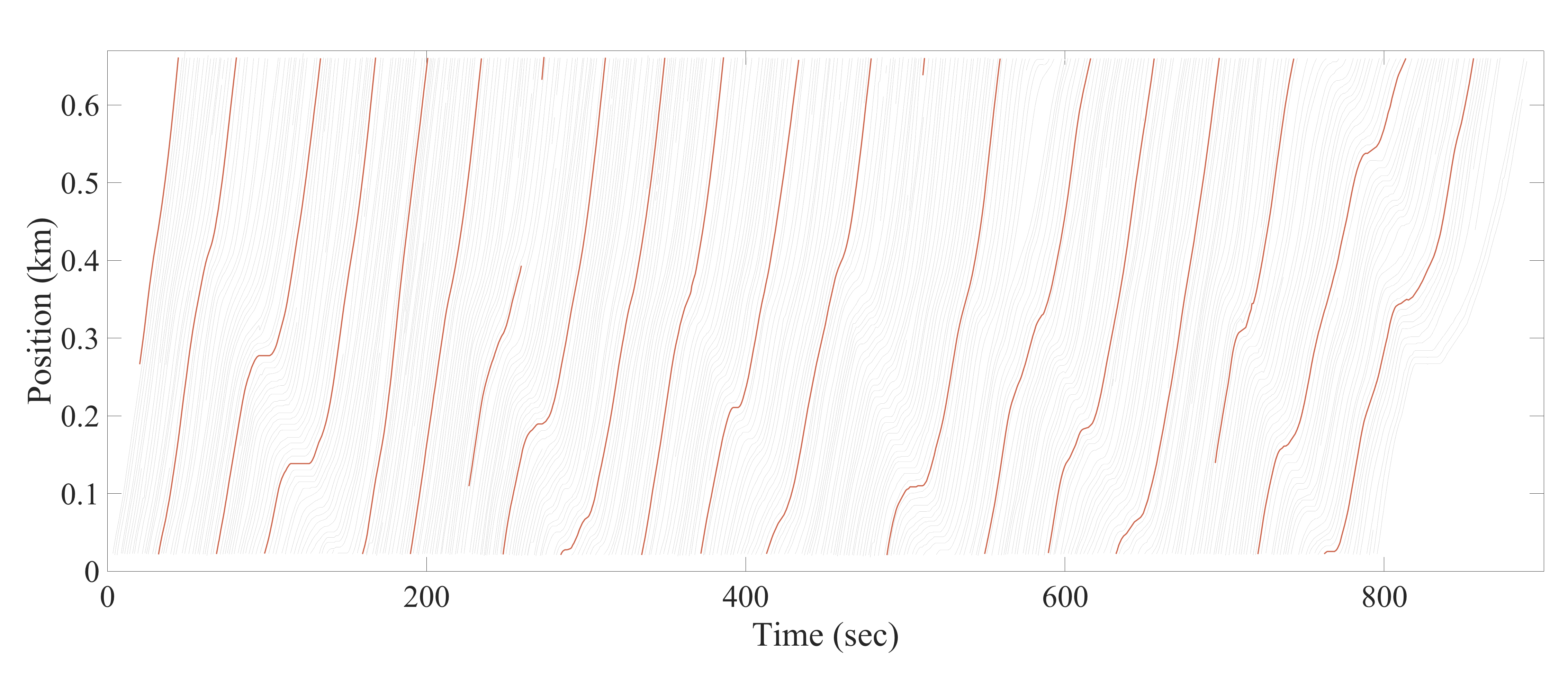}}
	\caption{Sampled vehicle trajectories data for 5\% penetration rate.} \label{f_i80trajSamples_5}
\end{figure}

\begin{figure}[h!]
	\centering
	\resizebox{0.65\textwidth}{!}{%
		\includegraphics{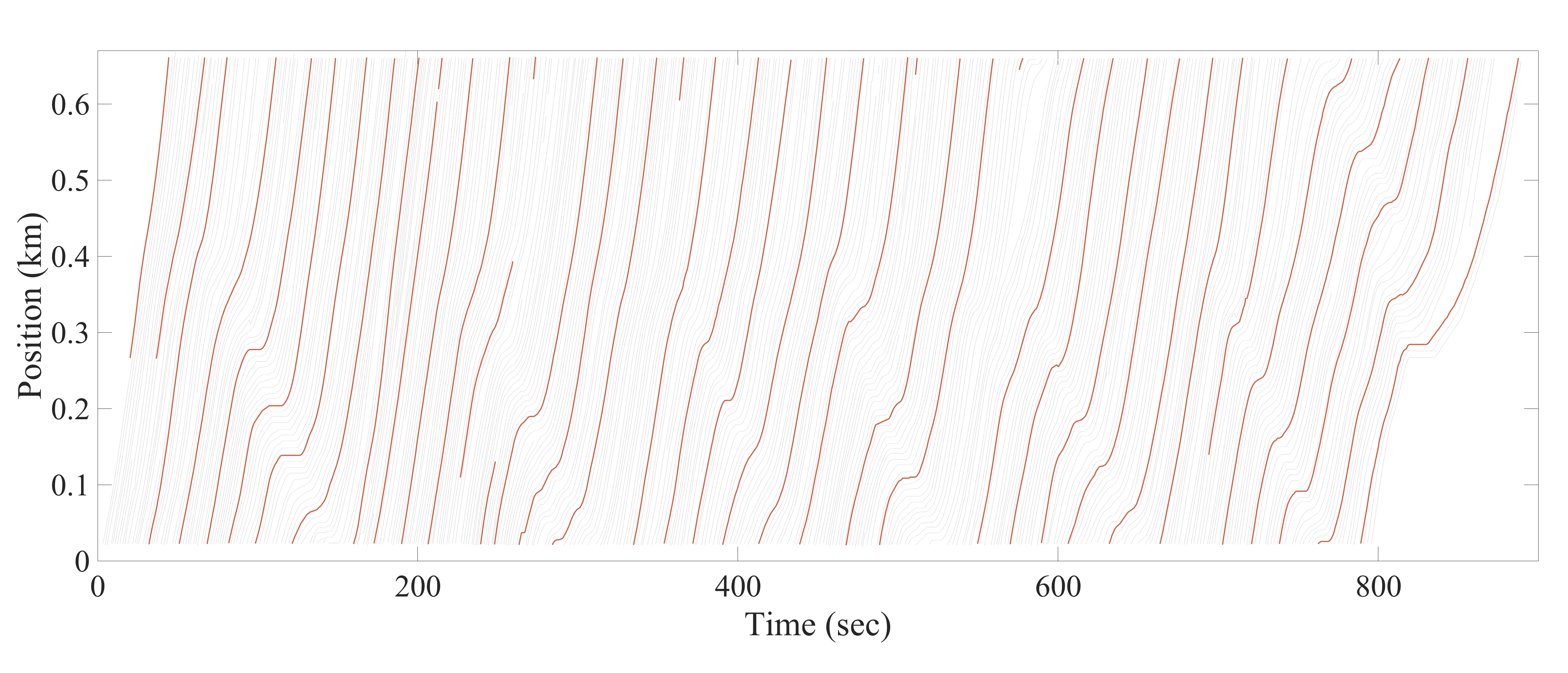}}
	\caption{Sampled vehicle trajectories data for 10\% penetration rate.} \label{f_i80trajSamples_10}
\end{figure}

\begin{figure}[h!]
	\centering
	\resizebox{0.65\textwidth}{!}{%
		\includegraphics{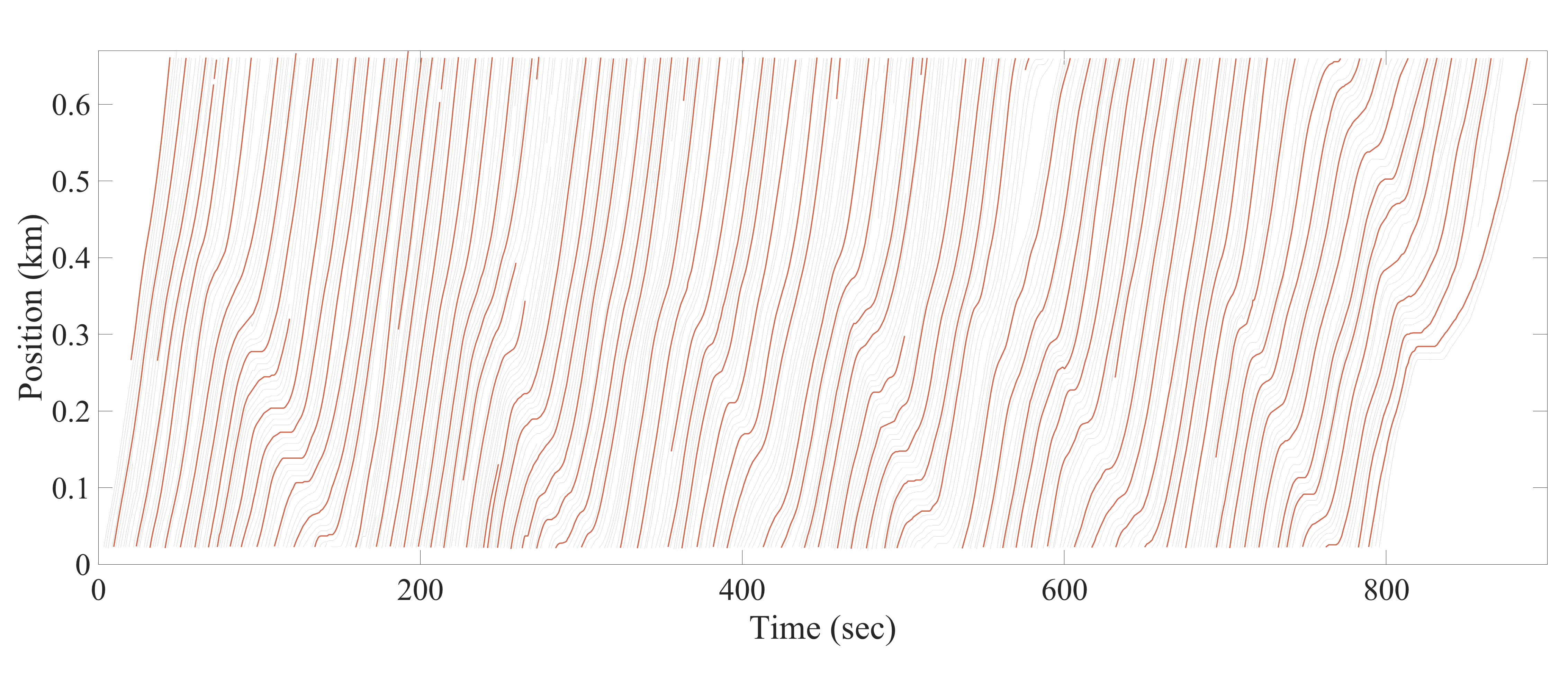}}
	\caption{Sampled vehicle trajectories data for 20\% penetration rate.} \label{f_i80trajSamples_20}
\end{figure}

\begin{figure}[h!]
	\centering
	\resizebox{0.65\textwidth}{!}{%
		\includegraphics{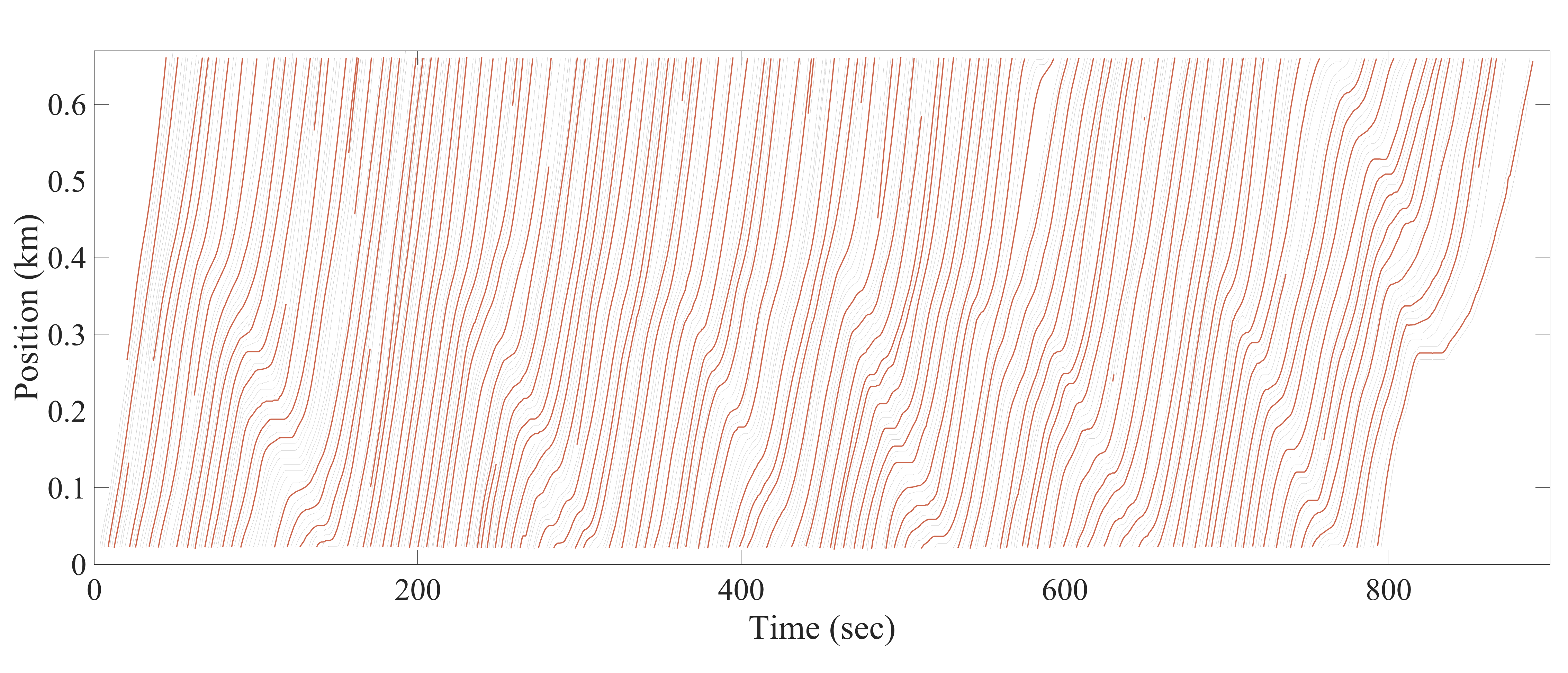}}
	\caption{Sampled vehicle trajectories data for 30\% penetration rate.} \label{f_i80trajSamples_30}
\end{figure}

\textbf{Data preparation}.  In order to test the performance of the proposed model and estimation approach with field data, we use the NGSIM data collected on eastbound I-80 in the San Francisco Bay area in Emeryville, CA. The study area is approximately 500 meters in length and we selected the vehicle trajectory data on the farthest left lane with time period of 15min between 4:00PM and 4:15PM on April 13, 2005 (see \autoref{f_i80traj}).  
\autoref{f_i80trajSamples_5} - \autoref{f_i80trajSamples_30} depict the sampled vehicle trajectory data for the four penetration rates. 

The parameters $v_{\ff}(\omega)$,  $d(\omega)$, and $c(\omega)$ are independent Beta distributed random variables with supports $[v_{\ff}^{\min},v_{\ff}^{\max}] = [95,105]$ km/hr, $[d^{\min},d^{\max}] = [5.9,7.7]$ meters, and $[c^{\min},c^{\max}] = [2340,3672]$ veh/hr.  These were fitted using ground truth data.  

\begin{figure}[h!]
	\centering
	\resizebox{0.9\textwidth}{!}{%
		\includegraphics{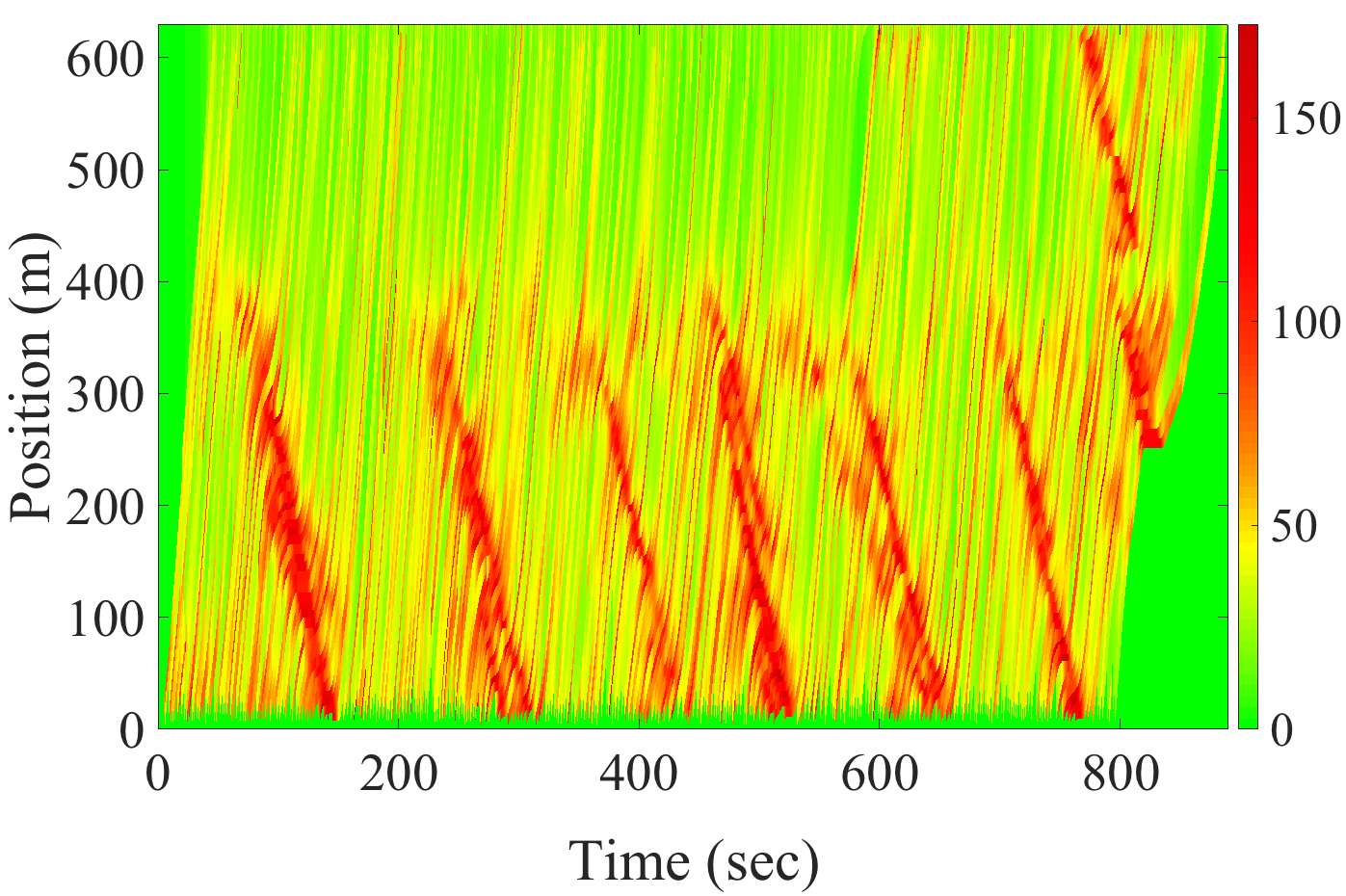}
		\includegraphics{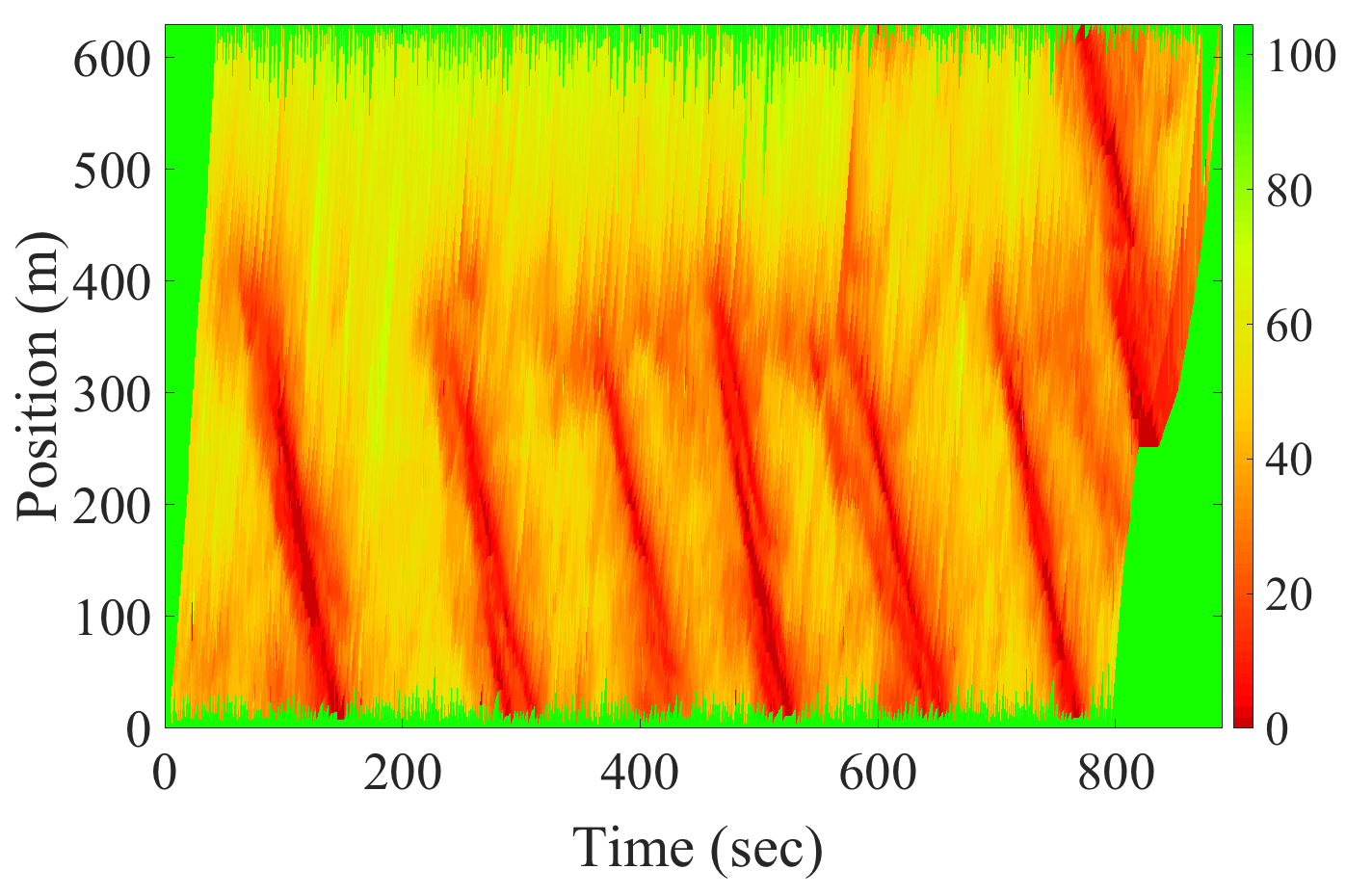}}
	
		(a) \hspace{2.2in} (b) 
	\caption{Ground truth from from NGSIM data: (a) densities (veh/km), (b) speeds (km/hr).} \label{f_gTruthNGSIM}
\end{figure}

\textbf{Estimation results}. We applied the proposed data assimilation approach to the second type of measurements as discussed in \autoref{SS_measurements} where we assume that spacing measurements are available, e.g., connected vehicles with their surrounding information available (both their leaders and followers) and spacings between the current vehicle and its immediate leader and follower can be measured (5 measurements).  \autoref{f_gTruthNGSIM} depicts the ground truth density and speed fields.
To see the performance of the new measurement equation and the impact of data availability on the uncertainty of traffic state estimation, we consider four penetration rates: 5\%, 10\%, 20\% and 30\%.

\begin{figure}[h!]
	\centering
	\small
	\resizebox{0.9\textwidth}{!}{%
		\includegraphics{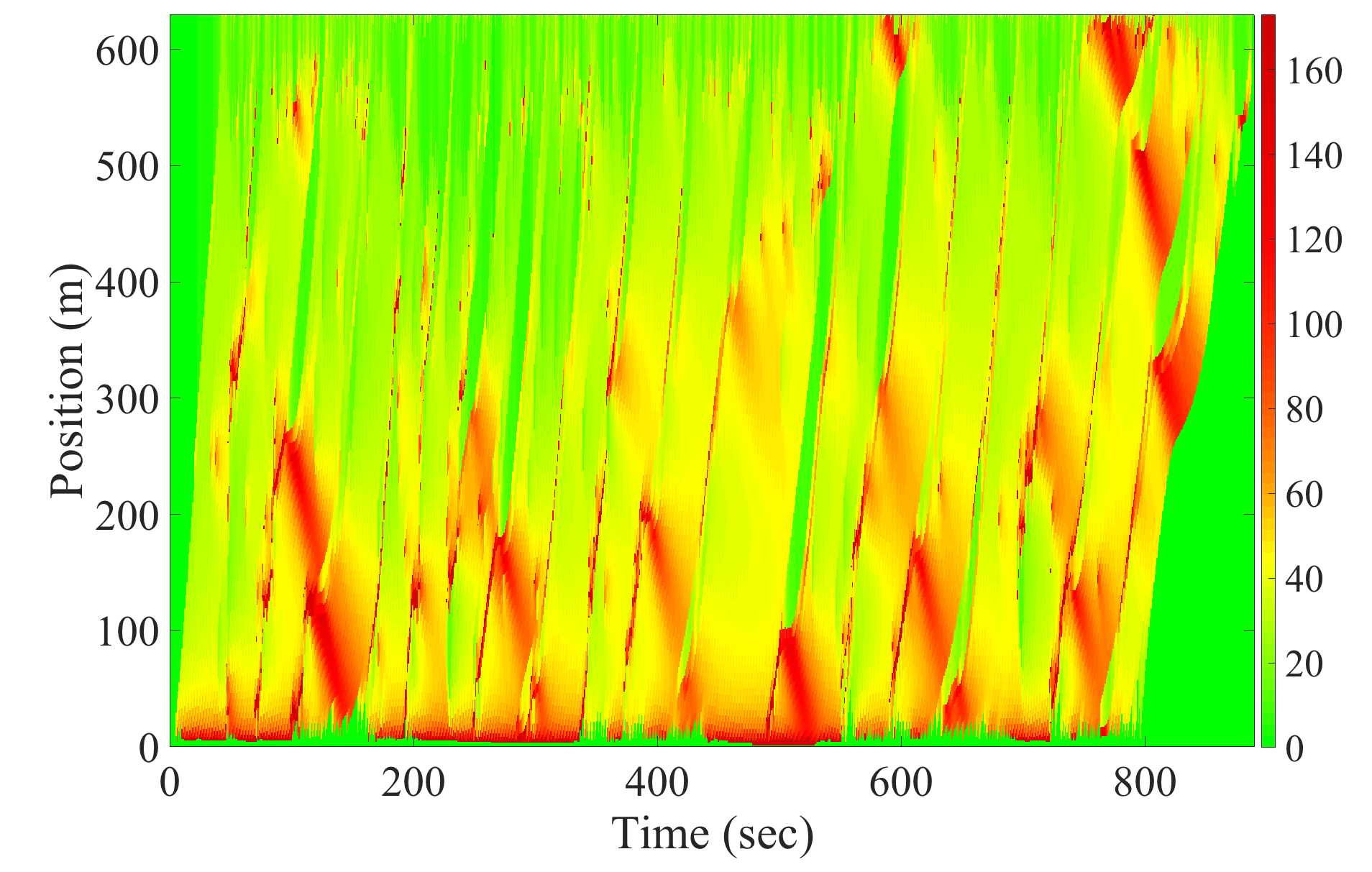}
		\includegraphics{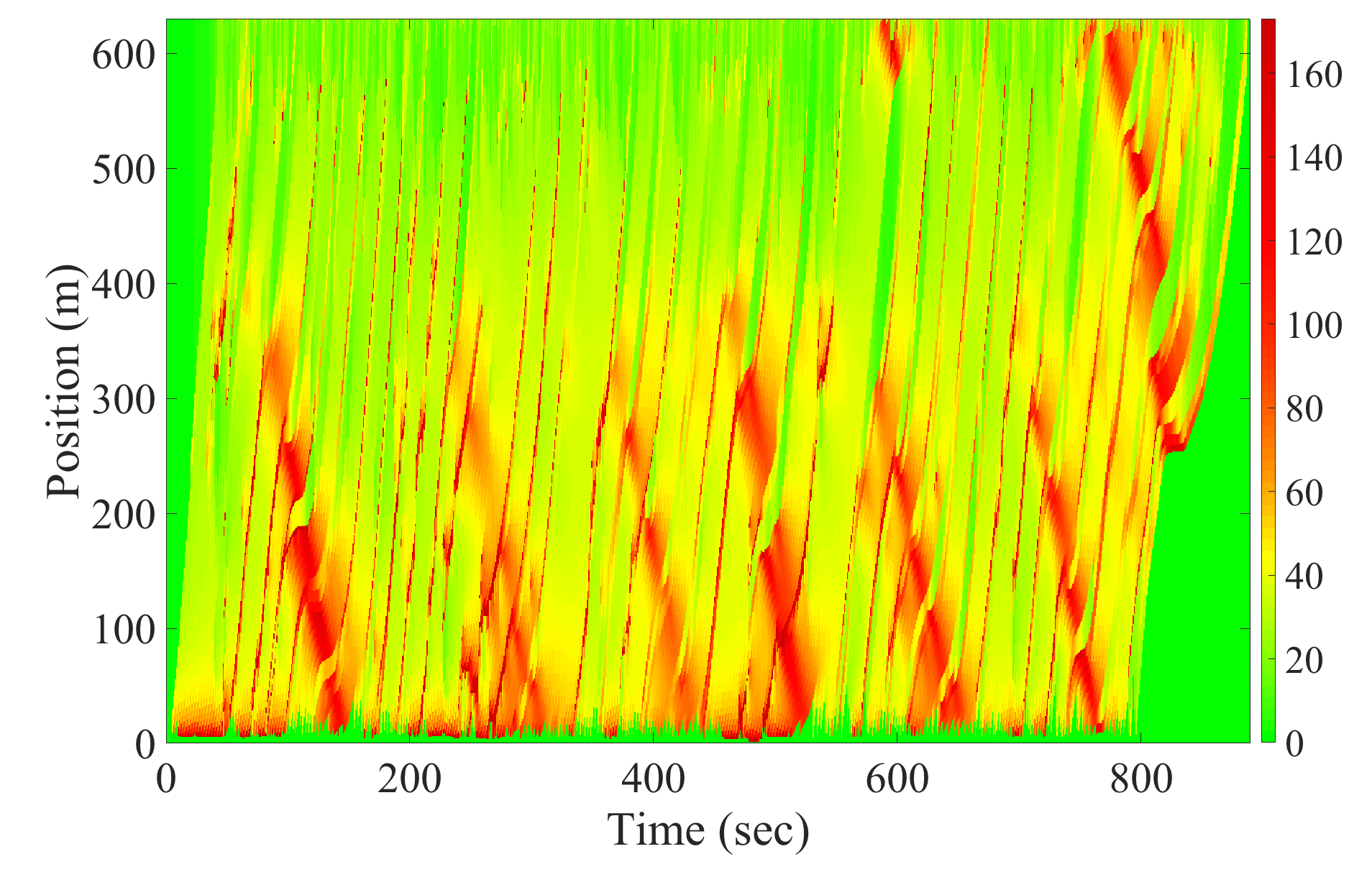}}
	
	(a) 5\% \hspace{1.7in} (b) 10\%
	
	\resizebox{0.9\textwidth}{!}{%
		\includegraphics{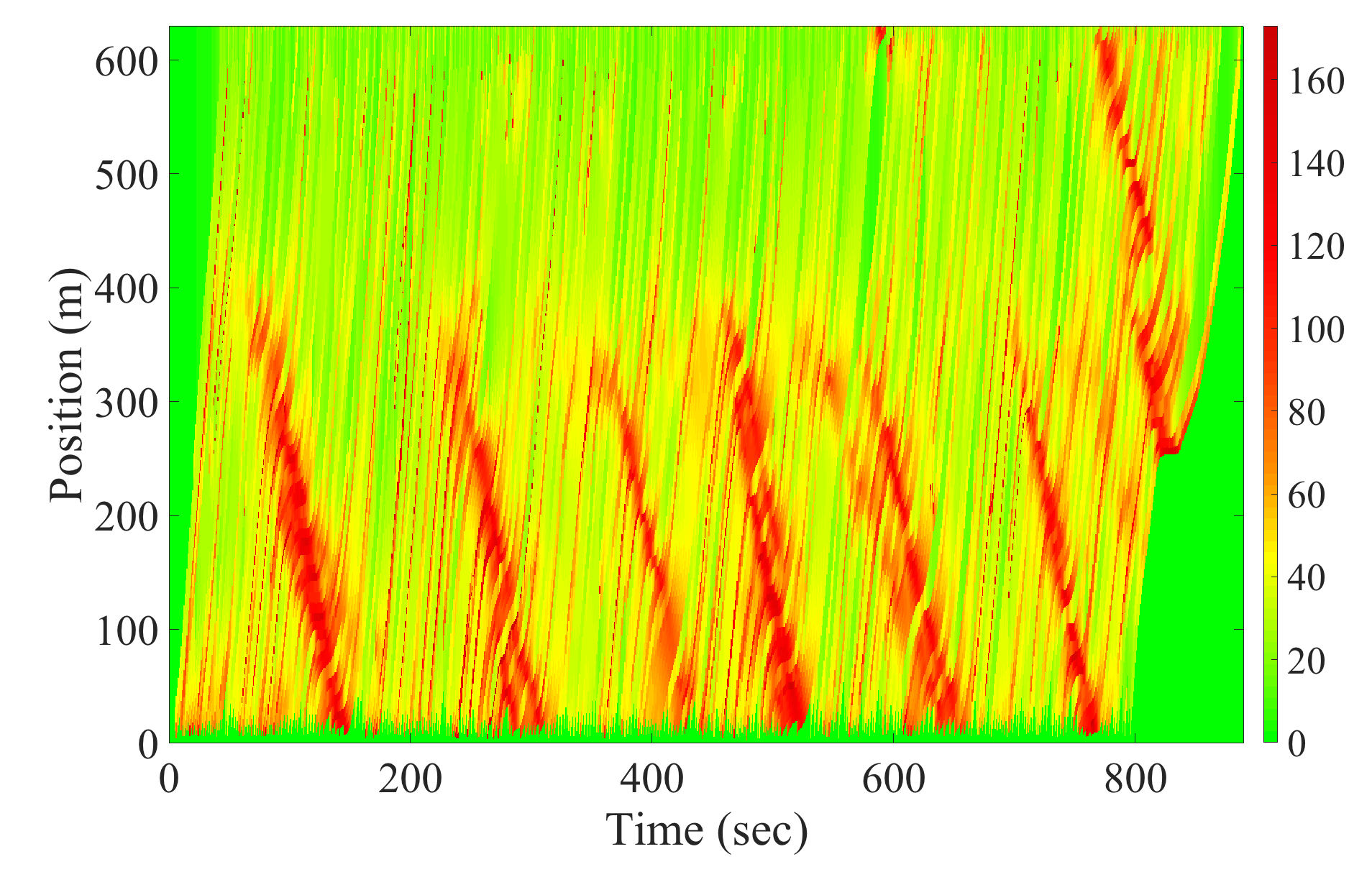}
		\includegraphics{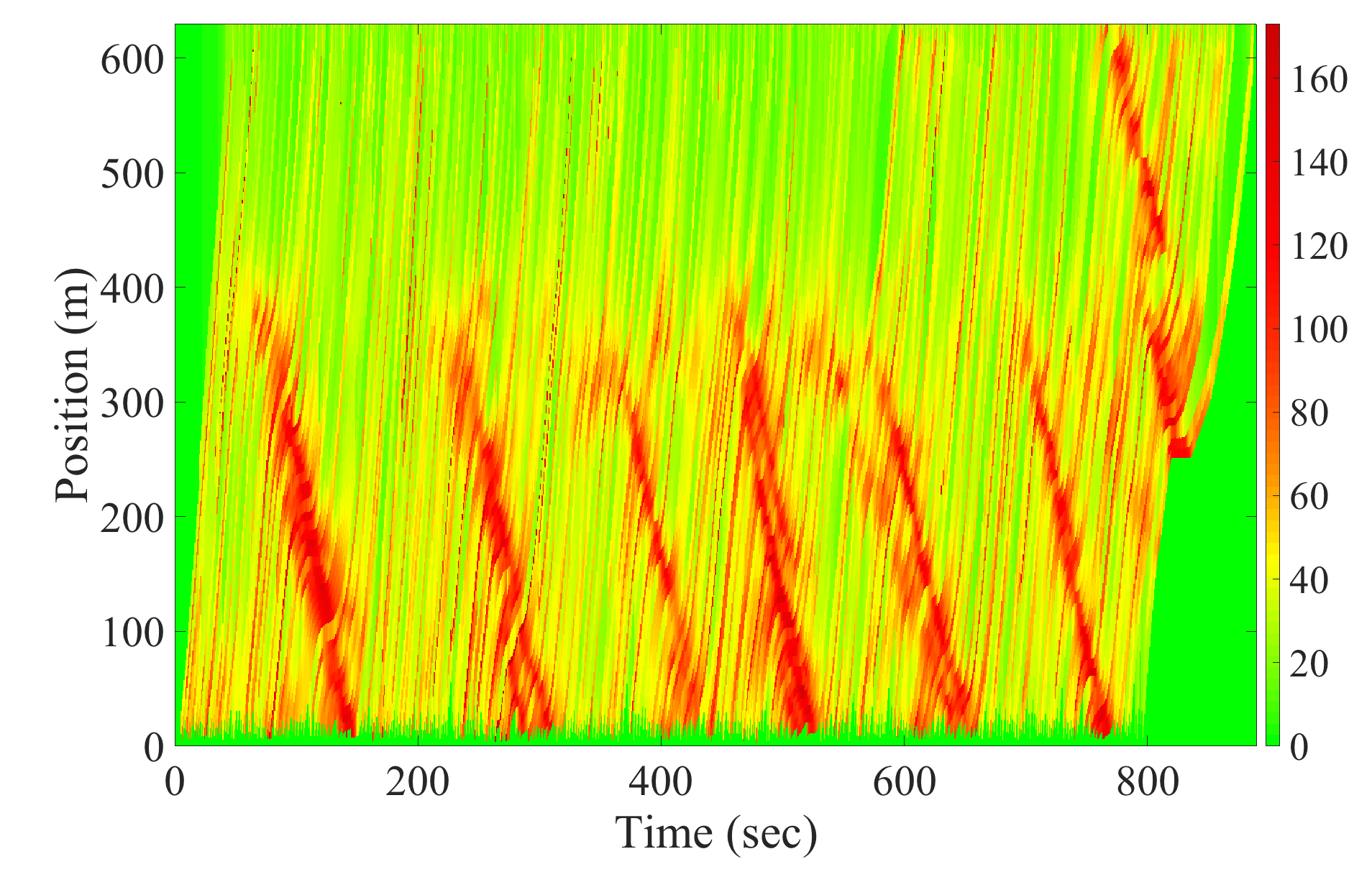}}
	
	(c) 20\% \hspace{1.6in} (d) 30\%
	
	\caption{Estimated densities (veh/km) at different penetration rates.} \label{f_densitiesNGSIM}
\end{figure}

\begin{figure}[h!]
	\centering
	\small
	\resizebox{0.9\textwidth}{!}{%
		\includegraphics{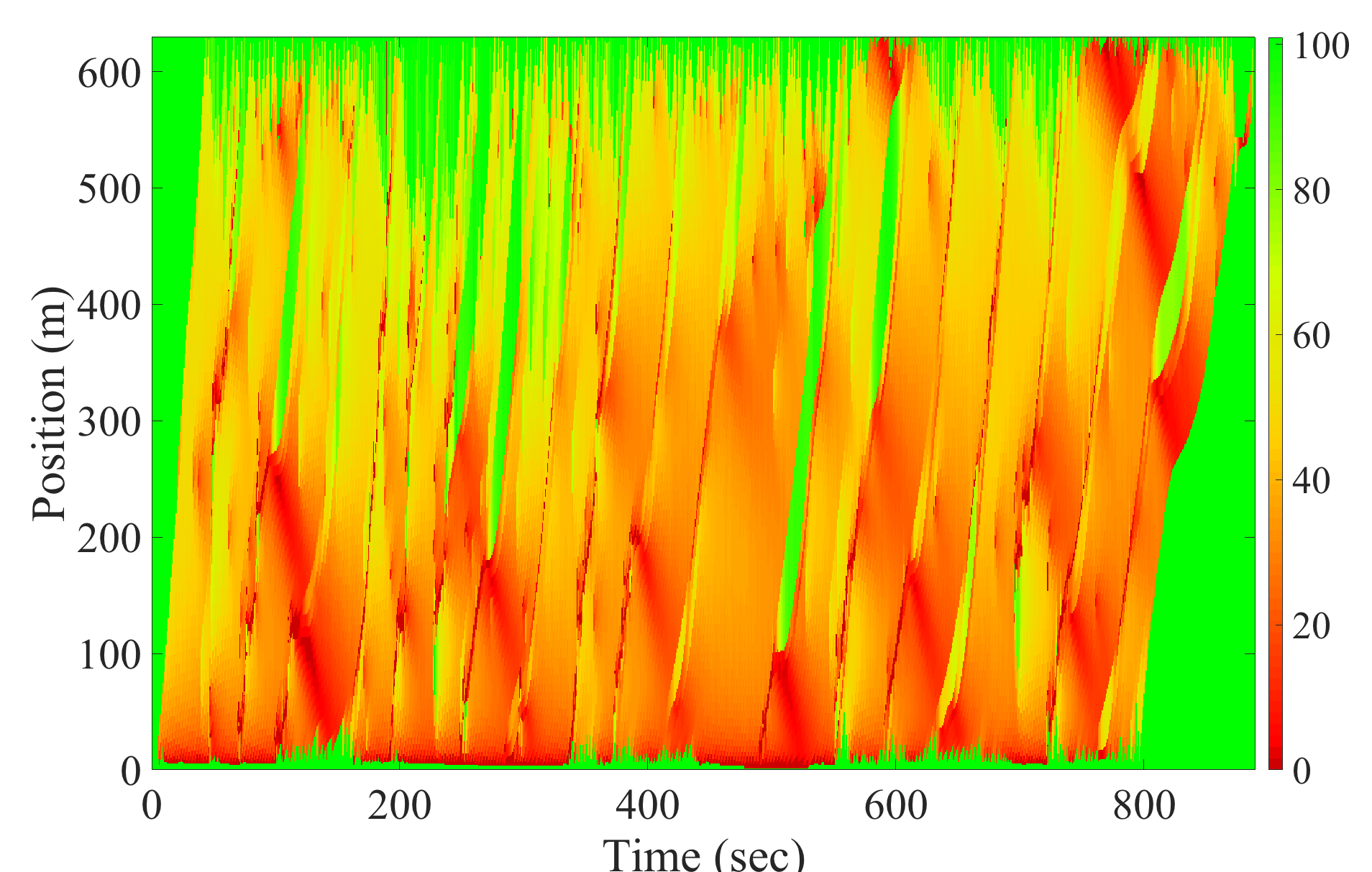}
		\includegraphics{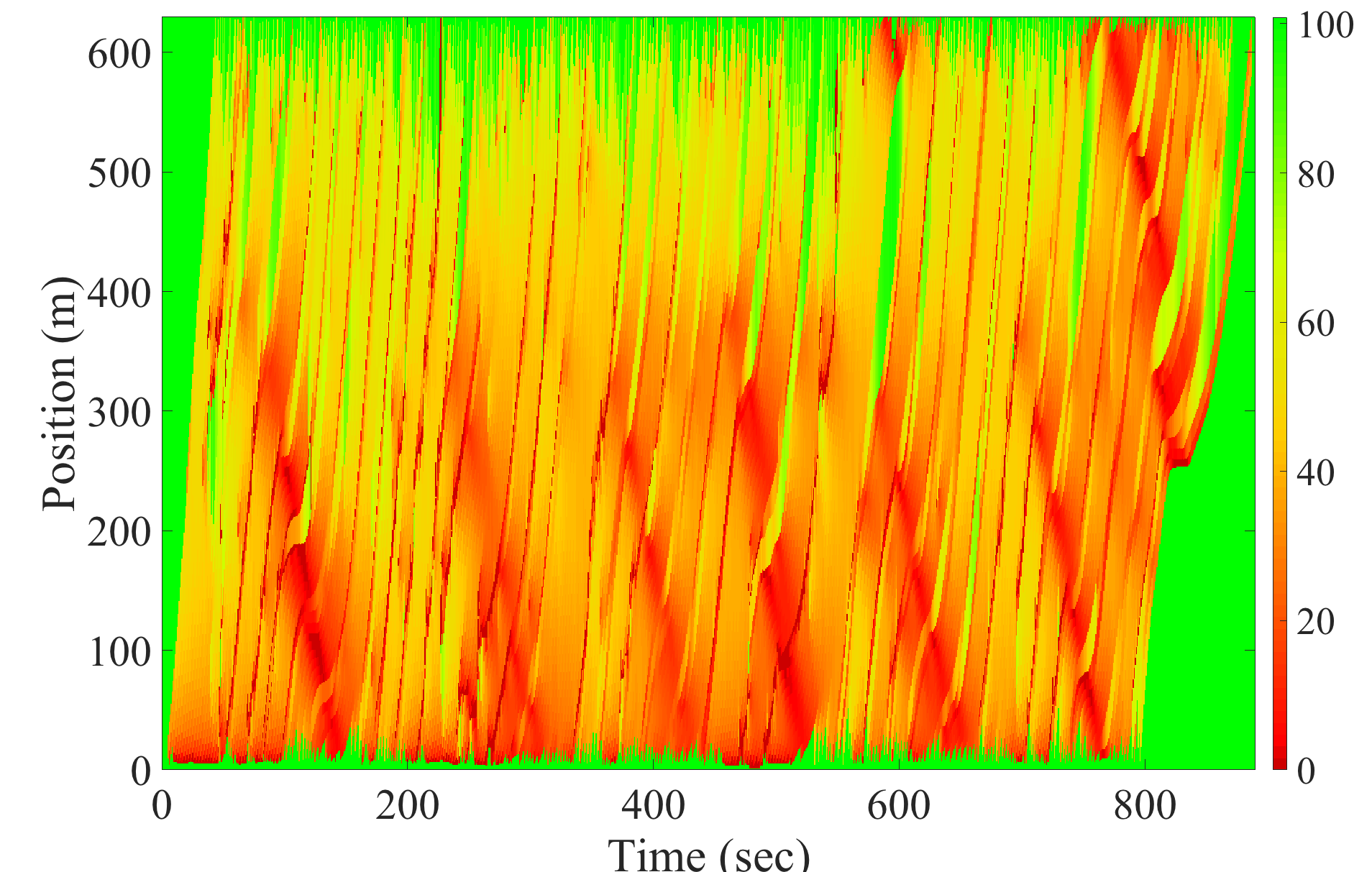}}
	
	(a) 5\% \hspace{1.7in} (b) 10\%
	
	\resizebox{0.9\textwidth}{!}{%
		\includegraphics{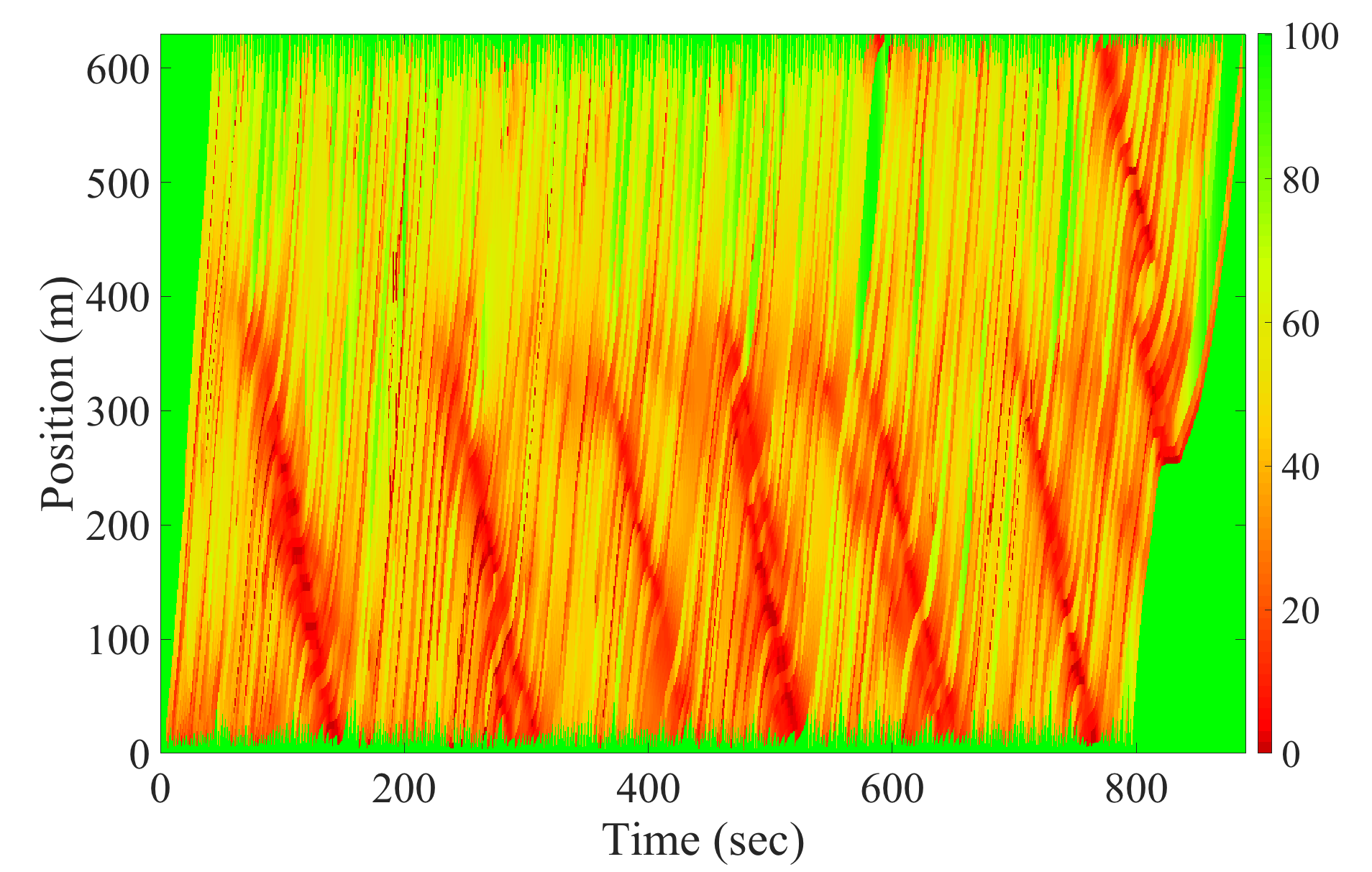}
		\includegraphics{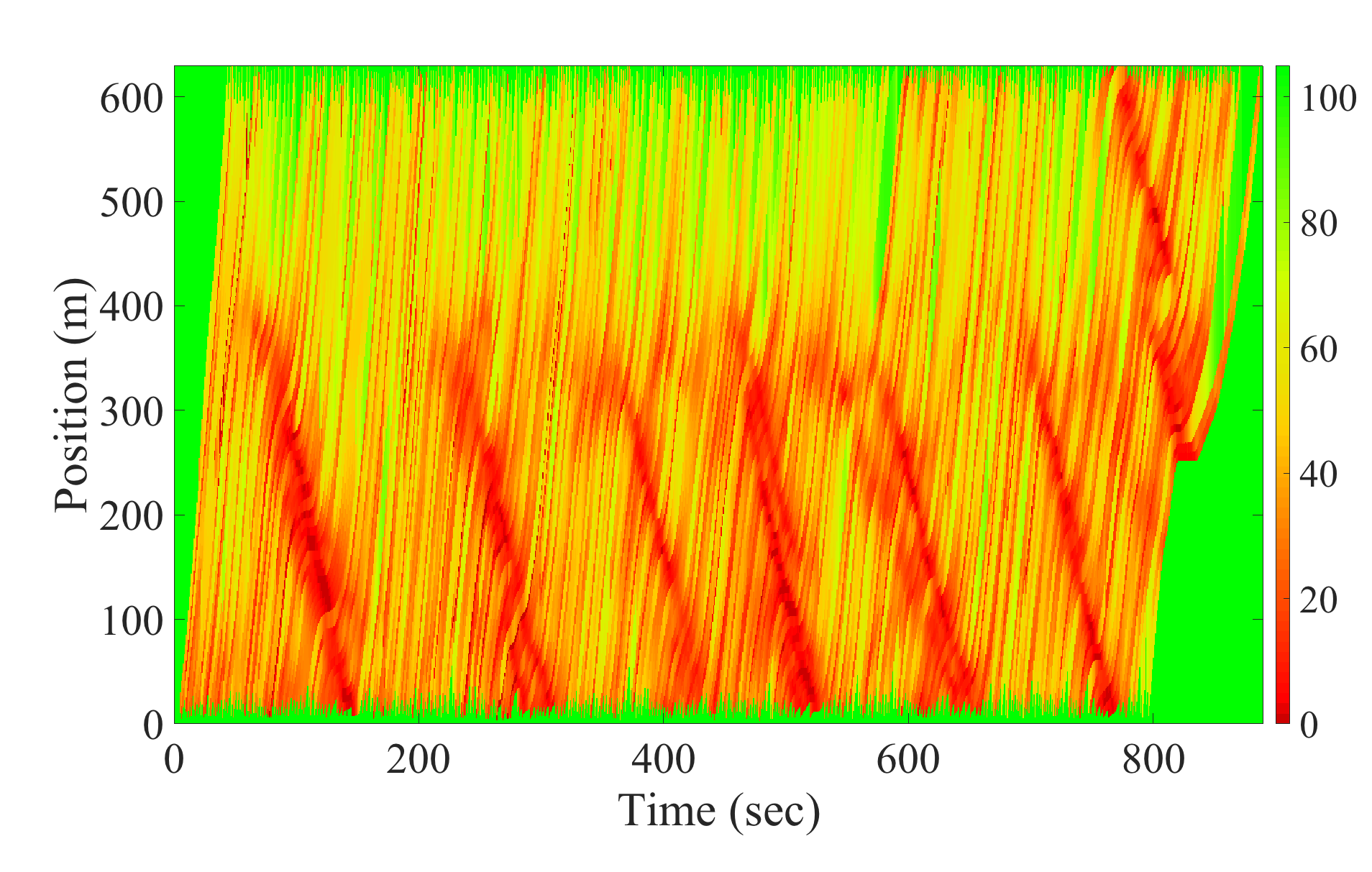}}
	
	(c) 20\% \hspace{1.6in} (d) 30\% 
	
	\caption{Estimated speeds (km/hr) at different penetration rates.} \label{f_speedsNGSIM}
\end{figure}

\autoref{f_densitiesNGSIM} depicts the estimated density and speed dynamics. When the penetration rate increases from 5\% to 10\%, there is a clear improvement of density estimation accuracy. The congestion shockwave can be well captured when the penetration rate increases to 10\%.  \autoref{f_speedsNGSIM} depicts the estimated speed dynamics. When the penetration rate increases to 20\%, the proposed approach is able to provide good estimation results in terms of congestion dynamics.

\begin{figure}[h!]
	\centering
	\resizebox{0.75\textwidth}{!}{%
		\includegraphics{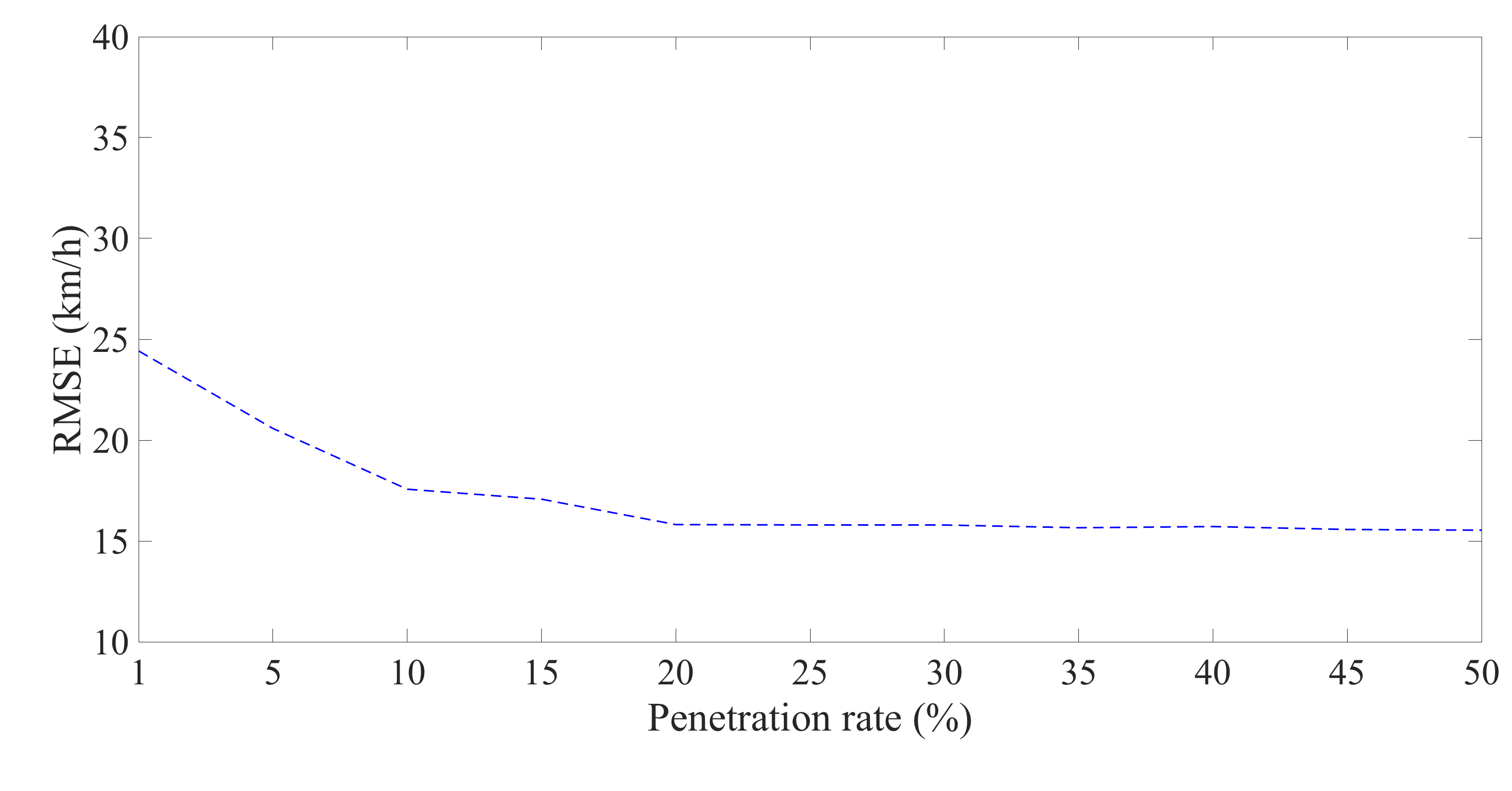}}
	\caption{RMSE in speed estimates vs. probe penetration rate.} \label{f_RMSE_SPEED_NGSIM}
\end{figure}

As a summary of the estimation accuracy, \autoref{f_RMSE_SPEED_NGSIM} plots the RMSEs in speed for the varying penetration rates.  The magnitudes of the RMSEs are comparable to most of the results in the literature, with the notable difference that our comparisons involve estimates of microscopic data.

\section{Conclusion}
\label{S:conc}
This paper proposes a second-order Gaussian approximation of a stochastic Lagrangian model. The Newell-Franklin speed-spacing relation is adopted and stochasticity is introduced by considering parametric uncertainties (by treating free flow speed, minimum safety distance and the slope of the speed-spacing relation when stationary as random variables). An ensemble averaged process is derived, which is consistent with traditional first-order Godunov schemes using a mean speed-spacing relation (as numerical flux), not a traditional equilibrium relation. The mean process is shown to converge to a conservation law in Lagrangian coordinates.  We then derive the covariance dynamics of the model by applying a Gaussian approximation. One important property of this covariance derivation is that it captures dependence of the covariance matrix on traffic state (namely, spacings) and it is much more tractable than covariance calculations in other data assimilation techniques such as particle filtering or ensemble Kalman filtering. We demonstrate the application of the proposed model in data assimilation. The system state dynamics are composed of both the mean and covariance dynamics. Two types of measurements are investigated: the first type of measurements includes positions and speeds of a sample set of vehicles. We assume that spacings are obtained as noisy measurements from measured speeds. The second type of measurements (with sensing and communication technologies) includes not only the positions and speeds of the probe vehicle, but also the positions and speeds of its surrounding vehicles (immediate leader and follower) and spacings between the immediate leader and the probe vehicle, as well as spacings between the probe vehicle and its immediate follower.

In order to demonstrate how this model works for traffic state estimation, we apply the model in a recursive estimation algorithm (Kalman-Bucy) using the derived ensemble mean and covariance dynamics and considering availability of vehicle trajectory data with different penetration rates. The numerical example shows that increasing the penetration rate (from 5\% to 50\%) results in a clear improvement in estimation accuracy both qualitatively and quantitatively. With the proposed stochastic model, the uncertainty of the estimation in terms of queue length is quantified (e.g., 95\% CI). In order to demonstrate the estimation capability of the proposed approach for more realistic settings, we provide two examples. In the first example we use vehicle trajectory data from a calibrated microscopic traffic simulation model of an arterial road in Ann Arbor City in Michigan. The estimation results show that traffic states, in terms of speed and density dynamics, can be well captured when the penetration rate increase to 20\%. In the second example we use NGSIM trajectory data along I-80 in the San Francisco Bay area in Emeryville, CA. The traffic state in terms of density dynamics can be well estimated with 10\% penetration rate.  The investigation of the speed estimation error in terms of RMSEs for different penetration rates illustrate that there is strong improvement as the penetration rates increase from 5\% to 15\%.



This study focuses on the derivation of the stochastic model and its application in traffic state estimation.  Future research could be carried out in various directions.  
For the model itself, we assume human driving vehicles which can be extended to consider mixed traffic flow conditions in which both human driving vehicles and autonomous vehicles exist. From an application point of view, the proposed model has the potential to be applied for the real-time traffic state estimation and traffic control considering reliability.


	
\section*{Acknowledgments}
\label{Ack}
This work was funded in part by the C$^2$SMART Center, a Tier 1 USDOT University Transportation Center, and National Science Foundation of China under project code NSFC 61673321.

\appendix
\makeatletter
\def\@seccntformat#1{\csname Pref@#1\endcsname \csname the#1\endcsname\quad}
\makeatother
{
	\numberwithin{equation}{section}
	\setcounter{equation}{0}
	
	\section{Algorithm for simulating a single sample path of the process} \label{sec:A}
	The algorithm below illustrates how to simulate a single sample path of the stochastic (Lagrangian) process.  Essentially, the algorithm randomly generates a realization of the parameters, one realization per vehicle, and then simulates a heterogeneous driving environment.
	

	\renewcommand{\algorithmicrequire}{\textbf{Initialize:}}
	\renewcommand{\algorithmicensure}{\textbf{Iterate:}} 
	\begin{algorithm}[h!]
		\caption{Simulating a single sample path of the process}
	\label{alg:singleSamplePath}
		\begin{algorithmic}[1]
			\REQUIRE $N$, $T$, $x_0(\cdot)$, $v_0(\cdot)$, $\{s_n(0)\}_{n=1}^N$, $F_{\theta}$, $k \mapsfrom 0$
			\FOR { $n \mapsfrom  1$ \TO $n\le N$}
			\STATE $U_1,U_2,U_3,~\sim$ Uniform[0,1]$^3$
			\STATE $(v_{n,\ff},d_n,c_n) \mapsfrom F_{\theta}^{-1}(U_1,U_2,U_3)$
			\ENDFOR	
			\STATE $\Delta t \mapsfrom  \underset{1 \le n \le N}{\min} \dfrac{\Delta n}{c_n}$
			\ENSURE
			\WHILE {$k \le \lfloor T/ \Delta t \rfloor$}
			\STATE $k \mapsfrom  k + 1$
			\FOR { $n  \mapsfrom 1$ \TO $n<N$ }
			\STATE $v_n(k \Delta t) \mapsfrom V_n\big( s_n(k \Delta t) \big)$
			\STATE $s_n(k\Delta t) \mapsfrom  s_n((k-1)\Delta t) + \Delta t \Big( v_{n-1}((k-1) \Delta t) - v_n((k-1) \Delta t)\Big)$
			\STATE $x_n(k\Delta t) \mapsfrom x_{n-1}(k\Delta t) - s_n(k\Delta t)$
			\ENDFOR
			\ENDWHILE
		\end{algorithmic}
	\end{algorithm}
}
	
	
	
	
	\bibliographystyle{plain}
	\bibliography{refs}

\end{document}